\documentclass[aps,twocolumn,prx,amsmath,nofootinbib,longbibliography,amssymb,superscriptaddress,10pt]{revtex4-2}
\usepackage[english]{babel}
\usepackage{graphicx}
\usepackage{ulem}
\usepackage{soul}
\usepackage{float} %added by Evyatar
\usepackage[colorlinks, linkcolor=red , citecolor= blue,breaklinks=true]{hyperref}

\usepackage{verbatim}
\usepackage{esint}
\usepackage{marginnote}
\newcommand{\RNum}[1]{\uppercase\expandafter{\romannumeral #1\relax}}

\def \beq {\begin{eqnarray}}
\def \eeq {\end{eqnarray}}

\begin{document}

\title{Strongly coupled phonon fluid and Goldstone modes in an anharmonic quantum solid: transport and chaos}

\author{Evyatar Tulipman}
\affiliation{Department of Condensed Matter Physics, Weizmann Institute of Science, Rehovot, 76100, Israel}
\author{Erez Berg}
\affiliation{Department of Condensed Matter Physics, Weizmann Institute of Science, Rehovot, 76100, Israel}
\begin{abstract}
We study properties of thermal transport and quantum many-body chaos in a lattice model with $N\to\infty$  oscillators per site, coupled by strong anharmonic terms. We first consider a model with only optical phonons. We find that the thermal diffusivity $D_{\rm th}$ and chaos diffusivity $D_L$ (defined as $D_L = v_B^2/ \lambda_L$, where $v_B$ and $\lambda_L$ are the butterfly velocity and the scrambling rate, respectively) satisfy $D_{\rm th} \approx \gamma D_L$ with $\gamma\gtrsim 1$. At intermediate temperatures, the model exhibits a ``quantum phonon fluid'' regime, where both diffusivities satisfy $D^{-1} \propto T$, and the thermal relaxation time and inverse scrambling rate are of the order the of Planckian timescale $\hbar/k_B T$. We then introduce acoustic phonons to the model and study their effect on transport and chaos. The long-wavelength acoustic modes remain long-lived even when the system is strongly coupled, due to Goldstone's theorem. As a result, for $d=1,2$, we find that $D_{\rm th}/D_L\to \infty$, while for $d=3$, $D_{\rm th}$ and $D_{L}$ remain comparable.    

\end{abstract}
\maketitle

\section{Introduction}
\label{intro}

Quantum many-body chaos has recently been put forward as a useful paradigm in the strive to understand thermalization and transport properties of strongly correlated systems. Namely, by relating transport coefficients to current relaxation times, a recurrent theme has emerged, in which the relaxation time satisfies $\tau \sim \tau_{\rm Pl}$, where $\tau_{\rm Pl} = \hbar / k_B T$ is the so-called Planckian timescale \cite{damle_nonzero-temperature_1997,zaanen_why_2004,sachdev_quantum_2011,zaanen_planckian_2019,hartnoll_planckian_2021}. The most canonical instances are `strange metals,' where the electrical current relaxation time exhibits striking universality with $\tau = \alpha \tau_{\rm Pl}$, such that $\alpha \approx 1$ in various setups, within the linear-in-$T$ regime of the resistivity \cite{bruin_similarity_2013,legros_universal_2019,polshyn_large_2019,cao_strange_2020,licciardello_electrical_2019,grissonnanche_linear-temperature_2021}. These observations, supported by evidence from solvable models, holography and systems near quantum critical points \cite{kovtun_viscosity_2005,sachdev_quantum_2011,shenker_black_2014,roberts_localized_2015,Kitaev_SYK_talk,davison_holographic_2014,zaanen_liu_sun_schalm_2015,davison_thermoelectric_2017,hartnoll_holographic_2018,song_strongly_2017,patel_magnetotransport_2018,chowdhury_translationally_2018,patel_theory_2019,chowdhury_sachdev-ye-kitaev_2021}, motivated the notion of a fundamental limit to (inelastic) relaxation times, saturated by $\tau_{\rm Pl}$, up to an unknown coefficient of order unity. Alongside, the establishment of quantum many-body chaos as a probe of thermalization \cite{Larkin1969,Kitaev_SYK_talk}, together with the celebrated bound on chaos \cite{maldacena_bound_2016}, where the quantum Lyapunov exponent was shown to obey $\lambda_L \leq 2\pi/\tau_{\rm Pl}$, readily led to the idea that the transport in strongly correlated systems might be connected to their quantum many-body chaotic dynamics.

Making a concrete connection between transport and quantum many-body chaos is a challenging task. Indeed, it is not a priori clear which physical quantities, if any, should be subjected to a fundamental bound. In particular, bounding relaxation times directly typically fail in the presence of elastic scattering processes. In an attempt to overcome this issue, it has been proposed that the proper way to formulate such a connection is by considering thermoelectric diffusivities, rather than the relaxation times directly \cite{blake_universal_2016,hartnoll_theory_2015,patel_quantum_2017-1,patel_quantum_2017}. Namely, the diffusivities $D_{\rm th,el}$ were suggested to obey $D\sim D_L \equiv v_B^2 \tau_L$, which implies that $D \gtrsim v_B^2 \tau_{\rm Pl}/2\pi$ due to the bound on chaos. While several works have demonstrated the lack of such connection for the electrical (charge) diffusivity \cite{lucas_charge_2016,davison_thermoelectric_2017,werman_quantum_2017,niu_diffusion_2017}, it appears that the \textit{thermal} diffusivity might be related to the chaos diffusivity in actuality, such that $D_{\rm th} \sim D_L$ in many generic cases\footnote{Ref. \cite{gu_energy_2017} constructed a model in which $D_{\rm th}/ D_L$ can arbitrarily small, necessitating a sharper formulation of the conditions for which this proposed relation is expected to hold.} \cite{guo_transport_2019,gu_local_2017,werman_quantum_2017,blake_thermal_2017,li_thermal_2019,blake_universal_2016,blake_thermal_2017,jeong_thermal_2018,werman_quantum_2017} .

Towards testing this hypothesis, Zhang et al. performed measurements of the thermal diffusivity in the intermediate-$T$ `bad metal' regime of a strongly correlated cuprate \cite{zhang_anomalous_2017}. Measuring the diffusivity allows one to extract the thermal relaxation time from the relation $D_{\rm th} = v^2 \tau_{\rm th}$, where the velocity scale $v$ is operationally defined based on the characteristic velocity in the system. Clearly, identifying the relevant $v$ is necessary in order to meaningfully define $\tau_{\rm th}$. However, these measurements showed pronounced phononic contributions, which led to an interpretation in terms of a strongly coupled incoherent electron-phonon ``soup'', where neither electrons nor phonons are well-defined quasiparticles. While being intriguing on its own, this interpretation has made it clear that relating the thermal diffusivity to a relaxation time in this scenario is particularly challenging since energy is carried by two degrees of freedom with different characteristic velocities. Therefore, it is highly desirable to find simpler setups where this hypothesis can be tested, where thermal transport can be assigned to a single degree of freedom.

To this end, insulators, in which lattice vibrations carry the thermal current, could serve as a much simpler platform for experimental and theoretical investigations. Remarkably, recent experimental studies of the thermal diffusivity in a wide class of insulating compounds identified emergence of a Planckian transport time at intermediate temperatures \cite{martelli_thermal_2018,behnia_lower_2019,zhang_thermalization_2019,martelli_thermal_2021}. These materials - e.g., complex oxides like SrTiO$_3$ - showed a wide range of temperatures, ranging from $\sim 50$K to well above room temperature, where the thermal diffusivity is inversely proportional to temperature. The thermal relaxation time was extracted according to $\tau_{\rm th} \equiv D_{\rm th}/v^2$, where $v$ was operationally defined as the averaged speed of sound. The relaxation time was found to obey $\tau_{\rm th} = \alpha \tau_{\rm Pl}$, with $\alpha\approx$1--3, in opposed to good thermal conductors such as diamond, where the same procedure yields $\alpha \sim 50$ \cite{zhang_thermalization_2019}. The appearance of the Planckian timescale at these elevated temperatures is particularly counter-intuitive as one would naively expect the dynamics to be essentially classical, with $\alpha \gg 1$ \cite{tulipman_strongly_2020}. In addition, the observation of such short transport times has the interesting implication that these materials might be described in terms of a strongly coupled ``phonon fluid'', similarly to the ``soup'' in \cite{zhang_anomalous_2017}.

\begin{figure}[t]
\centering

\includegraphics[width=\columnwidth]{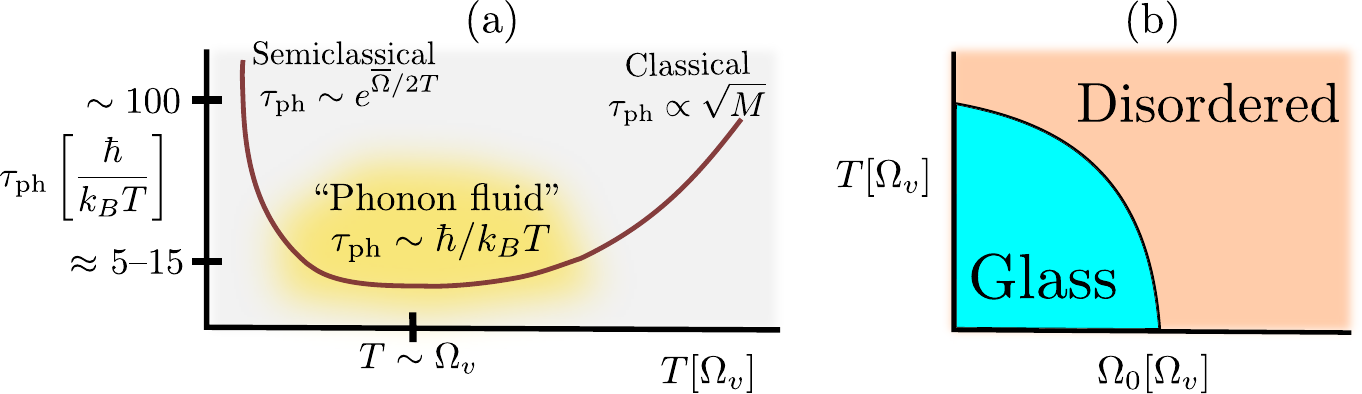}
\caption{{Summary of 0D model.} (a) Phonon lifetime against $T$ in a single-branch model. $\tau_{\rm ph}$ exhibits three dynamical regimes defined by their $T$-dependence. (b) Phase diagram of a single-branch model in the $(\Omega_0,T)$-plane: a glassy phase and a disordered phase, separated by a first-order transition.} 
\label{fig:0Dreview}
\end{figure}

These exciting observations motivated us to formulate a theoretical framework where thermal transport and many-body quantum chaos can be systematically studied in a model of strongly coupled phonons. In our previous work \cite{tulipman_strongly_2020}, we considered a zero-dimensional model of strongly coupled anharmonic oscillators. We demonstrated that the real-time dynamics, probed by the phonon lifetime, exhibits a ``phonon fluid'' regime at intermediate temperatures, where the phonon lifetime is of the order of the Planckian timescale. See Fig.~\ref{fig:0Dreview} for a schematic summary. In this work, we study properties of thermal transport and quantum many-body chaos in a lattice generalization of \cite{tulipman_strongly_2020}. 

This paper is organized as follows. In Sec.~\ref{review_of_0d}, we review the zero-dimensional model we studied in \cite{tulipman_strongly_2020}. In Sec.~\ref{sec:model}, we discuss a lattice generalization of \cite{tulipman_strongly_2020} and summarize our main results on transport and chaos. In Sec.~\ref{sec:td_of_lm}, we discuss the thermodynamics of the lattice model and describe the extension of the replica analysis to the lattice model. In Sec.~\ref{sec:th_transport}, we elaborate on the analysis of thermal transport properties. In Sec.~\ref{sec:chaos}, we describe our methods to study quantum many-body chaos and discuss some additional features of information scrambling. The correspondence between the chaos and thermal diffusivities in a three-dimensional system is discussed in Sec.~\ref{sec:3d}. The discussion and outlook are presented in Sec.~\ref{sec:disc}. Details on the imaginary- and real-time analysis are given in App.~\ref{Rep_app} and App.~\ref{SK_app}, respectively. In App.~\ref{high_d}, we define the generalization to higher dimensions, and in App.~\ref{Num_app}, we supply some details on the numerical methods.

\section{Review of zero-dimensional model}
\label{review_of_0d}

In this section, we highlight the main properties of the zero-dimensional (0D) model we studied in \cite{tulipman_strongly_2020}\footnote{Note that the 0D model is quantum mechanical, and hence can be thought of as a $d=0+1$ space-time dimensional model.}. This model serves as a single unit cell of the lattice model we consider in this work. As we will see later on, the lattice model inherits many of the properties of the 0D unit cell.   

The 0D model consists of $N$ coupled anharmonic oscillators, governed by the Hamiltonian 
\begin{eqnarray}
H &=& \sum_{i=1}^N \frac{\pi_i^2}{2} + \frac{\Omega_i^2}{2}  \phi_i^2 + \frac{1}{N}\sum_{i,j,k} {v}_{ijk} \phi_i \phi_j \phi_k \nonumber\\ 
&+& \frac{{u}}{4N} \left( \sum_{i=1}^N \phi_i^2 \right)^2,
\label{eq:H}
\end{eqnarray}
where $\phi_i$ is the displacement of the $i$th mode, dubbed the $i$th phonon field, $\pi_i$ is the conjugate momentum (such that $[\phi_i,\pi_j] = i\hbar \delta_{ij}$) and $\Omega_i$ is the frequency in the absence of anharmonicity. Note that, as in \cite{tulipman_strongly_2020}, we have eliminated the mass scale $M$ in (\ref{eq:H}) via the rescaling $\phi_i\to \sqrt{M}\phi_i,\pi_i \to \pi_i/\sqrt{M}$, and correspondingly $v_{ijk} \to M^{3/2}v_{ijk}$ and $u\to M^2 u$. Inspired by the Sachdev-Ye-Kitaev model \cite{sachdev_gapless_1993,Kitaev_SYK_talk,maldacena_remarks_2016}, the cubic couplings ${v}_{ijk}$ are chosen to be independent random Gaussian variables, each satisfying $\overline{{v}_{ijk}}=0$ and $\overline{{v}_{ijk}^2} = 2{v}^2$, where $\overline{(\cdot)}$ denotes averaging over realizations of ${v}_{ijk}$. The quartic interaction ${u}>0$ stabilizes the system (when ${u}=0$, the energy is not bounded from below, due to the cubic term).

In addition to the energy scale $\hbar \Omega_0$, we can define two energy scales associated with the anharmonic terms. The ${v}$ term defines an energy scale $\hbar \Omega_{v} = \hbar^{6/5}{v}^{2/5}$, while the ${u}$ term is associated with the scale $\hbar \Omega_{u} = \hbar^{4/3} {u}^{1/3}$. We set $\hbar=k_B = 1$ henceforth, unless stated otherwise. We focus on the strong coupling regime of the model, where $\Omega_0\sim v^{2/5} \sim u^{1/3}$.  
%In addition to the energy scale $\hbar \Omega_0$, we can define two energy scales associated with the anharmonic terms by dimensional analysis. The $\tilde{v}$ term defines an energy scale $\hbar \Omega_{v} = \left(\hbar^{2}/{M}\right)^{3/5}\tilde{v}^{2/5}$, while the $\tilde{u}$ term is associated with the scale $\hbar \Omega_{u} = \left(\hbar^{2}/M\right)^{2/3} \tilde{u}^{1/3}$. We set $\hbar=k_B = 1$ henceforth, unless stated otherwise. 

%Under the rescaling $\phi_i\to \sqrt{M}\phi_i,\pi_i \to \pi_i/\sqrt{M}$, the energy scales of the system are given by $\Omega_i,v^{2/5}$ and $u^{1/3}$, where the rescaled couplings are given by $u=\Omega_{u}^3$ and  $\overline{v_{ijk}^2}  \equiv 2v^2 = 2\Omega_{v}^5$. We focus on the strong coupling regime of the model, where $\Omega_0\sim v^{2/5} \sim u^{1/3}$. We use the rescaled formulation henceforth. 

 In the limit $N\to\infty$, the distribution of (bare) phonon frequencies $\Omega_i$ is defined as $W(\Omega) \equiv \sum_{i=1}^N \delta(\Omega - \Omega_i) \rightarrow N \rho(\Omega)$, where $\rho(\Omega)$ is a function normalized such that $\int d\Omega \rho(\Omega) = 1$. The support of $\rho(\Omega)$ extends from $\Omega_{\text{min}}$ to $\Omega_{\text{max}}$, where $\Omega_{\text{max}}-\Omega_{\text{min}}$ is the bandwidth of the model. In practice, we consider distributions of the form $\rho(\Omega) = \sum_{b=1}^{N_B} n_b \delta(\Omega-\Omega_b)$, where we have $N_B$ phonon branches with relative fractions $n_b\equiv N_b/N$, such that $\sum_{b=1}^{N_B} n_b = 1$. In the remaining of this section we consider the simplest case of $\rho(\Omega) = \delta(\Omega-\Omega_0)$, i.e., where all $\Omega_i = \Omega_0$, dubbed the single branch (SB) model. The more general case of multiple branches is discussed in \cite{tulipman_strongly_2020}, and will be further discussed in the following sections.

\subsection{Thermodynamics and phase diagram} 
We study the static properties of the system by considering the disorder-averaged free-energy density within the framework of the replica formalism. In the large-$N$ limit, saddle points of a functional integral over an effective action governs the thermodynamics. There are two stable saddle-points: a replica diagonal solution, corresponding to the disordered, self-averaging phase; and a one-step replica-symmetry breaking (1SRSB) solution, corresponding to an ordered, glassy phase. The analysis is similar to that of the quantum spherical $p$-spin-glass model \cite{cugliandolo_quantum_2001}.

The phase diagram in $(\Omega_0,T)$ plane (for fixed $v,u$) realizes a glassy (1SRSB) phase at low temperatures and small $\Omega_0$, while a disordered phase takes over for sufficiently high temperature and large $\Omega_0$ (see Fig.~\hyperref[fig:0Dreview]{\ref{fig:0Dreview}b}). The two phases are separated by a first-order transition. In addition, we study the specific heat $c(T)$ in the disordered phase, with which we diagnose the crossover to the classical limit of the model. At low-$T$, $c(T)$ is exponentially small since the phonons are gapped at $T= 0$; at high-$T$, $c(T) \to 3/4$, satisfying an anharmonic variant of the Dulong-Petit law and signaling the approach to the classical limit; and at intermediate-$T$, $c(T)$ extrapolates between these two regimes - emphasizing the non-classical nature of the dynamical ``phonon fluid'' regime discussed in the following subsection - and typically has a maximum near $T\approx \Omega_v$ \cite{tulipman_strongly_2020}. 

Before proceeding to consider the real-time dynamics, let us define the renormalized phonon frequency (or phonon stiffness) $\overline{\Omega}_0 \equiv \sqrt{\Omega_0^2 - \Pi(i\omega_n = 0)}$, $\Pi$ being the self-energy in Matsubara-frequency space. For generic parameters in the strong-coupling regime, $\overline{\Omega}_0$ is an increasing function of $T$. In particular, $\overline{\Omega}_0 \approx (uT)^{1/4}$ for $T/\Omega_v \gg 1$. The $T$-dependence of $\overline{\Omega}_0$ will be useful when we discuss the renormalized phonon velocities in the following sections. 

\subsection{Dynamics and the phonon lifetime}

We study the real-time dynamics in the disordered phase using the Keldysh formalism, along the lines of \cite{song_strongly_2017}. Similarly to the replica analysis, in the large-$N$ limit, the system is governed by a saddle point of a functional integral over an effective Keldysh action, corresponding to a set of self-consistent equations for the retarded and Keldysh Green's functions and the self-energy.

We focus on the phonon lifetime $\tau_{\rm ph}(T)$, defined by the late-time decay of the retarded Green's function $G_R(t) \propto e^{-t/\tau_{\rm ph}}$, as a probe to identify dynamical regimes as a function of temperature. We find that the system crosses over between three distinct regimes: a semiclassical regime with long-lived quasiparticle (phonon) excitations at low-$T$, where $\tau_\textnormal{ph} \sim e^{\overline{\Omega}_0/2T}$; a classical regime at high temperatures, where $\tau_\textnormal{ph}\propto \sqrt{M}$ and approaches a constant independent of $\hbar$; and an intermediate-$T$ strongly coupled ``phonon fluid'' regime, where phonons are not well-defined quasiparticles. In the latter regime, $\tau_\text{ph} = \alpha \tau_{\rm Pl}$, with $\alpha\sim 5 \textnormal{--} 15$ for generic parameters in the strongly coupled regime, See Fig.~\ref{fig:0Dreview}.

\section{Lattice Model and summary of results}
\label{sec:model}

In this section, we introduce a lattice generalization of (\ref{eq:H}) and highlight our main results for the thermal transport and quantum many-body chaos. The lattice model we consider henceforward has a few variants, each will serve us in a different inquiry. We will start by considering simpler case, where all phonon branches are optical (i.e., gapped), and later consider more a involved case where acoustic phonons are included. 

\subsection{Lattice model of optical phonons}
\label{sec:model_o_phonons}
The generalization of (\ref{eq:H}) to a $d$-dimensional square lattice ($d=1,2,3$) is given by the following Hamiltonian: $H = \sum_{\boldsymbol{r}} \left(H_{\boldsymbol{r},0} + H_{\boldsymbol{r},{\rm int}}\right)$, where 
\begin{eqnarray}
H_{\boldsymbol{r},0}  &=& \sum_{i=1}^{N}  \frac{\pi_{i,\boldsymbol{r}}^{2}}{2}+\frac{\Omega_{i}^{2}}{2}\phi_{i,\boldsymbol{r}}^{2}+\frac{\Omega_{\text{d},i}^{2}}{2} \sum_{\delta} \left(\phi_{i,\boldsymbol{r}+\boldsymbol{\delta}}-\phi_{i,\boldsymbol{r}}\right)^{2} ;\nonumber   \\ 
 H_{\boldsymbol{r},{\rm int}}
  &=& \frac{1}{N}\sum_{ijk}{v}_{ijk}\phi_{i,\boldsymbol{r}}\phi_{j,\boldsymbol{r}}\phi_{k,\boldsymbol{r}}+\frac{{u}}{4N}\left(\sum_{i=1}^{N}\phi_{i,\boldsymbol{r}}^{2}\right)^{2}.
  \label{H_LM}
\end{eqnarray} 
Here, $\boldsymbol{r}\in[-\mathcal{N}/2,...,\mathcal{N}/2]^d$ labels the lattice site, where $\mathcal{N}$ is the linear size of the system and the lattice constant is set to unity. $\boldsymbol{\delta}$ are unit vectors in each spatial direction. Different variants of (\ref{H_LM}) corresponds to different distributions of $\Omega_i$ and $\Omega_{\textnormal{d},i}$: $\rho(\boldsymbol{\Omega})$, defined similarly to Sec.~\ref{review_of_0d}, with $\delta(\Omega-\Omega_b)$ replaced by $\delta(\boldsymbol{\Omega} - \boldsymbol{\Omega}_b)$, where $\boldsymbol{\Omega}_b = (\Omega_b,\Omega_{{\rm d}b})$. Note that the $\Omega_{\text{d},i}$ is the only dispersive coupling in $H$, i.e., for $\Omega_{\text{d},i}=0$, $H$ is a sum of decoupled copies of (\ref{eq:H}). Note also that we choose our model to be translationally invariant and isotropic: for a given realization, the parameters $v_{ijk}
$ are identical for all sites, and $\Omega_\textnormal{d}$ is identical for all spatial directions.   

We begin by considering the simplest case of a single optical branch: $\rho(\boldsymbol{\Omega})=\delta(\boldsymbol{\Omega}-\boldsymbol{\Omega}_{\rm o})$, where $\boldsymbol{\Omega}_{\rm o} = (\Omega_{\rm o},\Omega_{\rm d})$, and we consider $d=1$. As in \cite{tulipman_strongly_2020}, the Green's functions in imaginary-time (real-time) are determined by the SPEs of the replica (Keldysh) effective action, respectively. For example, the imaginry-time SPEs of the single optical branch model in the disordered phase are given by 
\begin{eqnarray}
    {G}\left(i\omega_{m},k\right) &=& \frac{1}{ \omega_{m}^{2} + \Omega_{\rm o}^2 + 4\Omega_\textnormal{d}^2\sin^2(\frac{k}{2})  - {\Pi}\left(i\omega_{m},k\right)},  \label{eq:SPE_s_SOBM_imag1}
\\
\Pi\left(\tau,r\right) &=& v^{2}{{G}}\left(\tau,r\right)^2-u{{G}}\left(\tau,r\right)\delta\left(\tau\right)\delta\left(r\right).
\label{eq:SPE_s_SOBM_imag2}
\end{eqnarray}

Throughout this work, we will focus on the limit of weak dispersion, defined by $\Omega_\textnormal{d}\ll \Omega_{\rm o}$. This limit is particularly useful in that thermodynamical and dynamical properties of the single optical branch model are essentially identical to the 0D SB model, yet it allows us to consider transport and spatial aspects of chaos. In practice, the weakly dispersive limit corresponds to neglecting subleading corrections in $\Omega_\textnormal{d}$ by approximating ${\Pi}\left(i\omega_m,k\right) \approx{\Pi}\left( i\omega_m \right)$, where ${\Pi}\left( i\omega_m \right)$ is the self-energy of the corresponding 0D system (with $\Omega_\textnormal{d}=0$). Single-particle properties that are encoded in the self-energy (e.g., phonon lifetime, renormalization of the bare frequency $\Omega_{\rm o}$) can thus be directly inferred from the 0D model. 

Consider $v_\textnormal{o}$ - the speed associated with the optical branch. We define it as $v_\textnormal{o} \equiv \textnormal{max}_k \{|\partial_k \varepsilon_\textnormal{o}(k)| \}$, where $\varepsilon_\textnormal{o}$ is the dispersion of the optical branch. The dispersion of the gapped, optical modes is quadratic at small $k$, such that $\partial_k \varepsilon_\textnormal{o}(k)|_{k=0} = 0$ and the maximum is attained for some $k\ne0$ momentum. We identify the speed of the optical branch as $v_\textnormal{o} = \Omega_\textnormal{d}^2/\overline{\Omega}_\textnormal{o} $, $\overline{\Omega}_\textnormal{o} $ being the renormalized frequency. Importantly, $\overline{\Omega}_{\rm o}$ grows with increasing $T$, implying that $v_\textnormal{o}$ diminishes with increasing $T$. This is in contrast to acoustic modes, whose associated speed (the speed of sound) is proportional to $\overline{\Omega}_{\rm a}$ and thus grows with temperature, as we will see later on.

The generalization to multiple optical branches is given by $\rho(\boldsymbol{\Omega}) = \sum_b n_b\delta(\boldsymbol{\Omega}-\boldsymbol{\Omega}_b)$. Each branch then defines a Green's function $G_b$ with frequency $\Omega_b$, dispersive coupling $\Omega_{\textnormal{d}b}$ and relative fraction $n_b$, that satisfies (\ref{eq:SPE_s_SOBM_imag1}). The self-energy is branch-independent and is given by replacing the single optical branch Green's function $G$ in (\ref{eq:SPE_s_SOBM_imag2}) by $\mathcal{G} \equiv \sum_b n_b G_b$ - a weighted sum over all branches. In the remainder of this section, we will focus on the case of a single optical phonon (with or without acoustic phonons), whereas models with multiple optical branches will serve us in the following sections. 

\subsection{Adding acoustic phonons}
\label{sec:adding_a_phonons}

We consider a system with $N_\textnormal{a}$ acoustic phonons and $N_\textnormal{o} = N - N_\textnormal{a}$ optical phonons. Let us consider the case of $d=1$ for simplicity. Denote the flavor-subsets of acoustic and optical phonons by $I_\textnormal{a}\equiv\{1,...,N_\textnormal{a}\}$ and $I_\textnormal{o} \equiv \{N_\textnormal{a} + 1 , ..., N\}$, respectively. Then, the addition of acoustic phonons is done by letting $\Omega_{\textnormal{d}i}=0$ for $i\in I_\textnormal{a}$ and replacing the phonon fields $\phi_{i,{r}}$ in (\ref{H_LM}) with the generalized fields 
\begin{equation}
    \widetilde{\phi}_{i,{r}} =
    \begin{cases}
      {\phi}_{i,{r}+1} - {\phi}_{i,{r}} & \quad i\in I_\textnormal{a}, \\
      {\phi}_{i,{r}} &  \quad i\in I_\textnormal{o}.  \\
    \end{cases}
    \label{eq:generalized_phi}
\end{equation}
That is, we replace the phonon fields of the acoustic branches with discrete lattice derivatives. In this way, the Hamiltonian is invariant under a shift $\phi_{i\in I_a,r} \rightarrow \phi_{i\in I_a,r} + \rm{constant}$. The acoustic modes are the Goldstone modes associated with this continuous symmetry.

It is instructive to consider a system with a single acoustic branch and a single optical branch, defined by $\rho(\boldsymbol{\Omega}) =  n_{\rm a}\delta(\boldsymbol{\Omega}-\boldsymbol{\Omega}_{\rm a})+n_{\rm o} \delta(\boldsymbol{\Omega}-\boldsymbol{\Omega}_{\rm o})$ such that $\boldsymbol{\Omega}_{\rm a}=(\Omega_{\rm a},0)$ and $\boldsymbol{\Omega}_{\rm o}=(\Omega_{\rm o},\Omega_{\rm d})$, where a/o denotes the acoustic/optical branches, respectively. As before, the system is controlled by the SPEs of the real- and imaginary-time effective actions (see App.~\ref{Rep_app},\ref{SK_app}). For example, the disordered-phase SPEs in real-time are given by 

\begin{eqnarray}
{G}_{R\textnormal{o}}\left(\omega,k\right) &=& -\frac{1}{\omega^{2}-\Omega_{\textnormal{o}}^{2}-4\sin^2(\frac{k}{2})\Omega_\textnormal{d}^2 - {\Pi}_{R\textnormal{o}}\left(\omega,k\right)} \nonumber ; \\
{G}_{R\textnormal{a}}\left(\omega,k\right) &=& -\frac{1}{\omega^{2}-4\sin^2(\frac{k}{2})\Omega_\textnormal{a}^2 - {\Pi}_{R\textnormal{a}}\left(\omega,k\right)} \nonumber ; \\
 {G}_{Kb}\left(\omega,k\right)&=&2i\coth\left(\frac{\beta\omega}{2}\right)\text{Im}\left[{G}_{Rb}\left(\omega,k\right)\right],\quad  b=\textnormal{a,o};\nonumber   \\
\Pi_{R\textnormal{o}}\left(t,r\right) &=& iv^{2} \widetilde{\mathcal{G}}_{R}\left( t,r \right) \widetilde{\mathcal{G}}_{K}\left( t,r \right) -\frac{iu}{2}\widetilde{\mathcal{G}}_{K}\left(t,r\right)\delta(t)\delta(r) ; \nonumber \\
\Pi_{R\textnormal{a}}\left(\omega,k \right) &=& 4\sin^2\left(\frac{k}{2}\right) \Pi_{R\textnormal{o}}\left(\omega,k \right). \nonumber \\
 \label{eq:Keldysh_sc_eqs_AO_model}
\end{eqnarray}
Here the Green's functions $G$ are defined with respect to the $\phi$ fields (not $\widetilde{\phi}$). The self-energies, however, contain a weighted sum of the Green's functions of the generalized fields $\widetilde{\phi}$,  
$\widetilde{\mathcal{G}}\left(\omega,k\right) =  4 n_\textnormal{a} \sin^2\left(\frac{k}{2}\right)G_\textnormal{a}\left(\omega,k\right) + n_\textnormal{o}G_\textnormal{o}\left(\omega,k\right) $. Note that letting $n_\textnormal{a}\to 0$ collapses the SPEs (\ref{eq:Keldysh_sc_eqs_AO_model}) to those of a single optical branch. 

A few comments are in order. We first observe that, by construction, the acoustic branch becomes gapless and infinitely long-lived in the limit of $k\to 0$. Its momentum-dependent lifetime satisfies $\tau_\textnormal{a}(k) \sim 1/k^2 $ (see App.~\ref{SK_app}). In addition, the speed of sound $\overline{v}_{s}$ is determined by the renormalized acoustic frequency: $\overline{v}_{s}\equiv \partial_k\varepsilon_\textnormal{a}(k)|_{k=0} = \left( \Omega_\textnormal{a}^2 + \Pi_{R\textnormal{o}}\left(0,0 \right) \right)^{1/2} $, $\varepsilon_\textnormal{a}(k)$ being the dispersion relation of the acoustic branch. In particular, $\overline{v}_{s}$ is significantly enhanced by interactions for generic parameters in the strongly-coupled regime, where $\Pi_{R\textnormal{o}}\left(0,0 \right) > 0$. 

Despite its singular nature at small $k$, the acoustic branch has a non-singular contribution to the self-energy, in any dimension. This is due to the fact that the self-energy is a function of $\widetilde{G}_\textnormal{a}$, rather than $G_\textnormal{a}$, where $\widetilde{G}_\textnormal{a}\left(\omega,k\right) =  4  \sin^2\left(\frac{k}{2}\right)G_\textnormal{a}\left(\omega,k\right)$ is non-singular in the $k\to0$ limit. Importantly, the dependence on $\widetilde{G}_\textnormal{a}$ rather than ${G}_\textnormal{a}$ is also true for the free-energy. 
%That is to say, acoustic branches has a non-singular contribution to the thermodynamical quantities. 
This is in contrast to some dynamical quantities (e.g., thermal conductivity) in which the contribution of acoustic branches is singular in dimensions $d \leq 2$, as we will discuss later on.  

For analytical tractability, we will focus our attention on the limit $n_\textnormal{a} \ll n_\textnormal{o}$, where the small parameter $n_\textnormal{a}$ enables a controlled expansion about a system with $n_\textnormal{a}=0$. We will consider the first-order (linear) corrections in $n_\textnormal{a}$, for which our previous approximation, $\Pi_{R\textnormal{o}}\left(\omega,k \right) \approx \Pi_{R\textnormal{o}}\left(\omega \right)$, holds. Systems with larger $n_\textnormal{a}$, and in particular $n_\textnormal{a} = 1$, are also highly interesting, yet their analysis is more involved due to the strong momentum dependence of the acoustic modes, and we shall leave their treatment to future studies. Physically, the fact that acoustic phonons constitute a small fraction of the system means that the optical modes act essentially as a bath on the acoustic modes.

\begin{figure*}[t]
\centering

 \begin{minipage}[b]{\columnwidth}
     \includegraphics[width=\textwidth]{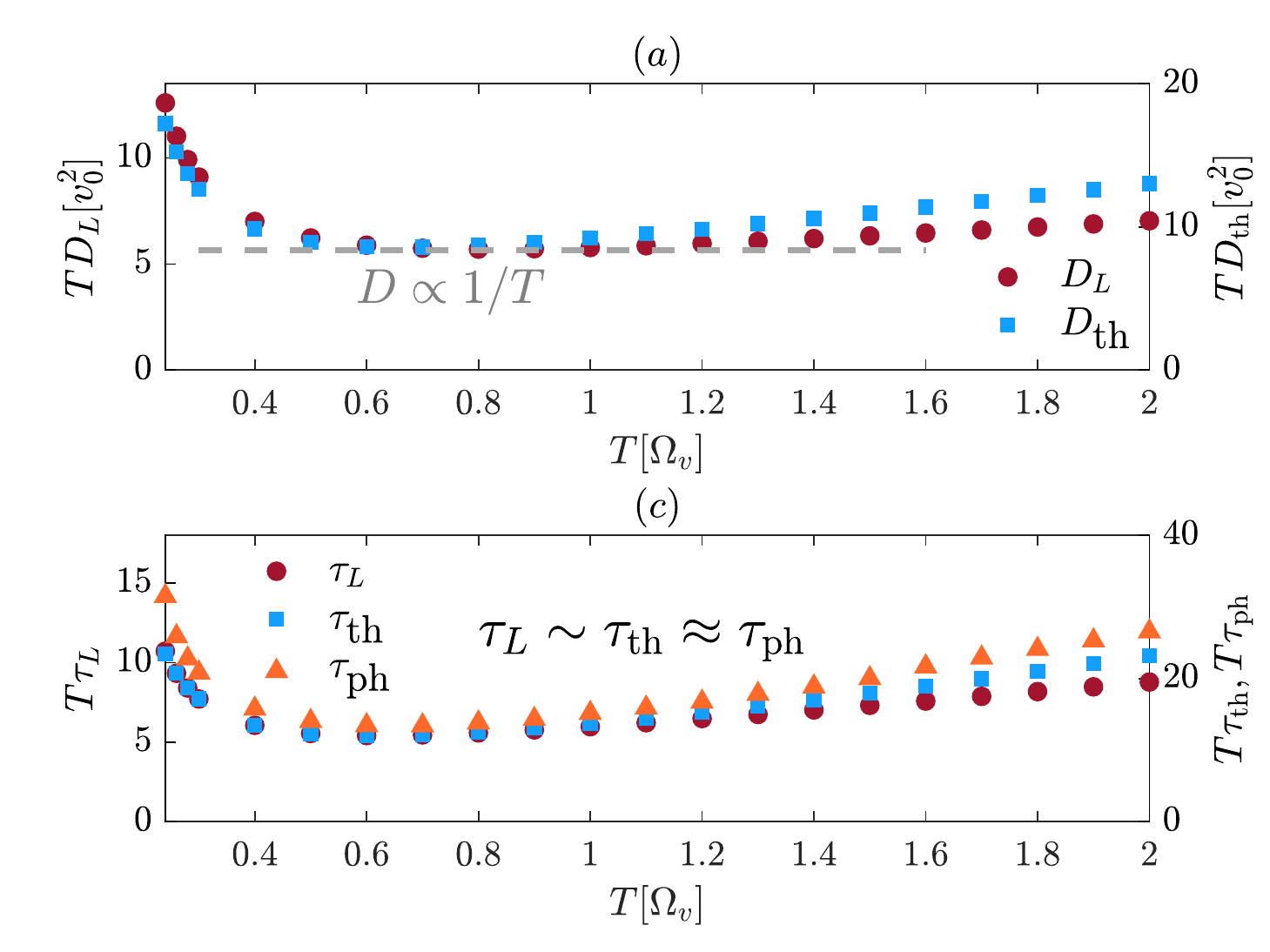}
  \end{minipage}
  \hfill
  \begin{minipage}[b]{\columnwidth}
    \includegraphics[width=\textwidth]{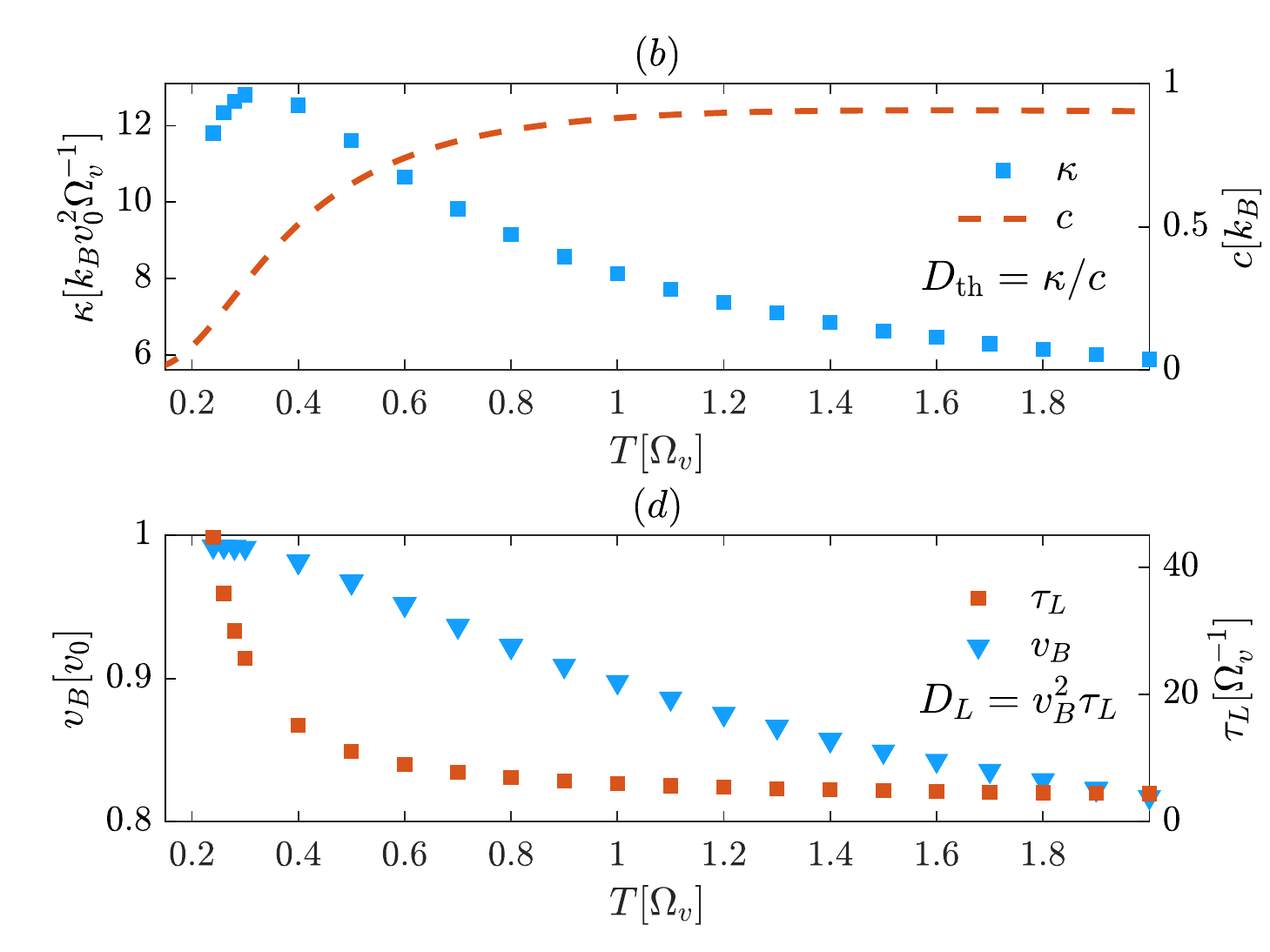}
  \end{minipage}
  
\caption{{Transport and chaos in a system containing a single highly degenerate optical phonon branch.} (a) Comparison between $TD_L$ and $TD_{\rm th}$ as a function of $T$. Flat regions of the curves correspond to $D\propto 1/T$. (b) Thermal conductivity $\kappa$ and specific heat $c$ (per mode) as a function of $T$. (c) Comparison between $T\tau_\textnormal{ph}, T\tau_\textnormal{th}$ and $T\tau_L$ as a function of $T$. Flat regions correspond to $\tau = \alpha / T$. (d) Lyapunov time ($\tau_L = 1/\lambda_L$) and butterfly velocity as a function of $T$. Data shown with $\Omega_{\rm o}/\Omega_v=1.1,\Omega_{\rm d}^2/\Omega_v^2 = 0.008,u/\Omega_v^3=1.4$, $v_0 \equiv \Omega_{\rm d}^2 / \Omega_{\rm o} = 0.0072\Omega_v$ and $d=1$.} 
\label{fig:results}
\end{figure*}
\subsection{Main results: transport and chaos}
\label{sec:main_results}
In this subsection, we highlight our main results on transport and chaos properties using the two simplest variants of the model: a single optical branch, and a single optical and a single acoustic branches. Variants with multiple phonon branches follow the trends we describe here and allow us to address more delicate questions regarding transport and many-body chaos, see Sec.~\ref{sec:th_transport} and Sec.~\ref{sec:chaos}. 

%We focus on the disordered phase, leaving the study of transport and chaos in the glassy phase, where some progress was recently made [glass sumilan,glass for holographers], to future studies.

\textbf{Optical phonons} ($n_{\rm a} = 0$). We characterize thermal transport with the $T$-dependence of the thermal diffusivity $D_{\rm th}$ and its associated thermal current relaxation time $\tau_{\rm th}$. $D_{\rm th}$ is obtained via the Einstein relation: $D_{\rm th} = \kappa/c$, where $\kappa$ and $c$ are the thermal conductivity and specific heat, respectively. Then, $\tau_{\rm th}$ is operationally defined as $\tau_{\rm th}\equiv  d D_{\rm th}/ v_{\rm o}^2$, where $v_{\rm o}$ is the phonon velocity - the only velocity scale in the system with $n_{\rm a}=0$ - and $d$ is the spatial dimension of the system. To characterize quantum many-body chaos, we consider the chaos diffusivity $D_L$, quantum Lyapunov time $\tau_L$ (i.e., the inverse scrambling rate $\tau_L=1/\lambda_L$) and butterfly velocity $v_B$. Unlike thermal transport, $D_L$ and $\tau_L$ are obtained directly from the Bethe-Salpeter equation (BSE) of the out-of-time-order correlator (OTOC), such that the butterfly velocity is determined by $v_B = \sqrt{dD_L/\tau_L}$. 

We find numerically that $D_L$ and $D_{\rm th}$ satisfy $D_{\rm th} \approx \gamma D_L$, with $\gamma > 1$ being a non-universal constant that depends on the system parameters. A representative set of results, with $\gamma \approx 1.5$, is found in Fig.~\hyperref[fig:results]{\ref{fig:results}a}. Typically, $\gamma \sim 1 \textnormal{--} 3$, such that $D_{\rm th} \gtrsim D_L$. Furthermore, we find that the diffusivities track a similar temperature trend to that of the phonon lifetime. In particular, generic parameters in the strongly coupled regime exhibits an intermediate-$T$ region where $D\propto 1/T$. This region is not parametrically large, yet we find that it expands in the case where optical phonons are spread over a finite bandwidth (see Sec.~\ref{sec:th_transport}). 
%To better understand this $T$-dependence we proceed to consider a comparison between the different timescales in the system. 

Remarkably, we find that all timescales follow the same $T$-dependence of the phonon lifetime. The thermal relaxation time satisfies $\tau_{\rm th} \approx \tau_{\rm ph}$, while the Lyapunov time is typically shorter than $\tau_{\rm ph}$ and obeys $\tau_L \approx \tau_{\rm ph}/\gamma$, $\gamma$ being the same coefficient as in the relation between the diffusivities, see  Fig.~\hyperref[fig:results]{\ref{fig:results}c}. In particular, it appears that the ``phonon fluid'' regime identified in \cite{tulipman_strongly_2020} corresponds to a Planckian dissipative regime, where $\tau_L$ and $\tau_{\rm th}$ are of the order of the Planckian timescale $\hbar / k_B T$. Note, however, that $\tau_L$ does not saturate the bound on chaos in our model. Moreover, unlike fermionic variants of the SYK model (e.g. \cite{song_strongly_2017,patel_theory_2019,gu_energy_2017,chowdhury_translationally_2018}), the appearance of the Planckian timescale in our model is not a result of an underlying quantum critical point \cite{giombi_bosonic_2017,tulipman_strongly_2020,benedetti_remarks_2021}. Hence, it is not clear whether any holographic interpretation \cite{kovtun_viscosity_2005, hartnoll_theory_2015, blake_quantum_2018,hartnoll_holographic_2018, baggioli_universal_2020,wu_universality_2021} may be applied.

Let us make a few more comments before proceeding to discuss the $n_{\rm a} >0$ case. Firstly, note that as we decrease $T$, $\kappa(T)$ is peaked at the temperature at which $\tau_{\rm ph}$ starts to increase rapidly, and $c$ starts to significantly drop ($T\sim 0.3$ in Fig.~\hyperref[fig:results]{\ref{fig:results}b,c}). At this temperature, the fact that the system is gapped (at $T=0$) starts to manifest. The decrease of $\kappa(T)$ as $T$ decreases further is associated with the fact that $c(T)$ decreases faster than $\tau_{\rm ph}(T)^{-1}$. Secondly, note that $v_B$ is weakly dependent on $T$ at low- to intermediate-$T$, in comparison to $\tau_L$, such that the dominant temperature dependence of $D_L$ is due to $\tau_L$, see Fig.~\hyperref[fig:results]{\ref{fig:results}d}. At high-$T$, however, $\tau_L$ saturates to a constant value, while $v_B(T) \sim 1/T^{1/4}$, such that $D_L \sim 1/\sqrt{T}$. Lastly, we find that $v_B \approx v_{\rm o}$. 
%, suggesting that our operational definition of $\tau_{\rm th}$ - where the value of $ v_{\rm o}$ is added by hand - is reasonable. 

\textbf{Acoustic and optical phonons} ($n_{\rm a} > 0$). We consider a system with $n_{\rm o}$ optical modes and $n_{\rm a}$ acoustic modes, such that $n_{\rm o} \gg n_{\rm a}$. This asymptotic case is handy in that our approximation that $\Pi_{R\rm{o}}$ in Eq. (\ref{eq:Keldysh_sc_eqs_AO_model}) is weakly momentum-dependent holds.  %and it allows us to ask how transport and chaos properties are affected by introducing acoustic phonons in the limit $n_{\rm a}\to 0$.

Consider thermal transport in the presence of acoustic phonons. It is well known that long-wavelength acoustic phonons may dominate the thermal conductivity in dimensions $d=1,2$, at any $T$, leading to $\kappa=\infty$ in the thermodynamic limit \cite{Ziman_2001,prosen_momentum_2000,gu_colloquium_2018}. Namely, any $n_{\rm a} > 0$ results in $\kappa$ and $D_{\rm th}$ being infinite for $d=1,2$, making the limit $n_{\rm a} \to 0$ highly singular. In $d\geq3$, however, $\kappa < \infty$ for any $n_{\rm a} > 0$. Furthermore, in $d=3$, we find that the velocity relevant for thermal transport of the acoustic branch is the renormalized speed of sound $\overline{v}_{s}$, which is typically much larger than $v_{\rm o}$ in the weakly dispersive limit. The imbalance between velocities is in competition with the phase space fraction, such that, roughly speaking, the ratio $r_{\rm{ao}}\equiv n_{\rm a} \overline{v}_s^2 / n_{\rm o}v_{\rm o}^2  $ determines whether acoustic ($r_{\rm{ao}}>1$) or optical ($r_{\rm{ao}}<1$) modes dominate transport in $d=3$. In the case of $r_{\rm{ao}}\ll 1$, thermal transport follows the trends described in Fig.~\ref{fig:results}. We comment on thermal transport in the general case in Sec.~\ref{sec:3d}. 

Next, we consider many-body quantum chaos in the presence of acoustic phonons. This setting is particularly interesting in that it enables us to study scrambling of a system where long-lived, weakly interacting Goldstone modes (i.e., acoustic phonons) are coexisting with a strongly interacting ``phonon fluid''. We find that the quantifiers of many-body chaos: $\lambda_L$ and $D_L$, are non-singular in the limit $n_{\rm a} \to 0$ in any dimension. Namely, $q_L$ with $q=\lambda,D$ admit an expansion of the form $q_L(n_{\rm a})=q_L(0)+n_{\rm a} \delta q_L+\mathcal{O}(n_{\rm a}^2)$, such that $\delta q_L$ is finite. Note that the existence of a smooth $n_{\rm a}\to0$ limit for $\lambda_L$ and $D_L$ is strikingly different than the singular $n_{\rm a} \to 0$ limit for $\kappa$ at low dimensions. In particular, in $d=1,2$ our model is an example of a system where $D_L\ll D_{\rm th} = \infty$.
%, suggesting that a universal relation $D_{\rm th} \sim D_L$ [mike blake universal incoherent BH papers] may hold only in $d>2$.  

\begin{figure}[t]
\centering

\includegraphics[width=\columnwidth]{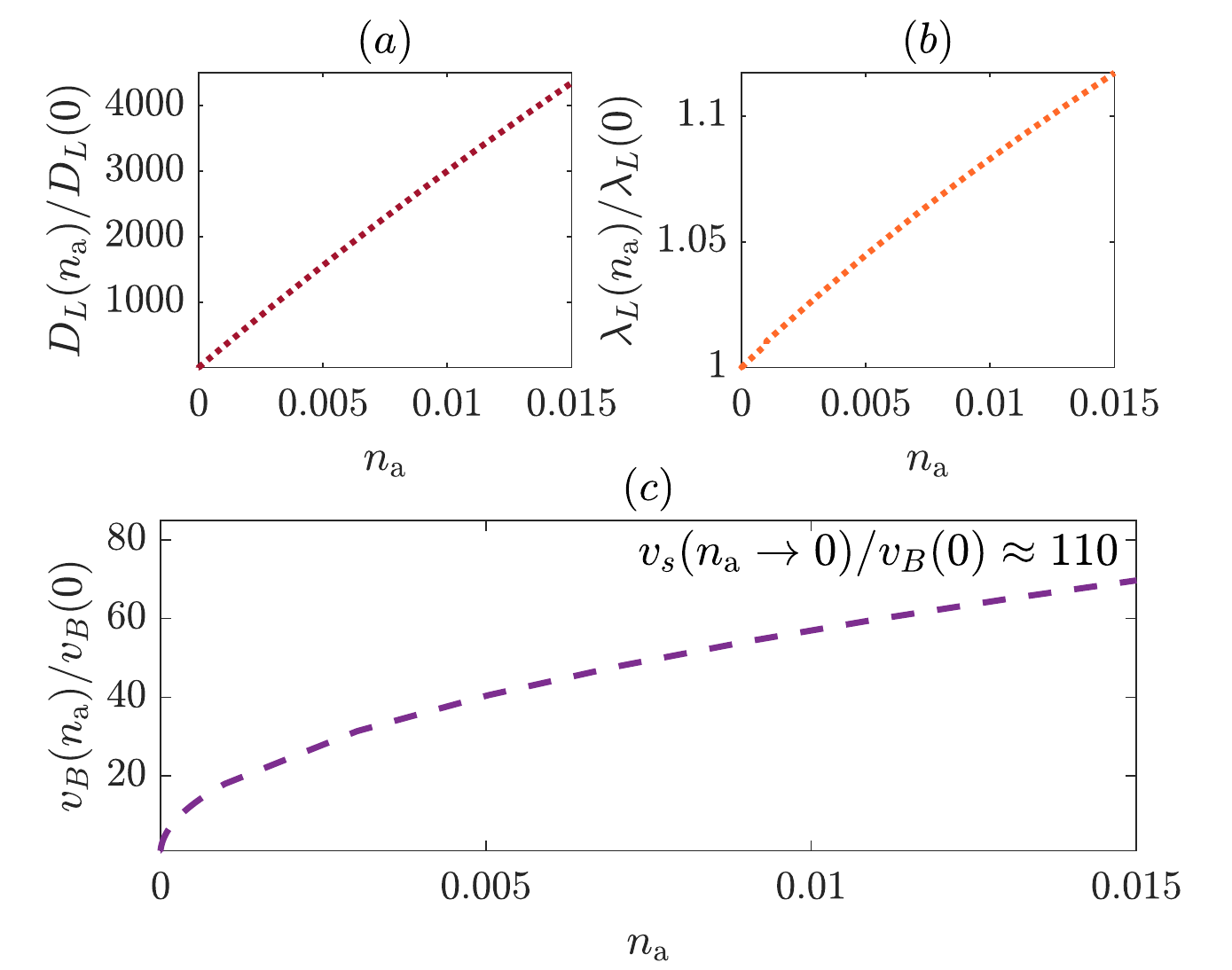}
\caption{{Chaos in the presence of acoustic phonons in the $n_{\rm a}\ll 1$ limit in $d=1$.} (a) $D_L(n_{\rm a})/D_L(0)$ as a function of $n_{\rm a}$. (b) $\lambda_L(n_{\rm a})/\lambda_L(0)$ as a function of $n_{\rm a}$. (c) $v_B(n_{\rm a})/v_B(0)$ as a function of $n_{\rm a}$. $\overline{v}_{s}(n_{\rm a} \to 0)\equiv \sqrt{\Omega_{\rm a}^2+\Pi_{R{\rm o}}(\omega=0)}$ with $\Pi_{R{\rm o}}(0)$ evaluated at $n_{\rm a}=0$. Data shown with $T/\Omega_v=1$, $\Omega_{\rm o}/\Omega_v =1.1$,$u/\Omega_v^3=1.4$,$\Omega_{\rm d}^2/\Omega_v^2=\Omega_{\rm a}^2/\Omega_v^2=0.008$, $v_B(0)=v_{\rm o}=\Omega_{\rm d}^2/\overline{\Omega}_{\rm o}=0.007\Omega_v$ and $\overline{v}_{s}(n_{\rm a}\to0)=0.77\Omega_v$.} 
\label{fig:acousticchaos}
\end{figure}

By numerically solving the BSEs, we find that $\lambda_L$ and $D_L$ are increasing functions of $n_{\rm a}$, see Fig.~\hyperref[fig:acousticchaos]{\ref{fig:acousticchaos}a,b}. The fact that $\lambda_L(n_{\rm a})$ grows with $n_{\rm a}$ may come as a surprise as one might suspect the long-lived modes associated with small $k$ would decrease the scrambling rate of the system. However, the contribution of these modes is strongly suppressed in the BSEs, such that fast-decaying operators dominate scrambling. Proceeding to consider $D_L$, we observe that the effect of acoustic phonons is much more significant, such that even a relatively small fraction of acoustic phonons have a considerable effect on $D_L$. The reason for this behavior is the fact that in the weakly dispersive limit, 
$\overline{v}_{s}\gg \overline{v}_{\rm o}$. 
%and $\overline{v}_{s}$, where the latter is typically greater by a factor of order 100 due to strong renormalization effects. 
For a sufficiently large $n_{\rm{a}}$ such that $n_{\rm a}\overline{v}_{s}^2 \gg n_{\rm o} \overline{v}_{\rm o}^2$, we find that $v_B \approx \sqrt{n_{\rm a}} \overline{v}_{s}$ (see Fig.~\hyperref[fig:acousticchaos]{\ref{fig:acousticchaos}c}). More details are given in Sec.~\ref{sec:chaos}. 
% This, in turn, enters $D_L$ as a competition between $n_{\rm a}\overline{v}_{s}^2$ and $n_{\rm o} v_{\rm o}^2$. In particular, this imbalance implies that $v_B \approx \sqrt{n_{\rm a}} \overline{v}_{s}$, for sufficiently large $n_{\rm a}$ where the speed of sound dominates the chaos diffusion, in agreement with our numerical findings, see Fig.~\hyperref[fig:acousticchaos]{\ref{fig:acousticchaos}c}. More details are given in Sec.~\ref{sec:chaos}. 

\section{Thermodynamics}
\label{sec:td_of_lm}
In this section, we discuss the thermodynamics of the lattice model. Our focus is two-fold. As our primary interest in this work is transport and chaos in the disordered, self-averaging phase of the model, we first map out the boundary of this phase. In addition, we compute the specific heat in the disordered phase (see Fig.~\hyperref[fig:results]{\ref{fig:results}b}) as it will allow us to evaluate the thermal diffusivity. Fortunately, much of the thermodynamic properties of the lattice model are inherited from the 0D model. 

%Let us clarify in what sense 
%the lattice model's thermodynamical properties are inherited from the 0D model. Recall that 
The thermodynamics properties of the system are controlled by saddle points of an effective replica action, whose replica-space structure characterizes different phases of the model, such that the off-diagonal terms in replica-space serve as order parameters \cite{mezard_spin_1986}. In the lattice model, apart from the replica-space structure, these off-diagonal terms may have spatial dependence. That is, the off-diagonal terms may probe replica-symmetry breaking between different lattice sites. Here, we consider \textit{on-site} order parameters to probe replica-symmetry breaking. This order parameter is particularly natural in light of the weakly dispersive limit considered and it relates the thermodynamics of the lattice model to the 0D model in a well-defined manner. 

In \cite{tulipman_strongly_2020}, we analyzed in detail the phase diagram of a system with a single phonon branch, while for systems with multiple branches, we relied on a simple physical argument to bound the boundary of the glassy phase. Here, we generalize the replica analysis to the case of multiple phonons, including acoustic modes. In practice, we consider distributions $\rho(\boldsymbol{\Omega})=\sum_b n_b \delta \left(\boldsymbol{\Omega}-\boldsymbol{\Omega}_b\right)$ with fixed bandwidth $\Delta\Omega$, fixed frequency spacing and fixed fractions $n_b$, and determine the transition as a function of $\Omega_{\rm min}$ in the limit $T\to 0$. We then restrict ourselves to systems that are in the disordered phase, away from the boundary to the glass phase. The technical details of the analysis are provided in App.~\ref{Rep_app}. 
%Loosely speaking, we are considering configurations to the right to of the $T=0$ glass boundary in Fig.~\hyperref[fig:0Dreview]{\ref{fig:0Dreview}b}. More details are found in App.~\ref{Rep_app}.  

\section{Thermal Transport}
\label{sec:th_transport}
In this section, we discuss thermal transport properties of the lattice model (\ref{H_LM}). In particular, we compute the thermal conductivity per mode, $\kappa$, the thermal diffusivity $D_{\rm th}$, and the thermal relaxation time $\tau_{\rm th}$. The diffusivity is exctracted using the Einstein relation, $D_{\rm th} = \kappa/c$. The thermal current relaxation time is related to $D_{\rm th}$ by operationally defining $D_{\rm th} \equiv v^2 \tau_{\rm th}/d$, where the relevant velocity scale $v$ is discussed below.  

\begin{figure}[t]
\centering

\includegraphics[width=\columnwidth]{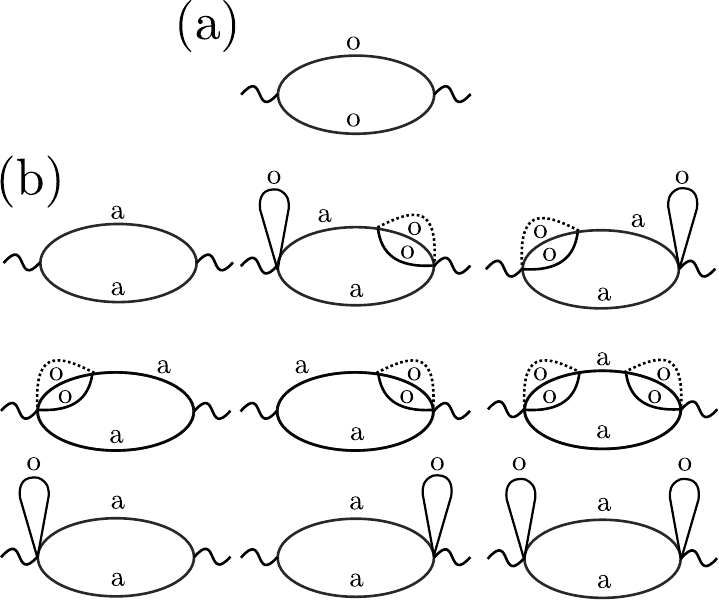}
\caption{{Leading contributions to current correlator for (a) optical and (b) acoustic branches.} Solid lines represent Green's functions, where a/o denotes Green's functions of acoustic/optical branches. Dashed lines denote averaging over realizations of $v_{ijk}$.} 
\label{fig:thcurrent}
\end{figure}

We compute $\kappa$ using the Kubo formula: 
\begin{eqnarray}
N\kappa = -\lim_{\omega \to 0}\frac{{\rm Im}[G_{R}^{J}(\omega)]}{T\omega},
\label{kappa_formula}
\end{eqnarray}
where $G_{R}^{J}$ is the retarded thermal current correlation function which we compute directly in real-time using the Keldysh formalism (see App.~\ref{SK_app}). We begin with the $n_{\rm a} = 0$ model and discuss the contribution of acoustic phonons later on. Note that we choose our model to by isotropic in any dimension, such that $\kappa_{\alpha\beta}= \delta_{\alpha\beta} \kappa $, where $\alpha,\beta$ are spatial indices.

\subsection{Optical phonons}
\label{op_results}
Consider a system with a single optical branch. In the weakly-dispersive limit, the vertex corrections to $G_R^J$ are suppressed by powers of $\Omega_{\rm d}/\Omega_{\rm o} \ll 1$. Therefore, $G_R^J$ is well approximated by $G_{R(0)}^J$, the correlation function without vertex corrections, depicted in Fig.~\hyperref[fig:thcurrent]{\ref{fig:thcurrent}a}. Using this approximation, we obtain that (see App.~\ref{SK_app})
\begin{eqnarray}
\kappa = \frac{\Omega_{\rm d}^{4}}{2}\int_{\nu,k}\sin^{2}k\mathcal{A_{\rm o}}\left(\nu,k\right)^{2}\left(\beta\nu\right)^{2}\text{csch}^{2}\left(\frac{\nu\beta}{2}\right). 
\label{eq:kappa_o}
\end{eqnarray}
Here, $\mathcal{A}_{\rm o}\left(\nu,k\right) = {\rm Im}[G_{R{\rm o}}(\nu,k)]$ is the spectral function of the optical branch. We use the shorthand notation $\int_{\nu,k}\equiv \int{\frac{{\rm d}^{d}k \rm{d}\nu}{(2\pi)^{d+1}}}$. Note that (\ref{eq:kappa_o}) essentially extends the familiar Boltzmann expression $\kappa = \int_k c_k v^2_k \tau_k$ beyond the quasiparticle regime, where $c_k,v_k$ and $\tau_k$ are the $k$-dependent specific heat, phonon velocity and phonon lifetime, respectively. 

In the 0D model, the phonon lifetime becomes much larger than their inverse frequency in both the low- and high-temperature limits~\cite{tulipman_strongly_2020}. This property carries over to the lattice model in the weakly dispersive limit. Hence, in these regimes, the Boltzmann expression for $\kappa$ is valid. Furthermore, $\kappa \sim c\overline{v}^2_{\rm o} \tau_{\rm ph}$, where the $k$-dependence of $c_k$ and $\tau_k$ can be neglected.  
%such that $\kappa \sim c\overline{v}^2_{\rm o} \tau_{\rm ph}$. 
We may use properties of the 0D model to understand the $T$ dependence of $\kappa$ in the different dynamical regimes. At high $T$, $c\to 3/4$ and $\tau_{\rm ph} \to {\rm constant}$, while $v_{\rm o} = \Omega_{\rm d}^2/\overline{\Omega}_{\rm o}\sim 1/(uT)^{1/4}$. Hence, $\kappa \sim 1/\sqrt{uT}$ for $T/\Omega_v\gg 1$. At low-$T$, $c\sim e^{-\beta \overline{\Omega}_{\rm o}}$, where $\beta=1/T$ (since the optical phonons are gapped), and $\tau_{\rm ph} \sim e^{\beta \overline{\Omega}_{\rm o}/2}$, while $v_{\rm o}\approx \Omega^2_{\rm d}/\overline{\Omega}_{\rm o}(T=0)$ as interaction have little effect on the frequency renormalization in the low-$T$ limit. We then expect $\kappa \sim e^{-\beta \overline{\Omega}_{\rm o}/2} \to 0 $ as $T/\Omega_v\to0$. In particular, notice that $\kappa$ increases as we approach from high- to intermediate-$T$ and eventually decrease at low-$T$. This implies that $\kappa$ attains a maximum at some intermediate temperature that is related to the onset of the low-$T$ behavior. Indeed, this maximum can be seen in the numerical solution of $\kappa(T)$, see Fig.~\hyperref[fig:results]{\ref{fig:results}b}.   

For systems with multiple optical branches, using the same approximation as above, the thermal conductivity is given by a weighted average over the contribution of the different branches: $\kappa = \sum_b n_b \kappa_b$ where $\kappa_b$ is defined by replacing $\Omega_{\rm d} \to \Omega_{{\rm d}b}$ and $\mathcal{A}_{\rm o} \to \mathcal{A}_b$ in (\ref{eq:kappa_o}). We find that systems with multiple optical branches have two interesting implications on thermal transport. 

\begin{figure}[t]
\centering

\includegraphics[width=\columnwidth]{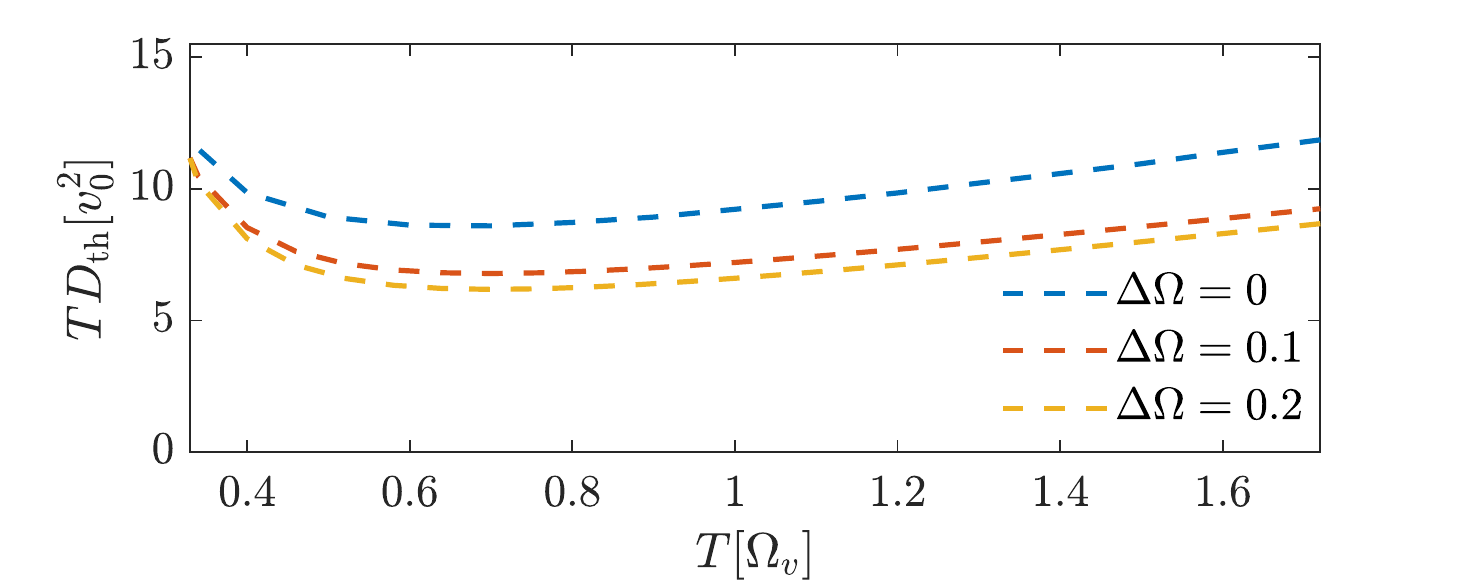}
\caption{{Thermal diffusion in systems containing multiple optical branches.} We plot $TD_{\rm th}$ as a function of $T$ for three models with different bandwidths. In all cases, $\Omega_{\rm max} = 1.1$, while $\Omega_{\rm min}=1.1,1,0.9$. Ten branches are distributed uniformly between $\Omega_{\rm min}$ and $\Omega_{\rm max}$. Data shown with $u/\Omega_v^3=1.4$ and $v_0\equiv\Omega_{\rm  d}^2/\Omega_{\rm max}= 0.007\Omega_v$.} 
\label{fig:MB_planckian_test}
\end{figure}

Firstly, we study the effect of broadening the bandwidth on the intermediate-$T$ ``phonon fluid'' regime, qualitatively identified as the region where diffusivity satisfies $D \propto1/T$. Namely, we consider a system with ten optical branches uniformly distributed between $\Omega_{\rm min}$ and $\Omega_{\rm max}$, where the bandwidth is $\Delta\Omega \equiv \Omega_{\rm max} - \Omega_{\rm min}$. We set $\Omega_{{\rm d}b}=\Omega_{\rm d}$ for all $b$, fix $\Omega_{\rm max}$ and evaluate $D_{\rm th}(T)$ as a function of $\Delta\Omega$ in the vicinity of the ``phonon fluid'' regime. We find that increasing $\Delta\Omega$ expands the ``phonon fluid'' regime and lowers the values of the diffusivity in this regime, making it more ``Planckian''. Indeed, the presence of a finite bandwidth actually increases the effective velocity associated with thermal transport, due to weaker renormalization of the phonon frequencies (compared to the $\Delta\Omega=0$ case). It then follows that the decrease in the value of $D_{\rm th}$ is due to shorter transport times. See Fig.~\ref{fig:MB_planckian_test} for $D_{\rm th}$ in a representative system with multiple optical branches. 

Secondly, notice that $D_{\rm th}$ can be made arbitrarily small in a model with multiple branches by phase space considerations \cite{wu_classical_2021}. Indeed, consider a model with two optical branches such that $\Omega_1=\Omega_2 \equiv \Omega_{\rm o}$ and $\Omega_{{\rm d}1}>\Omega_{{\rm d}2}=0 $. Namely, only $\phi_1$ carries heat in the system. Since $\kappa= n_1 \kappa_1$ and $c$ is independent of $\Omega_{\rm d}$ in the weak dispersion limit and independent of $n_1,n_2$ since $\Omega_1=\Omega_2$, we obtain that $D_{\rm th} = n_1\kappa_1/c$. Hence, by decreasing $n_1$ we effectively increase the specific heat of the system $c_{\rm eff} = c/n_1$, such that $D_{\rm th} \to 0$ as $n_1 \to 0$. This simplistic demonstration can be easily generalized to any number of optical branches or to systems containing acoustic branches (if $d>2$). However, this manipulation does not break down the correspondence between $D_{\rm th}$ and $D_L$. See further discussions in Sec~\ref{section:MB_details} and Sec.~\ref{sec:disc}.

\subsection{Acoustic phonons}

Consider the contribution of acoustic modes to thermal transport. Let us first recall the effect of long-wavelength acoustic modes on $\kappa$. For this purpose, it is useful to invoke the Boltzmann expression for the thermal conductivity $\kappa = \int_k c_k v_k^2 \tau_k$. Indeed, since $\tau_k\propto 1/k^2$, irrespective of the temperature, there will be a non-zero measure in $k$-space of acoustic modes that are sufficiently long-lived for which the Boltzmann expression holds. For these modes, since $v_k \approx \overline{v}_{s}$ and $c_k \sim 1$, we have that 
\begin{eqnarray}
\kappa_{\rm a} = n_{\rm a}c\overline{v}_{s}^2 \int_k \tau_k \sim \int {\rm d}k \frac{k^{d-1}}{k^2}\sim \begin{cases}
  \mathcal{V} & d=1 \\
  \log\mathcal{V} & d = 2 \\
  {\rm finite} & d \geq 3
\end{cases}
\label{eq:botlz_argument}
\end{eqnarray}
where $\mathcal{V}$ is the volume of the system. Namely, $\kappa \to \infty$ in the thermodynamic limit for $d=1,2$ due to the large phase-space of low-lying momentum states, whereas beyond the critical dimension, $d=2$, their contribution is suppressed due to the their limited phase space. 

Let us proceed to study the contribution of acoustic phonons to $\kappa$ for $d=3$. We focus on the $n_{\rm a} \ll n_{\rm o}$ limit.
%Our main purpose in the remainder of this subsection is to clarify under what conditions do optical phonons dominate thermal transport. We will see that even relatively small $n_{\rm a}$ may lead to a dominance of the acoustic modes due to the weakly dispersive limit. We postpone further discussion on this matter to Sec.~\ref{sec:3d}. 
Consider a system with a single acoustic branch and a single optical branch. Generalizing the following to multiple branches is straightforward. To leading order in $n_{\rm a}$ and $\Omega_{\rm d}/\overline{\Omega}_{\rm o}$, the thermal conductivity can be written as $\kappa = n_{\rm a} \kappa_{\rm a} + n_{\rm o}\kappa_{\rm o}$, where $\kappa_{\rm a/o}$ corresponds to the contributions from acoustic/optical phonons, respectively. $\kappa_{\rm o}$ is given by (\ref{eq:kappa_o}) with replacing $\sin^{2}k\mathcal{A_{\rm o}}\left(\nu,k\right)^{2}$ by $\epsilon\left(\boldsymbol{k}\right)\mathcal{A_{\rm o}}\left(\nu,\boldsymbol{k}\right)^{2}$, where $\epsilon\left(\boldsymbol{k}\right)\equiv\sin^{2}k_{x}+\sin^{2}k_{y}+\sin^{2}k_{z}$, and replacing the integration over $k$ to an integral over the 3-dimensional Brillouin zone. 

We have previously noted that the physically relevant velocity of the acoustic modes is the renormalized speed of sound. In particular, we expect the thermal current carried by acoustic modes to propagate with this velocity. Indeed, we find that the anharmonic contributions to the thermal current operator of the acoustic phonons sets the renormalized speed of sound as the velocity of the thermal current of the acoustic modes. Considering the retarded current correlation function of the acoustic phonons, these contribution correspond to the diagrams in Fig.~\hyperref[fig:thcurrent]{\ref{fig:thcurrent}b}. By computing their contribution to $\kappa_{\rm a}$, we obtain
\begin{eqnarray}
\kappa_{\text{a}}=\int_{\boldsymbol{k},\nu}\frac{\overline{\Omega}_{\text{a}}^{4}\left(\nu\right)}{2}\epsilon\left(\boldsymbol{k}\right)\mathcal{A}_{\text{a}}^2\left(\nu,\boldsymbol{k}\right)\left(\beta\nu\right)^{2}\text{csch}^{2}\left(\frac{\nu\beta}{2}\right)
\label{eq:kappa_a}
\end{eqnarray}
where $\overline{\Omega}_{\text{a}}^{2}\left(\nu\right)\equiv\Omega_{\text{a}}^{2}+\text{Re}\Pi_{R\text{o}}\left(\nu\right)$. The appearance of $\overline{\Omega}_{\text{a}}^{4}=\overline{v}_s^4$ as a prefactor confirms that thermal current carried by acoustic modes propagates at the renormalized speed of sound. This can be understood in analogy to the case of optical modes, where the prefactor $\Omega_{\rm d}^4$ correspond to the appearance of the renormalized velocity squared $\Omega_{\rm d}^4/\overline{\Omega}_{\rm o}^2 = \overline{v}_{\rm o}^2$. In the case of acoustic phonons, we instead find that $\overline{\Omega}_{\text{a}}^{4}(\nu=0)$ correspond to the velocity $\overline{\Omega}_{\text{a}}^{4}/\overline{\Omega}_{\text{a}}^{2}=\overline{\Omega}_{\text{a}}^{2}=\overline{v}_s^2$ (see App.~\ref{SK_app}). 

Note that in the weakly dispersive limit, $\overline{v}_s \gg \overline{v}_{\rm o}$ due to strong renormalization of the speed of sound. Hence, the small $n_{\rm a}$ limit, where optical modes dominate transport, should be taken such that $n_{\rm a} \overline{v}_s^2 \ll n_{\rm o} \overline{v}_{\rm o}^2$, which corresponds to $r_{\rm ao} \ll 1$ as we discussed in Sec.~\ref{sec:main_results}.

\section{Many-body quantum Chaos}
\label{sec:chaos}

In this section, we study many-body quantum chaotic properties of our system by analyzing the time and space dependence of the OTOC. We begin by considering a model with a single optical branch. We will then generalize our method to multiple optical branches. This generalization will enable us to address the case of scrambling in the presence of multiple time scales and multiple velocity scales, from which we will gain some intuition that will be helpful when we will finally consider scrambling in the presence of acoustic phonons. 

We diagnose chaos by considering the regularized OTOC\footnote{Dependence on the regularization of the OTOC \cite{liao_nonlinear_2018,grozdanov_kinetic_2019,romero-bermudez_regularization_2019} is not expected in our model \cite{kobrin_many-body_2021}.}, defined as 
\begin{eqnarray}
C\left(1,2\right) \equiv \frac{-1}{N^2} \sum_{i,j}{\rm Tr}\left( \sqrt{\rho}\left[\phi_{i}\left(1\right),\phi_{j}\right]\sqrt{\rho}\left[\phi_{i}\left(2\right),\phi_{j}\right] \right), \nonumber \\
\label{eq:otoc_definition}
\end{eqnarray}
where $1\equiv (t_1,\boldsymbol{r}_1)$, $\phi_j \equiv \phi_j(0,\boldsymbol{0})$ and $\rho = e^{-\beta H}/Z$ is the thermal density matrix. In practice, we shall study the behavior of $C(t,\boldsymbol{r})\equiv C(1,1)$, which is expected to satisfy
\begin{eqnarray}
C(t,\boldsymbol{r}) \sim \frac{f}{N} {\rm exp} \left(\lambda_L t - \frac{\boldsymbol{r}^2}{D_L t}\right),
\label{eq:generic_otoc_behavior}
\end{eqnarray}
where $\lambda_L = 1/\tau_L$ is the scrambling rate (or quantum Lyapunov exponent) that satisfies a universal bound $\lambda_L \leq 2\pi T$ \cite{maldacena_bound_2016}, and $D_L$ is the chaos diffusivity. The exponential growth in (\ref{eq:generic_otoc_behavior}) is expected to hold up to some intermediate time scale $ t \lesssim \tau_L \log N$, called the scrambling time $t_{\rm scr}$. This is the time over which information encoded in $\mathcal{O}(1)$ degrees of freedom spreads into $\mathcal{O}(N)$ degrees of freedom, within the unit-cell, and becomes essentially inaccessible. Moreover, the diffusive propagation in (\ref{eq:generic_otoc_behavior}) is expected to break down at distances $|\boldsymbol{r}| \sim r_*$, roughly defined by $C(t,\boldsymbol{r}_*)^{-1} \partial_t C(t,\boldsymbol{r}_*) \approx 2 \pi T$, since (\ref{eq:generic_otoc_behavior}) is contradictory to the bound on chaos for $|\boldsymbol{r}|>r_*$ \cite{chowdhury_onset_2017}. These effects are attributed to higher orders in $1/N$ and are beyond the scope of our work. 

In addition to $\lambda_L$ and $D_L$, we identify the butterfly velocity $v_B \equiv \sqrt{\lambda_L D_L}$ by equating the two arguments in the exponent in (\ref{eq:generic_otoc_behavior}). This velocity is associated with an emergent effective light cone of information scrambling, in which the OTOC is `space-filling'. The emergence of an effective light cone is typical in quantities governed by an unstable temporal exponential growth that spreads diffusively in space \cite{aleiner_microscopic_2016}. 

\begin{figure}[t]
\centering

\includegraphics[width=\columnwidth]{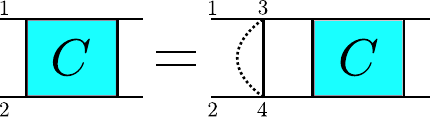}
\caption{{Diagrammatic representation of the Bethe-Salpeter equation for the OTOC (\ref{eq:kernel_eqn}). Dashed line denotes averaging over realizations of $v_{ijk}$.}} 
\label{fig:otoc_BSE}
\end{figure}

Let us proceed to consider the OTOC of a single optical branch, defined by $\Omega_{\rm o}$ and $\Omega_{\rm d}$. At order $1/N$, the exponential growth of the OTOC is governed by a BSE, represented diagrammatically in Fig.~\ref{fig:otoc_BSE}, 
\begin{eqnarray}
C(1,2) &=& \int_{3,4}\mathcal{K}(1,2,3,4)C(3,4), \nonumber \\
\mathcal{K}(1,2,3,4) &=& 2v^2 G_R(13)G_R(24)G_W(34),
\label{eq:kernel_eqn}
\end{eqnarray}
where $13\equiv (t_1 - t_3,\boldsymbol{r}_1 - \boldsymbol{r}_3)$, for example, and $G_W(t,\boldsymbol{r}) \delta_{ij} \equiv {\rm Tr}\left( \sqrt{\rho}\phi_i(t,\boldsymbol{r})\sqrt{\rho} \phi_j  \right)$ is the Wightman Green's function. $\mathcal{K}$ is the retarded ladder kernel, where rungs corresponds to a single Wightman function due to the cubic interaction, and rails are given by retarded Green's function. In the remaining of this section, we set $d=1$ for simplicity. Higher dimensions follow a similar treatment and do not change the physical picture. 

To proceed, we define the center of mass and relative coordinates: $a_+ \equiv \frac{a_1+a_2}{2},a_- \equiv a_1 - a_2, {\rm } a=t,r,k$ ($k$ denotes the lattice-momentum). Anticipating the behavior in (\ref{eq:generic_otoc_behavior}), we use the following ansatz, 
\begin{eqnarray}
C(1,2) = \int_{k_1,k_2} e^{ik_1r_1 + ik_2r_2}F(t_-,k_-)e^{\lambda_L(k_+)t_+}.
\label{eq:OTOC_ansatz_def}
\end{eqnarray}
Here, we assume the dominant momentum dependence is encoded in the chaos exponent $\lambda(k_+)$ and neglect the $k_+$-dependence of the coefficient $F$, assuming it is a non-singular function\footnote{Similarly to \cite{chowdhury_onset_2017}, but unlike some holographic theories, where the singular structure of the coefficient determines the spatial decay of the OTOC, see, e.g., \cite{shenker_stringy_2015,gu_energy_2017}.} that weakly depends on momentum. This assumption is motivated by the weakly dispersive limit, as our system is expected to be smoothly connected to a system with $\Omega_{\rm d}= 0$, which is completely momentum independent. Importantly, this assumption is invalid for systems with a sufficiently large fraction of (strongly momentum-dependent) acoustic phonons, as we will discuss later on. 

Note that (\ref{eq:generic_otoc_behavior}) is related to (\ref{eq:OTOC_ansatz_def}) by 
\begin{eqnarray}
C(t_+,r_+) = \int_{k_+,k_-} e^{2ik_+r_+}F(0,k_-)e^{\lambda_L(k_+)t_+}.
\end{eqnarray}
Hence, by expanding $\lambda_L(k_+) = \lambda_0 + \lambda_1 k_+^2 + \mathcal{O}(k_+^2)$ we may directly extract $D_L = -\lambda_1$. Note also that $\lambda_0 \equiv \lambda_L$ corresponds to the spatially averaged scrambling rate, related to the growth of $\int_r C(t,r)$. Furthermore, it will be convenient to define $f(t_-) \equiv \int_{k_-} F(t_-,k_-)$, with $f\equiv f(0)$ being the coefficient in (\ref{eq:generic_otoc_behavior}). We now proceed to describe the solution of (\ref{eq:kernel_eqn}). $\lambda_L$ and $D_L$ as a function of $T$ are given for a representative set of parameters in Fig.~\ref{fig:results}. 

\subsection{Scrambling rate of a single optical branch}
\label{section:scrambling_details}
We extract $\lambda_L$ numerically following a procedure in the spirit of Refs. \cite{maldacena_remarks_2016,banerjee_solvable_2017}. Namely, by substituting the ansatz (\ref{eq:OTOC_ansatz_def}) into (\ref{eq:kernel_eqn}), we recast (\ref{eq:kernel_eqn}) to an eigenvalue equation of the form $K|f\rangle = e(\lambda_L) |f\rangle$. Then, $\lambda_L$ is determined by demanding the eigenvalue $e(\lambda_L) = 1$. Explicitly, the eigenvalue equation is given by
\begin{eqnarray}
e \left( \lambda_L \right) f(t) &=& \int_{t'} K(t,t',k_+) f(t') , \label{eigen_value_eqn_for_f} \\
K(t,t',k_+) &=& 2v^2 \int_{k_-} h_{\lambda_L}(t-t',k_+) G_W(t') ,\nonumber \\
h_{\lambda_L}(t,k_+,k_-) &\equiv& \int_{\bar{t}}g_{R}\left(\bar{t},k_{+}+\frac{k_{-}}{2}\right)g_{A}\left(t-\bar{t},k_{+}-\frac{k_{-}}{2}\right) \nonumber
\end{eqnarray}
where $g_{R/A}(t,k) \equiv e^{\mp \lambda_L \frac{t}{2}} G_{R/A}(t,k)$. In the derivation of (\ref{eigen_value_eqn_for_f}) we used the weak dispersion limit to approximate $G_W(t,k) \approx G_W(t) = \int_k G_W(t,k)$.

This procedure allows us to extract the entire function $\lambda_L(k_+)$. The spatially averaged $\lambda_L$ is extracted by setting $k_+=0$, while $D_L$ can be extracted from a quadratic fit of $\lambda_L(k_+)\approx \lambda_L - D_L k_+^2 $ for sufficiently small values $k_+$. In the next subsection, we describe another simple semi-analytical approach that enables us to extract $D_L$ in a more physically transparent manner.   

\subsection{Chaos diffusivity of a single optical branch}
\label{sec:diffusion_details}

In this subsection we derive a simple perturbative approach to directly compute $D_L$. This approach will also be handy in the discussion on scrambling and chaos diffusion in the presence of acoustic phonons, as we will see later on.

Suppose we know the solution for the $k_+=0$ eigenvalue equation: $K_0|f_0\rangle = e |f_0\rangle$, i.e., $\lambda_L$ and the eigenvector $|f_0\rangle$, and that we want to know what is leading order correction for $k_+ \ne 0$. Let us drop the subscript of $k_+$ for brevity: $k_+ \to k$. We may expand the eigenvalue equation for small $k\ne0$:  
\begin{eqnarray}
\left( K_0 + \delta K \right) \left( |f_0\rangle + |\delta f\rangle \right) = (e+\delta e)\left( |f_0\rangle + |\delta f\rangle \right).
\end{eqnarray}
By multiplying the above with the \textit{left} eigenvector (note that $K$ is not symmetric - $\langle f_0 | \ne (|f_0\rangle)^{\dagger}$) we find that 
\begin{eqnarray}
\langle f_0 |\delta K |f_0\rangle = \delta e \langle f_0 |f_0\rangle + \mathcal{O}(\delta^2) .
\end{eqnarray}
In order to satisfy the kernel equation, we must require that $\delta e =0$, which is true when $\langle f_0 |\delta K |f_0\rangle  =0 $.

Expanding $K$ for small $k$ and $\delta \lambda_L$, we write $\delta K = \delta \lambda_L A + k^2 B$. Hence, $\delta \lambda_L \langle f_0 |A |f_0\rangle = -k^2 \langle f_0 |B |f_0\rangle 
$, from which one can identify
\begin{eqnarray}
D_L = \frac{\langle f_0 |B |f_0\rangle }{\langle f_0 |A |f_0\rangle }.
\label{eq:direct_D_L}
\end{eqnarray}
Here, $A \equiv \partial_{\lambda_L(k)} K|_{\lambda_L(k) = \lambda_L(0)} $ and $k^2B$ is obtained by expanding all terms in $K$ to leading (non-vanishing) order in $k$. The fact that the leading order correction is proportional to $k^2$ is a consequence of inversion symmetry. We supply explicit expressions for $A$ and $B$ in App.~\ref{SK_app}.   

We find that $D_L$ extracted from (\ref{eq:direct_D_L}) is identical to the straightforward extraction from the computation of $\lambda_L(k)$. Moreover, (\ref{eq:direct_D_L}) allows for some analytical control over $D_L$. For example, we observe that $D_L \propto \Omega_{\rm d}^4$ since $B \propto \Omega_{\rm d}^4$. Then, using $v_B = \sqrt{D_L \lambda_L}$ implies that $v_B \propto \Omega_{\rm d}^2$, allowing us to tune $v_B$. By dimensional arguments we may further argue that $v_B \sim \Omega_{\rm d}^2/\overline{\Omega}_o = v_{\rm o}$. Indeed, the numerical solution yields $v_B = v_{\rm o}$ with an error of roughly $5\%$. We will see more examples in the following subsections. 

\subsection{Generalization to multiple branches}
\label{section:MB_details}
Consider a system with $N_B$ optical branches, defined by $\rho(\boldsymbol{\Omega}) = \sum_b n_b \delta(\boldsymbol{\Omega} - \boldsymbol{\Omega}_b)$. We define the branch-dependent OTOC for branches $a$ and $b$ (no sum): 
\begin{eqnarray}
C_{ab}\left(1,2\right) \equiv -{\rm Tr}\left( \sqrt{\rho}\left[\phi_{a}\left(1\right),\phi_{b}\right]\sqrt{\rho}\left[\phi_{a}\left(2\right),\phi_{b}\right] \right).
\label{eq:MB_otoc_definition}
\end{eqnarray}

The BSEs now contains cross scrambling terms between the different branches:
\begin{eqnarray}
C_{ab}(1,2) = \sum_c n_c \int_{3,4} \mathcal{K}_{ac}(1,2,3,4) C_{cb}(3,4).
\label{eq:BSE_MB_otoc}
\end{eqnarray}
The kernels are given by $\mathcal{K}_{ab}(1,2,3,4) = 2v^2 G_{Ra}(13)G_{Ra}(24)G_W(34)$, where $G_W = \sum_b n_b G_{Wb}$. Note that $\mathcal{K}_{ab}$ depends only on $a$ since the external legs in $C_{ab}$ are fixed. The summation over branches in $G_W$ and $C_{cb}$ in the right-hand-side of (\ref{eq:BSE_MB_otoc}) comes from contractions with two interaction vertices.

Consider the case of $N_B=2$. The BSEs can be conveniently written as a matrix equation in the branch-space, where they factorize to blocks of size $N_B$ (in this case, 2): 
\begin{eqnarray}
\begin{pmatrix}
C_{11}  \\ 
C_{21}  
\end{pmatrix}
= 
\begin{pmatrix}
n_1 \mathcal{K}_{11} & n_2 \mathcal{K}_{12}  \\ 
n_1 \mathcal{K}_{21} & n_2 \mathcal{K}_{22} 
\end{pmatrix}
\begin{pmatrix}
C_{11}  \\ 
C_{21}  
\end{pmatrix}  ,
\label{eq:MB_otoc_before_ansatz}
\end{eqnarray}
and identically for $C_{12},C_{22}$, such that it is sufficient to solve the BSE of one block to characterize the chaos in the system. The identical BSEs of the two blocks hints that the scrambling rate is an intrinsic property of the system, rather than a property of the individual branches. We discuss this further later on. We proceed by assuming the system has a unique scrambling rate and generalize (\ref{eq:OTOC_ansatz_def}) to
\begin{eqnarray}
\begin{pmatrix}
C_{11}(1,2)  \\ 
C_{21}(1,2) 
\end{pmatrix}
=  \int_{k_1,k_2} e^{i k_1 r_1 + i k_2 r_2} e^{\lambda_L(k_+)t_+}
\begin{pmatrix}
F_1(t_-,k_-)  \\ 
F_2(t_-,k_-)
\end{pmatrix} , \nonumber\\ 
\label{eq:MB_OTOC_ansatz_def}
\end{eqnarray}
such that (\ref{eq:MB_otoc_before_ansatz}) can be recasted into 
\begin{eqnarray}
\begin{pmatrix}
|f_{1}\rangle   \\ 
|f_{2} \rangle
\end{pmatrix}
=  
\begin{pmatrix}
n_1 K_{1} & n_2 K_{1}  \\ 
n_1 K_{2} & n_2 K_{2} 
\end{pmatrix}
\begin{pmatrix}
|f_{1}\rangle   \\ 
|f_{2} \rangle
\end{pmatrix},
\label{eq:MB_otoc_after_ansatz}
\end{eqnarray}
 where $K_b$ is defined as in (\ref{eigen_value_eqn_for_f}) with replacing $g_{R/A}$ with $g_{R/Ab}$ of the corresponding branch in $h_{\lambda_L}$. At this point, we may solve for $\lambda_L$ and $D_L$ using the methods we described in subsections \ref{section:scrambling_details} and \ref{sec:diffusion_details}. The generalization to $N_B>2$ is straightforward.  
 
 \subsection{Chaos with multiple scales}
 \label{sec:multiplebranches}
In this subsection, we consider a system with multiple optical branches and make a few observations regarding many-body quantum chaos in a system with multiple time and velocity scales. The intuition we will gain here will help us understand the more involved case of chaos in the presence of acoustic and optical phonons. 

Whether $\lambda_L$ is intrinsic to the system or varies between different degrees of freedom is an interesting question. While $\lambda_L$ is intrinsic in most generic systems, exceptions may arise in cases where different operators belong to different sectors of the system \cite{lunts_many-body_2019}. In our system, the uniqueness of the scrambling rate is supported by the following observation. 

Consider a system with two optical phonon branches ($N_B=2$) and suppose that there exist two scrambling rates $\lambda_1 > \lambda_2$. The OTOCs can be written as 
\begin{eqnarray}
\begin{pmatrix}
C_{11}(t)   \\ 
C_{21}(t)
\end{pmatrix}
\sim 
\begin{pmatrix}
f_1e^{\lambda_1 t} \\ 
f_2e^{\lambda_2 t}
\end{pmatrix}.
\end{eqnarray}
Since the same BSE governs the dynamics of $C_{12}$ and $C_{22}$, it follows that $C_{12}(t) \sim e^{\lambda_1 t} \ne C_{21}$. However, this contradicts the fact that $C_{12}$ and $C_{21}$ grows with the \textit{same} rate, which is an immediate consequence of time-reversal symmetry. Hence, $\lambda_1 > \lambda_2$ is an inconsistent solution of the BSEs. Another argument can be made by writing the OTOCs as 
\begin{eqnarray}
\begin{pmatrix}
C_{11}(t)   \\ 
C_{21}(t)
\end{pmatrix}
=  
e^{\lambda_1 t}
\begin{pmatrix}
f_1 \\ 
e^{\Delta \lambda_L t}f_2
\end{pmatrix}\equiv 
e^{\lambda_1 t}
\begin{pmatrix}
f_1 \\ 
\tilde{f}_2(t)
\end{pmatrix},
\end{eqnarray}
where $\Delta \lambda_L = \lambda_2 - \lambda_1 < 0$. Notice that $\tilde{f_2}(t) \to 0$ after a sufficiently long time, such that $\phi_2$ effectively decouples itself from $\phi_1$. However, since $f_2>0$, some weight of $\phi_2$ has already been transferred to $\phi_1$, and this non-zero weight now grows with a rate $\lambda_1$, rendering this decoupling implausible. 

\begin{figure}[t]
\centering

\includegraphics[width=\columnwidth]{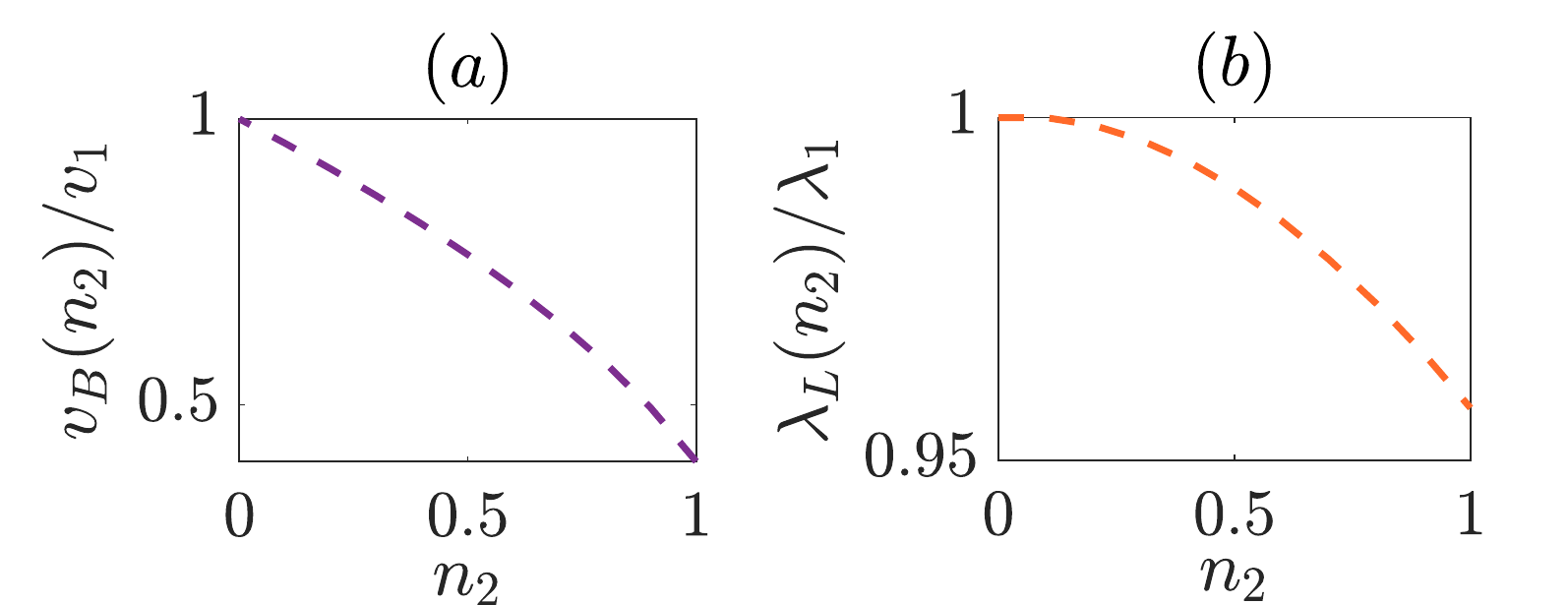}
\caption{{Chaos in a system containing two optical phonon branches.} (a) $v_B(n_2)/v_1$ as a function of $n_2$ for two optical branches with $\Omega_{{\rm d}1}^2/\Omega_{{\rm d}2}^2 = 0.4$. $v_B$ satisfies $v_B=\sqrt{n_1 v_1^2+n_2 v_2^2}$ as expected from the formula below (\ref{eq:D_L_2b_exp}). (b) $\lambda_L(n_2)/\lambda_1$ as a function of $n_2$. In (a), $\Omega_1=\Omega_2=1$ such that only the velocities are different between the two branches. In (b), $\Omega_1/\Omega_v=1<\Omega_2/\Omega_v=1.1$ such that $\tau_1 < \tau_2$ for all $n_1,n_2$. Note that $\lambda_1>\lambda_2$ as expected from $\lambda_L\propto 1/\tau_{\rm ph}$, where $\lambda_1=\lambda_L(n_1=1)$ and similarly for $\lambda_2$. Data shown with $T/\Omega_v = 1$ and $u/\Omega_v^3 = 1.35$. } 
\label{fig:2bgames}
\end{figure}

In Fig.~\hyperref[fig:2bgames]{\ref{fig:2bgames}b}, we plot $\lambda_L$ in a system with two optical branches as function of their relative fractions. This demonstrates that $\lambda_L$ is determined by all degrees of freedom. Indeed, consider the limit $n_1 \ll n_2$. To zeroth order in $n_1$, $\lambda_L$ is determined solely by the BSE of $\phi_2$: $|f_2\rangle = K_2 |f_2\rangle$ in (\ref{eq:MB_otoc_after_ansatz}), while the eigenvector related to cross scrambling with $\phi_1$ is given by $|f_1\rangle = K_1 |f_2 \rangle$. The same is true in the opposite limit, implying that for general $n_1,n_2$, $\lambda_L$ must extrapolate between the two cases.   
Notice, however, that while all degrees of freedom affect $\lambda_L$, the effect of operators with a short lifetime is more pronounced, see Fig.~\hyperref[fig:acousticchaos]{\ref{fig:acousticchaos}b}.

Let us further ask what determines $v_B$ in a system where different degrees of freedom propagate with distinct velocities. Intuitively one can think of quantum information as propagating in an `effective medium' composed of all degrees of freedom in the system. Each degree of freedom carries information with its corresponding velocity while scattering events average out the net velocity with which the information propagates, such that $v_B$ is set by the velocity of the `effective medium,' rather than the largest velocity scale. Indeed, consider the system from above with $\Omega_1 = \Omega_2=\Omega$, $n_1 =n_2 = 1/2$ and general $\Omega_{\rm d1},\Omega_{\rm d2}$. The velocities of the two branches are defined as $v_b = \Omega_{{\rm d}b}^2/\overline{\Omega}$ for $b=1,2$. In the weak dispersive limit, $\lambda_L$ and $|f_0\rangle$ are independent of $\Omega_d$ for the two branches. Then, we use (\ref{eq:direct_D_L}) to obtain 
\begin{eqnarray}
D_L = \frac{1}{2}\langle A \rangle_0 ^{-1} \left( \langle B_1 \rangle_0 + \langle B_2 \rangle_0 \right)
\end{eqnarray}
where $\langle \cdot \rangle_0 \equiv  \langle f_0| \cdot | f_0\rangle$. Recall that $B_b \propto \Omega_{{\rm d}b}^4$ and that in the case of a system with a single optical branch we have found that $D_L = \tau_L v_{\rm o}^2 = \tau_L \Omega_{\rm d}^4 / \overline{\Omega}^2$. This implies that we can write $\langle A \rangle_0^{-1} \langle B_b \rangle_0 = \tau_L v_b^2$, such that 
\begin{eqnarray}
D_L = \frac{1}{2} \tau_L (v_1^2 + v_2^2).
\label{eq:D_L_2b_exp}
\end{eqnarray}
Namely, $v_B = \sqrt{(v_1^2 + v_2^2)/2}$.

Following the same steps as above, one may conclude that in a system with $N_B$ optical branches, in the weak dispersion limit, $D_L = \tau_L \sum_b n_bv_b^2$, see Fig.~\hyperref[fig:2bgames]{\ref{fig:2bgames}a}. In particular, one can think of a case where only a small fraction of the system is dispersive. Let us denote this fraction by $n_{\rm d}$. For thermal diffusion, such a case implies that $D_{\rm th} \propto n_{\rm d}$ (see Sec.~\ref{op_results}). For chaos diffusion, we see that the same proportionality holds, $D_L \propto n_{\rm d}$. In particular, this trick does not provide a way around the observed correspondence, $D_{\rm th} \sim D_L$.

\subsection{Chaos with acoustic phonons}

Consider a system with a single optical branch and a single acoustic branch. The generalization to multiple acoustic and optical branches is straightforward. Let us focus on $d=1$ for simplicity. The OTOCs of the acoustic phonons are defined by (\ref{eq:MB_otoc_definition}), identically to the optical phonons. The difference in the BSEs is due to the fact that the interaction vertices in the retarded kernel now contain the generalized fields $\widetilde{\phi}$ defined in (\ref{eq:generalized_phi}). This leads to the following BSEs: 
\begin{eqnarray}
\begin{pmatrix}
C_{\rm aa}  \\ 
C_{\rm oa}  
\end{pmatrix}
= 
\begin{pmatrix}
n_{\rm a} \mathcal{K}_{\rm aa} & n_{\rm o} \mathcal{K}_{\rm ao}  \\ 
n_{\rm a} \mathcal{K}_{\rm oa} & n_{\rm o} \mathcal{K}_{\rm oo} 
\end{pmatrix}
\begin{pmatrix}
\widetilde{C}_{\rm aa}  \\ 
\widetilde{C}_{\rm oa}  
\end{pmatrix} ,
\label{eq:OTOC_for_AO_model}
\end{eqnarray}
where $ \mathcal{K}_{ab}(1,2,3,4) = 2v^2\hat{G}_{Ra}(13)\hat{G}_{Ra}(24)\widetilde{\mathcal{G}}_W(34)$, the generalized Green's function and OTOCs in the right-hand-side are defined as 

\begin{eqnarray}
\hat{G}_{a}(t,r)  &=&
    \begin{cases}
      {G}_{a}(t,r+1)- {G}_{a}(t,r) &  a\in I_\textnormal{a}, \\
      {G}_{a}(t,r) &  a\in I_\textnormal{o},  \\
    \end{cases}  \\
      \widetilde{C}_{ab}(1,2)  &=&
    \begin{cases}
      {C}_{ab}(1+e_r,2+e_r) + {C}_{ab}(1,2)  \\
      -{C}_{ab}(1+e_r,2)-{C}_{ab}(1,2+e_r) &  a\in I_\textnormal{a}, \\
      {C}_{ab}(1,2)&   a\in I_\textnormal{o},  \\
    \end{cases} \nonumber
        \label{eq:generalized_OTOCs}
\end{eqnarray}
and $1+e_r \equiv (t,r+1)$, for example. $\widetilde{\mathcal{G}}_W$ is defined below (\ref{eq:Keldysh_sc_eqs_AO_model}). We proceed by substituting our previous ansatz (\ref{eq:OTOC_ansatz_def}) for $C_{\rm aa},C_{\rm oa}$. Then, following similar steps to the derivation of (\ref{eigen_value_eqn_for_f}), we arrive at the following BSEs,
\begin{eqnarray}
\begin{pmatrix}
|f_{\rm a}\rangle   \\ 
|f_{\rm o} \rangle
\end{pmatrix}
=  
\begin{pmatrix}
n_{\rm a} K_{\rm aa} & n_{\rm o} K_{\rm ao}  \\ 
n_{\rm a} K_{\rm oa} & n_{\rm o} K_{\rm oo} 
\end{pmatrix}
\begin{pmatrix}
|f_{\rm a}\rangle   \\ 
|f_{\rm o} \rangle
\end{pmatrix},
\label{eq:AO_BSEs_after_ansatz}
\end{eqnarray}
where 
\begin{eqnarray}
K_{ab}(t,t',k_+) &=& 2v^2 \int_{k_-} s_{ab}(k_+,k_-) \nonumber \\
&&\quad \quad \quad \times h_{\lambda_L}^{ab}(t-t',k_+,k_-) \widetilde{\mathcal{G}}_W(t') ,\nonumber \\
h_{\lambda_L}^{ab}(t,k_+,k_-) &\equiv& \int_{\bar{t}}g_{Ra}\left(\bar{t},k_{+}+\frac{k_{-}}{2}\right) \nonumber \\ 
&& \quad\quad\quad\times g_{Aa}\left(t-\bar{t},k_{+}-\frac{k_{-}}{2}\right) \nonumber ,\\
s_{ab}&=&  
\begin{pmatrix}
s_1(k_+)s_2(k_+,k_-) & s_2(k_+,k_-)  \\ 
s_1(k_+) & 1 
\end{pmatrix}.
\label{eq:eigen_value_eqn_for_f_AO}
\end{eqnarray}
Here, $a,b={\rm a,o}$ are in correspondence with (\ref{eq:AO_BSEs_after_ansatz}), $\widetilde{\mathcal{G}}_W(t) \equiv \int_k \widetilde{\mathcal{G}}_W(t,k)$, and the functions $s_{ab}$ are related to the spatial couplings between the OTOCs in (\ref{eq:OTOC_for_AO_model}) and are defined as $s_1(k_+) \equiv \pi\left(\cos\left(k_{+}\right)+\cos\left(3k_{+}\right)\right)$ and $s_2(k_+,k_-) \equiv -4 \sin\left(\frac{k_{+}}{2}+\frac{k_{-}}{4}\right)\sin\left(\frac{k_{+}}{2}-\frac{k_{-}}{4}\right)$. Note that $h_{\lambda_L}^{ab}$ depends only on $a$. 

Note that the ansatz (\ref{eq:OTOC_ansatz_def}) relies on the weakly dispersive limit of the system. Here, we expect this ansatz to be valid in the limit $n_{\rm a}\to 0$, where acoustic phonons can be treated as a controlled perturbation to the BSEs with $n_{\rm a}$ being the small parameter that balances their strong momentum dependence. In this limit, the corrections to $\lambda_L$ and $D_L$ are linear in $n_{\rm a}$. In particular, the onset of nonlinear $n_{\rm a}$-dependence in the numerical evaluation of these quantities will signal the limit of validity for (\ref{eq:OTOC_ansatz_def}). 

In the remaining of this subsection we will show that $\lambda_L$ and $D_L$ are non-singular in the $n_{\rm a} \to 0$ limit, as we have claimed in Sec.~\ref{sec:main_results}. To do that, we assume that there exist an expansion of $\lambda_L$ and $D_L$ for small $n_{\rm a}$, such that $q_L(n_{\rm a}) = q_L(0) + n_{\rm a} \delta q_L + \mathcal{O}(n_{\rm a}^2)$, for $q=\lambda,D$. Then, we will show that the correction $\delta q_L$ is finite as it is expressed in terms of converging integrals of non-singular functions, which implies that our assumption is consistent. We also verify this argument numerically, where we find that $\lambda_L$ and $D_L$ are finite and grow linearly in $n_{\rm a}$ for sufficiently small values of $n_{\rm a}$.  

Consider the correction to $\lambda_L$. Following the steps of the derivation in Sec.~\ref{sec:diffusion_details} with $n_{\rm a}$ taking the role of the small perturbation (rather than $k_+$), we obtain that the correction to $\lambda_L$ is given by
\begin{eqnarray}
\delta \lambda_L \langle P \rangle _0 = - n_{\rm a} \langle Q \rangle _0,
\label{delta_lambda_corr}
\end{eqnarray}
where $P = \partial_{\lambda_L(n_{\rm a})} K|_{\lambda_L(n_{\rm a})=\lambda_L(0)}$, $Q = \partial_{n_{\rm a}} K|_{n_{\rm a} = 0}$. Here, $\langle \cdot \rangle _0$ are taken with respect to the $n_{\rm a}=0$ eigenvectors.  

Note that for $n_{\rm a}=0$, $\langle f_{\rm a}|=0$ because the top row in $K^T$ are proportional to $n_{\rm a}=0$ (see ($\ref{eq:AO_BSEs_after_ansatz}$)). This immediately implies that $\langle P \rangle _0$ is finite as it contains data of the optical branches alone: $ \langle P \rangle _0 = \langle f_{\rm o} |\partial_{\lambda_L} K_{\rm oo}| f_{\rm o} \rangle$. For  $ \langle Q \rangle _0$, we have that 
\begin{eqnarray}
\langle Q \rangle _0 &=& \langle f_{\rm o} |K_{\rm oa}| f_{\rm a} \rangle \nonumber \\
&=& \langle f_{\rm o} |K_{\rm oa} K_{\rm ao} | f_{\rm o} \rangle,
\end{eqnarray}
where we have used the fact that $| f_{\rm a} \rangle = K_{\rm ao} | f_{\rm o} \rangle$ for $n_{\rm a} = 0$. Since $K_{\rm oa}$ is composed out of non-singular exponentially decaying functions in time, the only issues may come from $K_{\rm ao}$, and in particular, from the function $h_{\lambda_L}^{\rm ao}(t,k_+,k_-)$ near $k_+,k_-=0$, since $\mathcal{G}_W$ is also non-singular. Namely, if $h_{\lambda_L}^{\rm ao}(t,k_+,k_-)$ is a non-singular function that decays sufficiently fast, we may conclude that $\langle Q \rangle _0$ is finite, showing that our assumption is consistent and thus the correction to $\lambda_L$ is finite. 

Indeed, note that 
\begin{eqnarray}
h_{\lambda_L}^{\rm ao}\left(\omega,k_+=0,k_-\right) &=&  \left|g_{R{\rm a}}\left(\omega,\frac{k_-}{4}\right)\right|^2 \nonumber \\
&=& \left|G_{R{\rm a}}\left(\omega + i\frac{\lambda_L}{2},\frac{k_-}{4}\right)\right|^2 
\end{eqnarray}
might be singular around $k_-=0$ only for $\lambda_L=0$. Since $\lambda_L(n_{\rm a}=0)>0$, these singularities are avoided. Furthermore, the fact that $\lambda_L>0$ also implies that $g_{R/A}$ - and therefore also $h_{\lambda_L}^{\rm ao}$ - are exponentially decaying in time. This concludes our discussion in the correction of $\lambda_L$.

Extending this argument to $D_L$ is immediate. Indeed, we may follow the steps above for $k_+ \ne 0$, with the minor modification that now the singularities of $h_{\lambda_L}^{\rm ao}$ in the $\lambda_L=0$ case are slightly shifted away from $k_-=0$. The fact that $\lambda_L(k_+\ne0,n_{\rm a}=0)>0$ implies that these singularities are avoided as in the previous case. Hence, the correction $\lambda_L(k_+\ne0)$ is finite, implying that the correction to $D_L$ is also finite because $\lambda_L(k_+) = \lambda_L - k_+^2 D_L+\mathcal{O}(k_+^2)$.  

Finally, let us comment on the correction to $D_L$ for $n_{\rm a}>0$. As demonstrated in Fig.~\ref{fig:acousticchaos}, this correction is rather violent, in comparison to the mild correction to $\lambda_L$. As we mentioned previously, this is due to the imbalance between $\overline{v}_{\rm o}$ and $\overline{v}_{s}$, which is an artifact of the weakly dispersive limit. This can be seen explicitly by considering the extension of (\ref{eq:direct_D_L}) to a system with acoustic phonons, where the sound velocity appears in the terms associated with the acoustic modes,
\begin{eqnarray}
D_L = \langle A \rangle_0^{-1} \left(\underbrace{ \langle B_{\rm aa} \rangle  +\langle  B_{\rm ao}\rangle }_{\sim n_{\rm a} \overline{v}_{s} ^2} + \underbrace{\langle B_{\rm oa}\rangle  + \langle B_{\rm oo}\rangle}_{\sim n_{\rm o} v_{\rm o}^2} \right),
\end{eqnarray}
where $\langle A \rangle _0$ is defined as in (\ref{eq:direct_D_L}) and $\langle B_{ab} \rangle\equiv \langle f_a| B_{ab} |f_b \rangle$ such that $a,b={\rm a,o}$ and $|f_a\rangle$ denotes the $n_{\rm a}>0,k_+=0$ eigenvectors, and similarly for $\langle f_a |$. 
We can now see that the large ratio $\overline{v}_{s}^2/v_{\rm o}^2$ accounts for the pronounced effect on the correction to $D_L$, and correspondingly on $v_B$ for relatively small values of $n_{\rm a}$. Explicit expressions are given in App.~\ref{SK_app}. 

\section{Diffusion in three dimensions}
\label{sec:3d}

In this section, we discuss the correspondence between thermal and chaos diffusion in a system with $n_{\rm a}>0$ in $d=3$. Let us first recall the physical picture in lower dimensions briefly. We have seen that the cases $n_{\rm a}=0$ and $n_{\rm a}>0$ are dramatically different in terms of transport, chaos, and their correspondence. In the absence of acoustic modes ($n_{\rm a}=0$) we found that $D_L\sim D_{\rm th}$, while for $n_{\rm a}>0$, this correspondence breaks down, since $D_{\rm th}$ diverges while $D_L$ remains finite. The divergence of the thermal diffusivity is rooted in the thermal conductivity that is dominated by long-lived, long-wavelength acoustic modes due to their relatively large phase space in lower dimensions. In $d=3$, the contribution of these modes is finite due to their rapidly vanishing phase space. Does this imply that correspondence $D_L \sim D_{\rm th}$ holds in $d=3$? 
We argue that, in a generic setting, it does, and demonstrate it explicitly in the $n_{\rm a} \ll n_{\rm o}$ limit under a spherical approximation of the three-dimensional Brillouin zone (BZ). However, a parametric violation is possible in a strongly anisotropic $3d$ system.

In principal, in order to study the diffusivities in $d=3$, one needs to solve the SPEs for the Green's functions and the BSE for the OTOC with a 3-dimensional BZ. A straightforward approach in this case is computationally costly. Here, we approximate $\boldsymbol{k} \in [-\pi,\pi]^3$ by $\boldsymbol{k} \in \mathcal{B}(k_{\rm reg})$ where $\mathcal{B}(k)\equiv \{\boldsymbol{k}\in\mathbb{R}^3;|\boldsymbol{k}|^2<k^2\}$, such that the dispersion is approximated to be isotropic (i.e., to be a function of $k\equiv|\boldsymbol{k}|$) in two approximation schemes. In the first scheme, $k_{\rm reg}$ is determined such that the volume of the BZ is preserved, with replacing dispersive terms by their $k\to0$ (``continuum'') limit. In the second scheme, we let $k_{\rm reg}=\pi$, and use the $1d$ expressions (i.e. starting with the first approximation scheme and replacing ${k}^2\mapsto 4\sin^2\left(\frac{k}{2}\right)$). The two approximations give qualitatively similar results, suggesting that they capture the physical picture in $d=3$.   

\begin{figure}[t]
\centering

\includegraphics[width=\columnwidth]{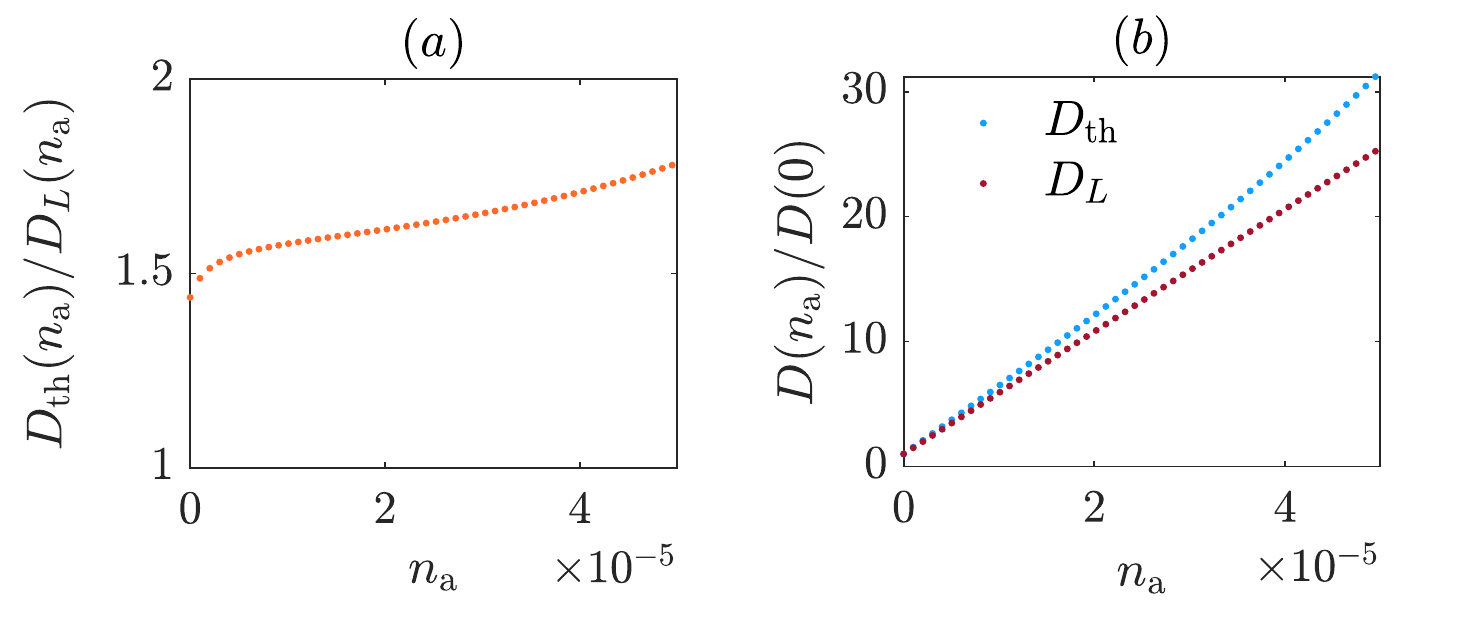}
\caption{Diffusion in three dimensions as a function of $n_{\rm a}$ for fixed $T$. (a) The ratio $D_{\rm th}/D_L$ as a function of $n_{\rm a}$. (b) Diffusivities, normalized with respect to $n_{\rm a} = 0$ value, as a function of $n_{\rm a}$. Data shown with $\Omega_{\rm o}/\Omega_v = 1.1,u/\Omega_v^3 = 1.4, T/\Omega_v = 1, \Omega_{\rm a,d}/\Omega_v = 0.09$.} 
\label{fig:3dex}
\end{figure}

We find that, for sufficiently small values of $n_{\rm a}$, $\gamma \equiv D_{\rm th} / D_L$ stays roughly constant, smoothly interpolating from its $n_{\rm a}=0$ value, such that that the relation $D_{\rm th} \sim D_L$ holds also for $n_{\rm a} > 0$. The diffusivities as a function of $n_{\rm a}$ for a fixed $T$ of a representative system is presented in Fig.~\ref{fig:3dex}. Notice that we restrict ourselves to rather small values of $n_{\rm a}$. This is because, in the weakly dispersive limit, the imbalance between $\overline{v}_{\rm s}$ and $\overline{v}_{\rm o}$ dictates the limit for which we expect the small $n_{\rm a}$ approximation, where the diffusivities are linear-in-$n_{\rm a}$, to be valid. As a rough estimate, we demand that $n_{\rm a}/n_{\rm o} \ll v^2_{\rm o}/\overline{v}_s^2$ ($\approx 10^{-4}$ in Fig.~\ref{fig:3dex}). In particular, when this inequality starts breaking down, we arrive at a situation where optical phonons serve as a bath to acoustic modes, but acoustic modes dominate the diffusivities due to their large renormalized speed of sound. To overcome this artifact, one has to go beyond the weakly dispersive limit. 

While our study is restricted to the weakly dispersive limit, we expect the correspondence $D_{\rm th} \sim D_L$ to hold in a more generic, isotropic $3d$ system. This is since the short-wavelength acoustic modes behave essentially as optical modes, for which we already know that $D_{\rm th} \sim D_L$ (even in a system with multiple optical phonon branches). 
%Moreover, the breakdown of this relation in lower dimensions is related to the large phase-space of long-lived, long-wavelength acoustic modes. These modes are weakly coupled and long-lived in any dimension due to Goldstone's theorem. However, 
In $d=3$, the long-wavelength acoustic modes do not dominate transport properties due to their vanishing phase space. Hence, we expect that in $d=3$ and at sufficiently low temperature, $D_{\rm th} \sim D_L$, as we have demonstrated above in the $n_{\rm{a}}\to 0$ limit. 
%Since no other such singularities are expected manifest beyond the weakly dispersive limit, we expect the relation $D_{\rm th} \sim D_L$ to hold generically in $d=3$. 

Nevertheless, we comment that a parametric violation of the correspondence between $D_{\rm th}$ and $D_L$ is possible in $d=3$ at finite temperature, if the system is strongly anisotropic. Consider, for example, a three-dimensional system that consists of an array of weakly coupled $1d$ chains. We define the dimensional crossover temperature $T^*$ as the energy at which the contours of equal acoustic phonon frequency change from elipsoids around $k=0$ to open surfaces. At temperatures $T>T^*$, we expect that $D_{\rm{th}}$ is parametrically large in the anisotropic limit (where the coupling between the chains vanishes), whereas $D_L$ remains finite in this limit. Hence, $D_{\rm{th}}\gg D_{L}$. At temperatures below $T^*$, the relation $D_{\rm{th}}\sim D_{\rm{th}}$ may be recovered, although we leave a detailed analysis of this case to future work.

\section{Discussion and outlook}

\label{sec:disc}

In this work, we have studied properties of thermal transport and quantum many-body chaos in a lattice model of $N\to \infty$ strongly coupled phonon modes per unit cell. In the absence of acoustic phonons, we found that the thermal and chaos diffusivities obey $D_{\rm th} \approx \gamma D_{L}$ with $\gamma \gtrsim 1$. In a system with a single optical branch, the thermal relaxation rate and scrambling rate track the inverse phonon lifetime, and the butterfly velocity is close to the maximum velocity of the optical phonons. Furthermore, the intermediate-$T$ ``phonon fluid'' regime we identified in \cite{tulipman_strongly_2020} corresponds to a linear-in-$T$ dependence of the inverse diffusivities, where timescales associated with transport and chaos are of the order of the Planckian timescale $\hbar/k_B T$ at strong coupling.

Introducing acoustic phonons to our system significantly changes the physical picture in low dimensions. We found that long-wavelength acoustic modes dominate transport but not chaos, breaking the relationship between the diffusivities. In particular, in $d=1$ and $2$, $D_{\rm{th}}$ diverges even at non-zero temperatures, while $D_L$ remains finite. Intuitively, this can be understood from the fact that the transport properties tend to be dominated by the longest-lived degrees of freedom, while the scrambling is determined by the fastest-growing operator, which is often related to the shortest-lived degrees of freedom. 
In three dimensions, for sufficiently small values of the relative fraction of the number of acoustic modes, $n_{\rm a}$, we showed that the correspondence between the diffusivities reappears. We expect this relation to persist beyond the weakly dispersive limit considered here for sufficiently isotropic systems. For strongly anisotropic systems, we expect the relation $D_{\rm th} \sim D_L$ to break down at temperatures above a crossover scale determined by the degree of anisotropy, but to be recovered at asymptotically low temperatures. 

Our interest in systems of strongly coupled phonons was primarily ignited by measurements of thermal diffusion in a broad class of insulating three-dimensional materials \cite{martelli_thermal_2018,behnia_lower_2019,zhang_thermalization_2019,martelli_thermal_2021}. Despite the fact that our simple model does not attempt to describe any specific material in detail, it is capable of producing a wide linear-in-$T$ ``Planckian'' regime for $D_{\rm th}^{-1}$ at intermediate temperatures. This suggests that a strongly coupled ``quantum phonon fluid'' is a generic property of strongly anharmonic quantum oscillators at intermediate temperatures.
%While our work is not meant to model any particular material realistically, we believe a few lessons can be drawn about the mentioned experiments. It appears that 
%The extraction of a single timescale that characterizes thermal transport requires assuming an `effective medium' that averages out all degrees of freedom in the system, including both acoustic and optical phonons. 
%Namely, the contribution of dispersive optical modes to transport cannot be neglected if their phase space is sufficiently large. 
Importantly, we note that the ``phonon fluid'' regime in our model emerges \textit{before} the system crosses over to its classical limit. This indicates that one may observe quantum mechanical signatures in transport even at naively ``high'' temperatures if a given system hosts sufficiently high-energy modes. 

As a step towards a more realistic model, it would be interesting to relax the assumption of weakly dispersive optical phonons. Upon introducing acoustic phonons, our study was restricted to the limit $n_{\rm a}\ll 1$, whereas for real insulators, $n_{\rm a}$ and $n_{\rm o}$ are comparable and, in particular, optical modes may be as dispersive as acoustic modes. Another natural direction would be to extend our analysis into to the glassy phase \cite{bera_quantum_2021,anous_quantum_2021}, especially in light of the fact that the minimal phonon lifetime, and correspondingly $\tau_L$ and $\tau_{\rm th}$, are attained near the phase boundary \cite{tulipman_strongly_2020}. 

Finally, it would be exciting to embed electrons in the strongly coupled lattice we studied here. The ``phonon fluid'' regime might serve as a breeding ground to theoretically study transport and chaos properties of a strongly coupled electron-phonon quantum ``soup'' \cite{zhang_anomalous_2017}.

\acknowledgements 
We thank Ehud Altman and Boris Spivak for useful discussions throughout this work. 
 EB was supported by the European Research Council (ERC) under grant HQMAT (grant no. 817799), by the US-Israel Binational Science Foundation (BSF), and by a research grant from Irving and Cherna Moskowitz.

\bibliography{refs}

\begin{thebibliography}{73}%
\makeatletter
\providecommand \@ifxundefined [1]{%
 \@ifx{#1\undefined}
}%
\providecommand \@ifnum [1]{%
 \ifnum #1\expandafter \@firstoftwo
 \else \expandafter \@secondoftwo
 \fi
}%
\providecommand \@ifx [1]{%
 \ifx #1\expandafter \@firstoftwo
 \else \expandafter \@secondoftwo
 \fi
}%
\providecommand \natexlab [1]{#1}%
\providecommand \enquote  [1]{``#1''}%
\providecommand \bibnamefont  [1]{#1}%
\providecommand \bibfnamefont [1]{#1}%
\providecommand \citenamefont [1]{#1}%
\providecommand \href@noop [0]{\@secondoftwo}%
\providecommand \href [0]{\begingroup \@sanitize@url \@href}%
\providecommand \@href[1]{\@@startlink{#1}\@@href}%
\providecommand \@@href[1]{\endgroup#1\@@endlink}%
\providecommand \@sanitize@url [0]{\catcode `\\12\catcode `\$12\catcode
  `\&12\catcode `\#12\catcode `\^12\catcode `\_12\catcode `\%12\relax}%
\providecommand \@@startlink[1]{}%
\providecommand \@@endlink[0]{}%
\providecommand \url  [0]{\begingroup\@sanitize@url \@url }%
\providecommand \@url [1]{\endgroup\@href {#1}{\urlprefix }}%
\providecommand \urlprefix  [0]{URL }%
\providecommand \Eprint [0]{\href }%
\providecommand \doibase [0]{https://doi.org/}%
\providecommand \selectlanguage [0]{\@gobble}%
\providecommand \bibinfo  [0]{\@secondoftwo}%
\providecommand \bibfield  [0]{\@secondoftwo}%
\providecommand \translation [1]{[#1]}%
\providecommand \BibitemOpen [0]{}%
\providecommand \bibitemStop [0]{}%
\providecommand \bibitemNoStop [0]{.\EOS\space}%
\providecommand \EOS [0]{\spacefactor3000\relax}%
\providecommand \BibitemShut  [1]{\csname bibitem#1\endcsname}%
\let\auto@bib@innerbib\@empty
%</preamble>
\bibitem [{\citenamefont {Damle}\ and\ \citenamefont
  {Sachdev}()}]{damle_nonzero-temperature_1997}%
  \BibitemOpen
  \bibfield  {author} {\bibinfo {author} {\bibfnamefont {K.}~\bibnamefont
  {Damle}}\ and\ \bibinfo {author} {\bibfnamefont {S.}~\bibnamefont
  {Sachdev}},\ }\bibfield  {title} {\bibinfo {title} {Nonzero-temperature
  transport near quantum critical points},\ }\href
  {https://doi.org/10.1103/PhysRevB.56.8714} {\bibfield  {journal} {\bibinfo
  {journal} {Physical Review B}\ }\textbf {\bibinfo {volume} {56}},\ \bibinfo
  {pages} {8714}}\BibitemShut {NoStop}%
\bibitem [{\citenamefont {Zaanen}(2004)}]{zaanen_why_2004}%
  \BibitemOpen
  \bibfield  {author} {\bibinfo {author} {\bibfnamefont {J.}~\bibnamefont
  {Zaanen}},\ }\bibfield  {title} {{\selectlanguage {english}\bibinfo {title}
  {Why the temperature is high}},\ }\href {https://doi.org/10.1038/430512a}
  {\bibfield  {journal} {\bibinfo  {journal} {Nature}\ }\textbf {\bibinfo
  {volume} {430}},\ \bibinfo {pages} {512} (\bibinfo {year}
  {2004})}\BibitemShut {NoStop}%
\bibitem [{\citenamefont {Sachdev}(2011)}]{sachdev_quantum_2011}%
  \BibitemOpen
  \bibfield  {author} {\bibinfo {author} {\bibfnamefont {S.}~\bibnamefont
  {Sachdev}},\ }\href {https://doi.org/10.1017/CBO9780511973765} {\emph
  {\bibinfo {title} {Quantum Phase Transitions}}},\ \bibinfo {edition} {2nd}\
  ed.\ (\bibinfo  {publisher} {Cambridge University Press},\ \bibinfo {year}
  {2011})\BibitemShut {NoStop}%
\bibitem [{\citenamefont {Zaanen}(2019)}]{zaanen_planckian_2019}%
  \BibitemOpen
  \bibfield  {author} {\bibinfo {author} {\bibfnamefont {J.}~\bibnamefont
  {Zaanen}},\ }\bibfield  {title} {\bibinfo {title} {Planckian dissipation,
  minimal viscosity and the transport in cuprate strange metals},\ }\href
  {https://doi.org/10.21468/SciPostPhys.6.5.061} {\bibfield  {journal}
  {\bibinfo  {journal} {SciPost Physics}\ ,\ \bibinfo {pages} {061}} (\bibinfo
  {year} {2019})},\ \bibinfo {note} {arXiv: 1807.10951}\BibitemShut {NoStop}%
\bibitem [{\citenamefont {Hartnoll}\ and\ \citenamefont
  {Mackenzie}()}]{hartnoll_planckian_2021}%
  \BibitemOpen
  \bibfield  {author} {\bibinfo {author} {\bibfnamefont {S.~A.}\ \bibnamefont
  {Hartnoll}}\ and\ \bibinfo {author} {\bibfnamefont {A.~P.}\ \bibnamefont
  {Mackenzie}},\ }\bibfield  {title} {\bibinfo {title} {Planckian dissipation
  in metals},\ }\href {http://arxiv.org/abs/2107.07802} {\bibfield  {journal}
  {\bibinfo  {journal} {{arXiv}:2107.07802 [cond-mat, physics:hep-th]}\
  }}\Eprint {https://arxiv.org/abs/2107.07802} {2107.07802} \BibitemShut
  {NoStop}%
\bibitem [{\citenamefont {Bruin}\ \emph {et~al.}(2013)\citenamefont {Bruin},
  \citenamefont {Sakai}, \citenamefont {Perry},\ and\ \citenamefont
  {Mackenzie}}]{bruin_similarity_2013}%
  \BibitemOpen
  \bibfield  {author} {\bibinfo {author} {\bibfnamefont {J.~a.~N.}\
  \bibnamefont {Bruin}}, \bibinfo {author} {\bibfnamefont {H.}~\bibnamefont
  {Sakai}}, \bibinfo {author} {\bibfnamefont {R.~S.}\ \bibnamefont {Perry}},\
  and\ \bibinfo {author} {\bibfnamefont {A.~P.}\ \bibnamefont {Mackenzie}},\
  }\bibfield  {title} {{\selectlanguage {english}\bibinfo {title} {Similarity
  of {Scattering} {Rates} in {Metals} {Showing} {T}-{Linear} {Resistivity}}},\
  }\href {https://doi.org/10.1126/science.1227612} {\bibfield  {journal}
  {\bibinfo  {journal} {Science}\ }\textbf {\bibinfo {volume} {339}},\ \bibinfo
  {pages} {804} (\bibinfo {year} {2013})}\BibitemShut {NoStop}%
\bibitem [{\citenamefont {Legros}\ \emph {et~al.}(2019)\citenamefont {Legros},
  \citenamefont {Benhabib}, \citenamefont {Tabis}, \citenamefont {Laliberté},
  \citenamefont {Dion}, \citenamefont {Lizaire}, \citenamefont {Vignolle},
  \citenamefont {Vignolles}, \citenamefont {Raffy}, \citenamefont {Li},
  \citenamefont {Auban-Senzier}, \citenamefont {Doiron-Leyraud}, \citenamefont
  {Fournier}, \citenamefont {Colson}, \citenamefont {Taillefer},\ and\
  \citenamefont {Proust}}]{legros_universal_2019}%
  \BibitemOpen
  \bibfield  {author} {\bibinfo {author} {\bibfnamefont {A.}~\bibnamefont
  {Legros}}, \bibinfo {author} {\bibfnamefont {S.}~\bibnamefont {Benhabib}},
  \bibinfo {author} {\bibfnamefont {W.}~\bibnamefont {Tabis}}, \bibinfo
  {author} {\bibfnamefont {F.}~\bibnamefont {Laliberté}}, \bibinfo {author}
  {\bibfnamefont {M.}~\bibnamefont {Dion}}, \bibinfo {author} {\bibfnamefont
  {M.}~\bibnamefont {Lizaire}}, \bibinfo {author} {\bibfnamefont
  {B.}~\bibnamefont {Vignolle}}, \bibinfo {author} {\bibfnamefont
  {D.}~\bibnamefont {Vignolles}}, \bibinfo {author} {\bibfnamefont
  {H.}~\bibnamefont {Raffy}}, \bibinfo {author} {\bibfnamefont {Z.~Z.}\
  \bibnamefont {Li}}, \bibinfo {author} {\bibfnamefont {P.}~\bibnamefont
  {Auban-Senzier}}, \bibinfo {author} {\bibfnamefont {N.}~\bibnamefont
  {Doiron-Leyraud}}, \bibinfo {author} {\bibfnamefont {P.}~\bibnamefont
  {Fournier}}, \bibinfo {author} {\bibfnamefont {D.}~\bibnamefont {Colson}},
  \bibinfo {author} {\bibfnamefont {L.}~\bibnamefont {Taillefer}},\ and\
  \bibinfo {author} {\bibfnamefont {C.}~\bibnamefont {Proust}},\ }\bibfield
  {title} {{\selectlanguage {english}\bibinfo {title} {Universal {T} -linear
  resistivity and {Planckian} dissipation in overdoped cuprates}},\ }\href
  {https://doi.org/10.1038/s41567-018-0334-2} {\bibfield  {journal} {\bibinfo
  {journal} {Nature Physics}\ }\textbf {\bibinfo {volume} {15}},\ \bibinfo
  {pages} {142} (\bibinfo {year} {2019})}\BibitemShut {NoStop}%
\bibitem [{\citenamefont {Polshyn}\ \emph {et~al.}(2019)\citenamefont
  {Polshyn}, \citenamefont {Yankowitz}, \citenamefont {Chen}, \citenamefont
  {Zhang}, \citenamefont {Watanabe}, \citenamefont {Taniguchi}, \citenamefont
  {Dean},\ and\ \citenamefont {Young}}]{polshyn_large_2019}%
  \BibitemOpen
  \bibfield  {author} {\bibinfo {author} {\bibfnamefont {H.}~\bibnamefont
  {Polshyn}}, \bibinfo {author} {\bibfnamefont {M.}~\bibnamefont {Yankowitz}},
  \bibinfo {author} {\bibfnamefont {S.}~\bibnamefont {Chen}}, \bibinfo {author}
  {\bibfnamefont {Y.}~\bibnamefont {Zhang}}, \bibinfo {author} {\bibfnamefont
  {K.}~\bibnamefont {Watanabe}}, \bibinfo {author} {\bibfnamefont
  {T.}~\bibnamefont {Taniguchi}}, \bibinfo {author} {\bibfnamefont {C.~R.}\
  \bibnamefont {Dean}},\ and\ \bibinfo {author} {\bibfnamefont {A.~F.}\
  \bibnamefont {Young}},\ }\bibfield  {title} {\bibinfo {title} {Large
  linear-in-temperature resistivity in twisted bilayer graphene},\ }\href
  {https://doi.org/10.1038/s41567-019-0596-3} {\bibfield  {journal} {\bibinfo
  {journal} {Nature Physics}\ }\textbf {\bibinfo {volume} {15}},\ \bibinfo
  {pages} {1011} (\bibinfo {year} {2019})}\BibitemShut {NoStop}%
\bibitem [{\citenamefont {Cao}\ \emph {et~al.}(2020)\citenamefont {Cao},
  \citenamefont {Chowdhury}, \citenamefont {Rodan-Legrain}, \citenamefont
  {Rubies-Bigorda}, \citenamefont {Watanabe}, \citenamefont {Taniguchi},
  \citenamefont {Senthil},\ and\ \citenamefont
  {Jarillo-Herrero}}]{cao_strange_2020}%
  \BibitemOpen
  \bibfield  {author} {\bibinfo {author} {\bibfnamefont {Y.}~\bibnamefont
  {Cao}}, \bibinfo {author} {\bibfnamefont {D.}~\bibnamefont {Chowdhury}},
  \bibinfo {author} {\bibfnamefont {D.}~\bibnamefont {Rodan-Legrain}}, \bibinfo
  {author} {\bibfnamefont {O.}~\bibnamefont {Rubies-Bigorda}}, \bibinfo
  {author} {\bibfnamefont {K.}~\bibnamefont {Watanabe}}, \bibinfo {author}
  {\bibfnamefont {T.}~\bibnamefont {Taniguchi}}, \bibinfo {author}
  {\bibfnamefont {T.}~\bibnamefont {Senthil}},\ and\ \bibinfo {author}
  {\bibfnamefont {P.}~\bibnamefont {Jarillo-Herrero}},\ }\bibfield  {title}
  {\bibinfo {title} {Strange {Metal} in {Magic}-{Angle} {Graphene} with near
  {Planckian} {Dissipation}},\ }\href
  {https://doi.org/10.1103/PhysRevLett.124.076801} {\bibfield  {journal}
  {\bibinfo  {journal} {Physical Review Letters}\ }\textbf {\bibinfo {volume}
  {124}},\ \bibinfo {pages} {076801} (\bibinfo {year} {2020})}\BibitemShut
  {NoStop}%
\bibitem [{\citenamefont {Licciardello}\ \emph {et~al.}()\citenamefont
  {Licciardello}, \citenamefont {Buhot}, \citenamefont {Lu}, \citenamefont
  {Ayres}, \citenamefont {Kasahara}, \citenamefont {Matsuda}, \citenamefont
  {Shibauchi},\ and\ \citenamefont {Hussey}}]{licciardello_electrical_2019}%
  \BibitemOpen
  \bibfield  {author} {\bibinfo {author} {\bibfnamefont {S.}~\bibnamefont
  {Licciardello}}, \bibinfo {author} {\bibfnamefont {J.}~\bibnamefont {Buhot}},
  \bibinfo {author} {\bibfnamefont {J.}~\bibnamefont {Lu}}, \bibinfo {author}
  {\bibfnamefont {J.}~\bibnamefont {Ayres}}, \bibinfo {author} {\bibfnamefont
  {S.}~\bibnamefont {Kasahara}}, \bibinfo {author} {\bibfnamefont
  {Y.}~\bibnamefont {Matsuda}}, \bibinfo {author} {\bibfnamefont
  {T.}~\bibnamefont {Shibauchi}},\ and\ \bibinfo {author} {\bibfnamefont
  {N.~E.}\ \bibnamefont {Hussey}},\ }\bibfield  {title} {\bibinfo {title}
  {Electrical resistivity across a nematic quantum critical point},\ }\href
  {https://doi.org/10.1038/s41586-019-0923-y} {\bibfield  {journal} {\bibinfo
  {journal} {Nature}\ }\textbf {\bibinfo {volume} {567}},\ \bibinfo {pages}
  {213}}\BibitemShut {NoStop}%
\bibitem [{\citenamefont {Grissonnanche}\ \emph {et~al.}()\citenamefont
  {Grissonnanche}, \citenamefont {Fang}, \citenamefont {Legros}, \citenamefont
  {Verret}, \citenamefont {Laliberté}, \citenamefont {Collignon},
  \citenamefont {Zhou}, \citenamefont {Graf}, \citenamefont {Goddard},
  \citenamefont {Taillefer},\ and\ \citenamefont
  {Ramshaw}}]{grissonnanche_linear-temperature_2021}%
  \BibitemOpen
  \bibfield  {author} {\bibinfo {author} {\bibfnamefont {G.}~\bibnamefont
  {Grissonnanche}}, \bibinfo {author} {\bibfnamefont {Y.}~\bibnamefont {Fang}},
  \bibinfo {author} {\bibfnamefont {A.}~\bibnamefont {Legros}}, \bibinfo
  {author} {\bibfnamefont {S.}~\bibnamefont {Verret}}, \bibinfo {author}
  {\bibfnamefont {F.}~\bibnamefont {Laliberté}}, \bibinfo {author}
  {\bibfnamefont {C.}~\bibnamefont {Collignon}}, \bibinfo {author}
  {\bibfnamefont {J.}~\bibnamefont {Zhou}}, \bibinfo {author} {\bibfnamefont
  {D.}~\bibnamefont {Graf}}, \bibinfo {author} {\bibfnamefont {P.~A.}\
  \bibnamefont {Goddard}}, \bibinfo {author} {\bibfnamefont {L.}~\bibnamefont
  {Taillefer}},\ and\ \bibinfo {author} {\bibfnamefont {B.~J.}\ \bibnamefont
  {Ramshaw}},\ }\bibfield  {title} {\bibinfo {title} {Linear-in temperature
  resistivity from an isotropic planckian scattering rate},\ }\href
  {https://doi.org/10.1038/s41586-021-03697-8} {\bibfield  {journal} {\bibinfo
  {journal} {Nature}\ }\textbf {\bibinfo {volume} {595}},\ \bibinfo {pages}
  {667}}\BibitemShut {NoStop}%
\bibitem [{\citenamefont {Kovtun}\ \emph {et~al.}(2005)\citenamefont {Kovtun},
  \citenamefont {Son},\ and\ \citenamefont
  {Starinets}}]{kovtun_viscosity_2005}%
  \BibitemOpen
  \bibfield  {author} {\bibinfo {author} {\bibfnamefont {P.~K.}\ \bibnamefont
  {Kovtun}}, \bibinfo {author} {\bibfnamefont {D.~T.}\ \bibnamefont {Son}},\
  and\ \bibinfo {author} {\bibfnamefont {A.~O.}\ \bibnamefont {Starinets}},\
  }\bibfield  {title} {\bibinfo {title} {Viscosity in strongly interacting
  quantum field theories from black hole physics},\ }\href
  {https://doi.org/10.1103/PhysRevLett.94.111601} {\bibfield  {journal}
  {\bibinfo  {journal} {Physical Review Letters}\ }\textbf {\bibinfo {volume}
  {94}},\ \bibinfo {pages} {111601} (\bibinfo {year} {2005})}\BibitemShut
  {NoStop}%
\bibitem [{\citenamefont {Shenker}\ and\ \citenamefont
  {Stanford}(2014)}]{shenker_black_2014}%
  \BibitemOpen
  \bibfield  {author} {\bibinfo {author} {\bibfnamefont {S.~H.}\ \bibnamefont
  {Shenker}}\ and\ \bibinfo {author} {\bibfnamefont {D.}~\bibnamefont
  {Stanford}},\ }\bibfield  {title} {{\selectlanguage {english}\bibinfo {title}
  {Black holes and the butterfly effect}},\ }\href
  {https://doi.org/10.1007/JHEP03(2014)067} {\bibfield  {journal} {\bibinfo
  {journal} {Journal of High Energy Physics}\ }\textbf {\bibinfo {volume}
  {2014}},\ \bibinfo {pages} {67} (\bibinfo {year} {2014})}\BibitemShut
  {NoStop}%
\bibitem [{\citenamefont {Roberts}\ \emph {et~al.}(2015)\citenamefont
  {Roberts}, \citenamefont {Stanford},\ and\ \citenamefont
  {Susskind}}]{roberts_localized_2015}%
  \BibitemOpen
  \bibfield  {author} {\bibinfo {author} {\bibfnamefont {D.~A.}\ \bibnamefont
  {Roberts}}, \bibinfo {author} {\bibfnamefont {D.}~\bibnamefont {Stanford}},\
  and\ \bibinfo {author} {\bibfnamefont {L.}~\bibnamefont {Susskind}},\
  }\bibfield  {title} {{\selectlanguage {english}\bibinfo {title} {Localized
  shocks}},\ }\href {https://doi.org/10.1007/JHEP03(2015)051} {\bibfield
  {journal} {\bibinfo  {journal} {Journal of High Energy Physics}\ }\textbf
  {\bibinfo {volume} {2015}},\ \bibinfo {pages} {51} (\bibinfo {year}
  {2015})}\BibitemShut {NoStop}%
\bibitem [{\citenamefont {Kitaev}()}]{Kitaev_SYK_talk}%
  \BibitemOpen
  \bibfield  {author} {\bibinfo {author} {\bibfnamefont {A.}~\bibnamefont
  {Kitaev}},\ }\href@noop {} {}\bibinfo {howpublished} {A simple model of
  quantum holography.
  \url{http://online.kitp.ucsb.edu/online/entangled15/kitaev/} \newline
  \url{http://online.kitp.ucsb.edu/online/entangled15/kitaev2/}}\BibitemShut
  {NoStop}%
\bibitem [{\citenamefont {Davison}\ \emph {et~al.}(2014)\citenamefont
  {Davison}, \citenamefont {Schalm},\ and\ \citenamefont
  {Zaanen}}]{davison_holographic_2014}%
  \BibitemOpen
  \bibfield  {author} {\bibinfo {author} {\bibfnamefont {R.~A.}\ \bibnamefont
  {Davison}}, \bibinfo {author} {\bibfnamefont {K.}~\bibnamefont {Schalm}},\
  and\ \bibinfo {author} {\bibfnamefont {J.}~\bibnamefont {Zaanen}},\
  }\bibfield  {title} {{\selectlanguage {english}\bibinfo {title} {Holographic
  duality and the resistivity of strange metals}},\ }\href
  {https://doi.org/10.1103/PhysRevB.89.245116} {\bibfield  {journal} {\bibinfo
  {journal} {Physical Review B}\ }\textbf {\bibinfo {volume} {89}},\ \bibinfo
  {pages} {245116} (\bibinfo {year} {2014})}\BibitemShut {NoStop}%
\bibitem [{\citenamefont {Zaanen}\ \emph {et~al.}(2015)\citenamefont {Zaanen},
  \citenamefont {Liu}, \citenamefont {Sun},\ and\ \citenamefont
  {Schalm}}]{zaanen_liu_sun_schalm_2015}%
  \BibitemOpen
  \bibfield  {author} {\bibinfo {author} {\bibfnamefont {J.}~\bibnamefont
  {Zaanen}}, \bibinfo {author} {\bibfnamefont {Y.}~\bibnamefont {Liu}},
  \bibinfo {author} {\bibfnamefont {Y.-W.}\ \bibnamefont {Sun}},\ and\ \bibinfo
  {author} {\bibfnamefont {K.}~\bibnamefont {Schalm}},\ }\href
  {https://doi.org/10.1017/CBO9781139942492} {\emph {\bibinfo {title}
  {Holographic Duality in Condensed Matter Physics}}}\ (\bibinfo  {publisher}
  {Cambridge University Press},\ \bibinfo {year} {2015})\BibitemShut {NoStop}%
\bibitem [{\citenamefont {Davison}\ \emph {et~al.}(2017)\citenamefont
  {Davison}, \citenamefont {Fu}, \citenamefont {Georges}, \citenamefont {Gu},
  \citenamefont {Jensen},\ and\ \citenamefont
  {Sachdev}}]{davison_thermoelectric_2017}%
  \BibitemOpen
  \bibfield  {author} {\bibinfo {author} {\bibfnamefont {R.~A.}\ \bibnamefont
  {Davison}}, \bibinfo {author} {\bibfnamefont {W.}~\bibnamefont {Fu}},
  \bibinfo {author} {\bibfnamefont {A.}~\bibnamefont {Georges}}, \bibinfo
  {author} {\bibfnamefont {Y.}~\bibnamefont {Gu}}, \bibinfo {author}
  {\bibfnamefont {K.}~\bibnamefont {Jensen}},\ and\ \bibinfo {author}
  {\bibfnamefont {S.}~\bibnamefont {Sachdev}},\ }\bibfield  {title}
  {{\selectlanguage {english}\bibinfo {title} {Thermoelectric transport in
  disordered metals without quasiparticles: {The} {Sachdev}-{Ye}-{Kitaev}
  models and holography}},\ }\bibfield  {journal} {\bibinfo  {journal}
  {Physical Review B}\ }\textbf {\bibinfo {volume} {95}},\ \href
  {https://doi.org/10.1103/PhysRevB.95.155131} {10.1103/PhysRevB.95.155131}
  (\bibinfo {year} {2017})\BibitemShut {NoStop}%
\bibitem [{\citenamefont {Hartnoll}\ \emph {et~al.}(2018)\citenamefont
  {Hartnoll}, \citenamefont {Lucas},\ and\ \citenamefont
  {Sachdev}}]{hartnoll_holographic_2018}%
  \BibitemOpen
  \bibfield  {author} {\bibinfo {author} {\bibfnamefont {S.~A.}\ \bibnamefont
  {Hartnoll}}, \bibinfo {author} {\bibfnamefont {A.}~\bibnamefont {Lucas}},\
  and\ \bibinfo {author} {\bibfnamefont {S.}~\bibnamefont {Sachdev}},\ }\href
  {https://mitpress.mit.edu/books/holographic-quantum-matter} {\emph {\bibinfo
  {title} {Holographic Quantum Matter}}}\ (\bibinfo  {publisher} {{MIT}
  Press},\ \bibinfo {year} {2018})\BibitemShut {NoStop}%
\bibitem [{\citenamefont {Song}\ \emph {et~al.}(2017)\citenamefont {Song},
  \citenamefont {Jian},\ and\ \citenamefont {Balents}}]{song_strongly_2017}%
  \BibitemOpen
  \bibfield  {author} {\bibinfo {author} {\bibfnamefont {X.-Y.}\ \bibnamefont
  {Song}}, \bibinfo {author} {\bibfnamefont {C.-M.}\ \bibnamefont {Jian}},\
  and\ \bibinfo {author} {\bibfnamefont {L.}~\bibnamefont {Balents}},\
  }\bibfield  {title} {\bibinfo {title} {A strongly correlated metal built from
  {Sachdev}-{Ye}-{Kitaev} models},\ }\href
  {https://doi.org/10.1103/PhysRevLett.119.216601} {\bibfield  {journal}
  {\bibinfo  {journal} {Physical Review Letters}\ }\textbf {\bibinfo {volume}
  {119}},\ \bibinfo {pages} {216601} (\bibinfo {year} {2017})},\ \bibinfo
  {note} {arXiv: 1705.00117}\BibitemShut {NoStop}%
\bibitem [{\citenamefont {Patel}\ \emph {et~al.}(2018)\citenamefont {Patel},
  \citenamefont {McGreevy}, \citenamefont {Arovas},\ and\ \citenamefont
  {Sachdev}}]{patel_magnetotransport_2018}%
  \BibitemOpen
  \bibfield  {author} {\bibinfo {author} {\bibfnamefont {A.~A.}\ \bibnamefont
  {Patel}}, \bibinfo {author} {\bibfnamefont {J.}~\bibnamefont {McGreevy}},
  \bibinfo {author} {\bibfnamefont {D.~P.}\ \bibnamefont {Arovas}},\ and\
  \bibinfo {author} {\bibfnamefont {S.}~\bibnamefont {Sachdev}},\ }\bibfield
  {title} {{\selectlanguage {english}\bibinfo {title} {Magnetotransport in a
  {Model} of a {Disordered} {Strange} {Metal}}},\ }\bibfield  {journal}
  {\bibinfo  {journal} {Physical Review X}\ }\textbf {\bibinfo {volume} {8}},\
  \href {https://doi.org/10.1103/PhysRevX.8.021049} {10.1103/PhysRevX.8.021049}
  (\bibinfo {year} {2018})\BibitemShut {NoStop}%
\bibitem [{\citenamefont {Chowdhury}\ \emph {et~al.}(2018)\citenamefont
  {Chowdhury}, \citenamefont {Werman}, \citenamefont {Berg},\ and\
  \citenamefont {Senthil}}]{chowdhury_translationally_2018}%
  \BibitemOpen
  \bibfield  {author} {\bibinfo {author} {\bibfnamefont {D.}~\bibnamefont
  {Chowdhury}}, \bibinfo {author} {\bibfnamefont {Y.}~\bibnamefont {Werman}},
  \bibinfo {author} {\bibfnamefont {E.}~\bibnamefont {Berg}},\ and\ \bibinfo
  {author} {\bibfnamefont {T.}~\bibnamefont {Senthil}},\ }\bibfield  {title}
  {\bibinfo {title} {Translationally invariant non-{Fermi} liquid metals with
  critical {Fermi}-surfaces: {Solvable} models},\ }\href
  {https://doi.org/10.1103/PhysRevX.8.031024} {\bibfield  {journal} {\bibinfo
  {journal} {Physical Review X}\ }\textbf {\bibinfo {volume} {8}},\ \bibinfo
  {pages} {031024} (\bibinfo {year} {2018})},\ \bibinfo {note} {arXiv:
  1801.06178}\BibitemShut {NoStop}%
\bibitem [{\citenamefont {Patel}\ and\ \citenamefont
  {Sachdev}(2019)}]{patel_theory_2019}%
  \BibitemOpen
  \bibfield  {author} {\bibinfo {author} {\bibfnamefont {A.~A.}\ \bibnamefont
  {Patel}}\ and\ \bibinfo {author} {\bibfnamefont {S.}~\bibnamefont
  {Sachdev}},\ }\bibfield  {title} {\bibinfo {title} {Theory of a {Planckian}
  metal},\ }\href {https://doi.org/10.1103/PhysRevLett.123.066601} {\bibfield
  {journal} {\bibinfo  {journal} {Physical Review Letters}\ }\textbf {\bibinfo
  {volume} {123}},\ \bibinfo {pages} {066601} (\bibinfo {year} {2019})},\
  \bibinfo {note} {arXiv: 1906.03265}\BibitemShut {NoStop}%
\bibitem [{\citenamefont {Chowdhury}\ \emph {et~al.}()\citenamefont
  {Chowdhury}, \citenamefont {Georges}, \citenamefont {Parcollet},\ and\
  \citenamefont {Sachdev}}]{chowdhury_sachdev-ye-kitaev_2021}%
  \BibitemOpen
  \bibfield  {author} {\bibinfo {author} {\bibfnamefont {D.}~\bibnamefont
  {Chowdhury}}, \bibinfo {author} {\bibfnamefont {A.}~\bibnamefont {Georges}},
  \bibinfo {author} {\bibfnamefont {O.}~\bibnamefont {Parcollet}},\ and\
  \bibinfo {author} {\bibfnamefont {S.}~\bibnamefont {Sachdev}},\ }\bibfield
  {title} {\bibinfo {title} {{Sachdev-Ye-Kitaev Models and Beyond: A Window
  into Non-Fermi Liquids}},\ }\href {http://arxiv.org/abs/2109.05037}
  {\bibfield  {journal} {\bibinfo  {journal} {{arXiv}:2109.05037 [cond-mat,
  physics:hep-th, physics:quant-ph]}\ }}\Eprint
  {https://arxiv.org/abs/2109.05037} {2109.05037} \BibitemShut {NoStop}%
\bibitem [{\citenamefont {Larkin}\ and\ \citenamefont
  {Ovchinnikov}(1969)}]{Larkin1969}%
  \BibitemOpen
  \bibfield  {author} {\bibinfo {author} {\bibfnamefont {A.}~\bibnamefont
  {Larkin}}\ and\ \bibinfo {author} {\bibfnamefont {Y.~N.}\ \bibnamefont
  {Ovchinnikov}},\ }\bibfield  {title} {\bibinfo {title} {Quasiclassical method
  in the theory of superconductivity},\ }\href@noop {} {\bibfield  {journal}
  {\bibinfo  {journal} {Sov Phys JETP}\ }\textbf {\bibinfo {volume} {28}},\
  \bibinfo {pages} {1200} (\bibinfo {year} {1969})}\BibitemShut {NoStop}%
\bibitem [{\citenamefont {Maldacena}\ \emph {et~al.}(2016)\citenamefont
  {Maldacena}, \citenamefont {Shenker},\ and\ \citenamefont
  {Stanford}}]{maldacena_bound_2016}%
  \BibitemOpen
  \bibfield  {author} {\bibinfo {author} {\bibfnamefont {J.}~\bibnamefont
  {Maldacena}}, \bibinfo {author} {\bibfnamefont {S.~H.}\ \bibnamefont
  {Shenker}},\ and\ \bibinfo {author} {\bibfnamefont {D.}~\bibnamefont
  {Stanford}},\ }\bibfield  {title} {\bibinfo {title} {A bound on chaos},\
  }\href {https://doi.org/10.1007/JHEP08(2016)106} {\bibfield  {journal}
  {\bibinfo  {journal} {Journal of High Energy Physics}\ }\textbf {\bibinfo
  {volume} {2016}},\ \bibinfo {pages} {106} (\bibinfo {year}
  {2016})}\BibitemShut {NoStop}%
\bibitem [{\citenamefont {Blake}(2016)}]{blake_universal_2016}%
  \BibitemOpen
  \bibfield  {author} {\bibinfo {author} {\bibfnamefont {M.}~\bibnamefont
  {Blake}},\ }\bibfield  {title} {\bibinfo {title} {Universal {Charge}
  {Diffusion} and the {Butterfly} {Effect} in {Holographic} {Theories}},\
  }\href {https://doi.org/10.1103/PhysRevLett.117.091601} {\bibfield  {journal}
  {\bibinfo  {journal} {Physical Review Letters}\ }\textbf {\bibinfo {volume}
  {117}},\ \bibinfo {pages} {091601} (\bibinfo {year} {2016})}\BibitemShut
  {NoStop}%
\bibitem [{\citenamefont {Hartnoll}(2015)}]{hartnoll_theory_2015}%
  \BibitemOpen
  \bibfield  {author} {\bibinfo {author} {\bibfnamefont {S.~A.}\ \bibnamefont
  {Hartnoll}},\ }\bibfield  {title} {{\selectlanguage {english}\bibinfo {title}
  {Theory of universal incoherent metallic transport}},\ }\href
  {https://doi.org/10.1038/nphys3174} {\bibfield  {journal} {\bibinfo
  {journal} {Nature Physics}\ }\textbf {\bibinfo {volume} {11}},\ \bibinfo
  {pages} {54} (\bibinfo {year} {2015})}\BibitemShut {NoStop}%
\bibitem [{\citenamefont {Patel}\ and\ \citenamefont
  {Sachdev}(2017)}]{patel_quantum_2017-1}%
  \BibitemOpen
  \bibfield  {author} {\bibinfo {author} {\bibfnamefont {A.~A.}\ \bibnamefont
  {Patel}}\ and\ \bibinfo {author} {\bibfnamefont {S.}~\bibnamefont
  {Sachdev}},\ }\bibfield  {title} {\bibinfo {title} {Quantum chaos on a
  critical {Fermi} surface},\ }\href {https://doi.org/10.1073/pnas.1618185114}
  {\bibfield  {journal} {\bibinfo  {journal} {Proceedings of the National
  Academy of Sciences}\ }\textbf {\bibinfo {volume} {114}},\ \bibinfo {pages}
  {1844} (\bibinfo {year} {2017})},\ \bibinfo {note} {arXiv:
  1611.00003}\BibitemShut {NoStop}%
\bibitem [{\citenamefont {Patel}\ \emph {et~al.}(2017)\citenamefont {Patel},
  \citenamefont {Chowdhury}, \citenamefont {Sachdev},\ and\ \citenamefont
  {Swingle}}]{patel_quantum_2017}%
  \BibitemOpen
  \bibfield  {author} {\bibinfo {author} {\bibfnamefont {A.~A.}\ \bibnamefont
  {Patel}}, \bibinfo {author} {\bibfnamefont {D.}~\bibnamefont {Chowdhury}},
  \bibinfo {author} {\bibfnamefont {S.}~\bibnamefont {Sachdev}},\ and\ \bibinfo
  {author} {\bibfnamefont {B.}~\bibnamefont {Swingle}},\ }\bibfield  {title}
  {\bibinfo {title} {Quantum butterfly effect in weakly interacting diffusive
  metals},\ }\bibfield  {journal} {\bibinfo  {journal} {Physical Review X}\
  }\textbf {\bibinfo {volume} {7}},\ \href
  {https://doi.org/10.1103/PhysRevX.7.031047} {10.1103/PhysRevX.7.031047}
  (\bibinfo {year} {2017}),\ \bibinfo {note} {arXiv: 1703.07353}\BibitemShut
  {NoStop}%
\bibitem [{\citenamefont {Lucas}\ and\ \citenamefont
  {Steinberg}(2016)}]{lucas_charge_2016}%
  \BibitemOpen
  \bibfield  {author} {\bibinfo {author} {\bibfnamefont {A.}~\bibnamefont
  {Lucas}}\ and\ \bibinfo {author} {\bibfnamefont {J.}~\bibnamefont
  {Steinberg}},\ }\bibfield  {title} {\bibinfo {title} {Charge diffusion and
  the butterfly effect in striped holographic matter},\ }\href
  {https://doi.org/10.1007/JHEP10(2016)143} {\bibfield  {journal} {\bibinfo
  {journal} {Journal of High Energy Physics}\ }\textbf {\bibinfo {volume}
  {2016}},\ \bibinfo {pages} {143} (\bibinfo {year} {2016})}\BibitemShut
  {NoStop}%
\bibitem [{\citenamefont {Werman}\ \emph {et~al.}(2017)\citenamefont {Werman},
  \citenamefont {Kivelson},\ and\ \citenamefont {Berg}}]{werman_quantum_2017}%
  \BibitemOpen
  \bibfield  {author} {\bibinfo {author} {\bibfnamefont {Y.}~\bibnamefont
  {Werman}}, \bibinfo {author} {\bibfnamefont {S.~A.}\ \bibnamefont
  {Kivelson}},\ and\ \bibinfo {author} {\bibfnamefont {E.}~\bibnamefont
  {Berg}},\ }\bibfield  {title} {{\selectlanguage {english}\bibinfo {title}
  {Quantum chaos in an electron-phonon bad metal}},\ }\href
  {http://arxiv.org/abs/1705.07895} {\bibfield  {journal} {\bibinfo  {journal}
  {arXiv:1705.07895 [cond-mat]}\ } (\bibinfo {year} {2017})}\BibitemShut
  {NoStop}%
\bibitem [{\citenamefont {Niu}\ and\ \citenamefont
  {Kim}(2017)}]{niu_diffusion_2017}%
  \BibitemOpen
  \bibfield  {author} {\bibinfo {author} {\bibfnamefont {C.}~\bibnamefont
  {Niu}}\ and\ \bibinfo {author} {\bibfnamefont {K.-Y.}\ \bibnamefont {Kim}},\
  }\bibfield  {title} {\bibinfo {title} {Diffusion and butterfly velocity at
  finite density},\ }\href {https://doi.org/10.1007/JHEP06(2017)030} {\bibfield
   {journal} {\bibinfo  {journal} {Journal of High Energy Physics}\ }\textbf
  {\bibinfo {volume} {2017}},\ \bibinfo {pages} {30} (\bibinfo {year}
  {2017})}\BibitemShut {NoStop}%
\bibitem [{\citenamefont {Gu}\ \emph {et~al.}(2017{\natexlab{a}})\citenamefont
  {Gu}, \citenamefont {Lucas},\ and\ \citenamefont {Qi}}]{gu_energy_2017}%
  \BibitemOpen
  \bibfield  {author} {\bibinfo {author} {\bibfnamefont {Y.}~\bibnamefont
  {Gu}}, \bibinfo {author} {\bibfnamefont {A.}~\bibnamefont {Lucas}},\ and\
  \bibinfo {author} {\bibfnamefont {X.-L.}\ \bibnamefont {Qi}},\ }\bibfield
  {title} {\bibinfo {title} {Energy diffusion and the butterfly effect in
  inhomogeneous sachdev-ye-kitaev chains},\ }\href
  {https://doi.org/10.21468/SciPostPhys.2.3.018} {\bibfield  {journal}
  {\bibinfo  {journal} {{SciPost} Physics}\ }\textbf {\bibinfo {volume} {2}},\
  \bibinfo {pages} {018} (\bibinfo {year} {2017}{\natexlab{a}})}\BibitemShut
  {NoStop}%
\bibitem [{\citenamefont {Guo}\ \emph {et~al.}(2019)\citenamefont {Guo},
  \citenamefont {Gu},\ and\ \citenamefont {Sachdev}}]{guo_transport_2019}%
  \BibitemOpen
  \bibfield  {author} {\bibinfo {author} {\bibfnamefont {H.}~\bibnamefont
  {Guo}}, \bibinfo {author} {\bibfnamefont {Y.}~\bibnamefont {Gu}},\ and\
  \bibinfo {author} {\bibfnamefont {S.}~\bibnamefont {Sachdev}},\ }\bibfield
  {title} {\bibinfo {title} {Transport and chaos in lattice sachdev-ye-kitaev
  models},\ }\href {https://doi.org/10.1103/PhysRevB.100.045140} {\bibfield
  {journal} {\bibinfo  {journal} {Physical Review B}\ }\textbf {\bibinfo
  {volume} {100}},\ \bibinfo {pages} {045140} (\bibinfo {year}
  {2019})}\BibitemShut {NoStop}%
\bibitem [{\citenamefont {Gu}\ \emph {et~al.}(2017{\natexlab{b}})\citenamefont
  {Gu}, \citenamefont {Qi},\ and\ \citenamefont {Stanford}}]{gu_local_2017}%
  \BibitemOpen
  \bibfield  {author} {\bibinfo {author} {\bibfnamefont {Y.}~\bibnamefont
  {Gu}}, \bibinfo {author} {\bibfnamefont {X.-L.}\ \bibnamefont {Qi}},\ and\
  \bibinfo {author} {\bibfnamefont {D.}~\bibnamefont {Stanford}},\ }\bibfield
  {title} {{\selectlanguage {english}\bibinfo {title} {Local criticality,
  diffusion and chaos in generalized {Sachdev}-{Ye}-{Kitaev} models}},\ }\href
  {https://doi.org/10.1007/JHEP05(2017)125} {\bibfield  {journal} {\bibinfo
  {journal} {Journal of High Energy Physics}\ }\textbf {\bibinfo {volume}
  {2017}},\ \bibinfo {pages} {125} (\bibinfo {year}
  {2017}{\natexlab{b}})}\BibitemShut {NoStop}%
\bibitem [{\citenamefont {Blake}\ \emph {et~al.}(2017)\citenamefont {Blake},
  \citenamefont {Davison},\ and\ \citenamefont {Sachdev}}]{blake_thermal_2017}%
  \BibitemOpen
  \bibfield  {author} {\bibinfo {author} {\bibfnamefont {M.}~\bibnamefont
  {Blake}}, \bibinfo {author} {\bibfnamefont {R.~A.}\ \bibnamefont {Davison}},\
  and\ \bibinfo {author} {\bibfnamefont {S.}~\bibnamefont {Sachdev}},\
  }\bibfield  {title} {\bibinfo {title} {Thermal diffusivity and chaos in
  metals without quasiparticles},\ }\href
  {https://doi.org/10.1103/PhysRevD.96.106008} {\bibfield  {journal} {\bibinfo
  {journal} {Physical Review D}\ }\textbf {\bibinfo {volume} {96}},\ \bibinfo
  {pages} {106008} (\bibinfo {year} {2017})}\BibitemShut {NoStop}%
\bibitem [{\citenamefont {Li}\ \emph {et~al.}(2019)\citenamefont {Li},
  \citenamefont {Lin},\ and\ \citenamefont {Mei}}]{li_thermal_2019}%
  \BibitemOpen
  \bibfield  {author} {\bibinfo {author} {\bibfnamefont {W.}~\bibnamefont
  {Li}}, \bibinfo {author} {\bibfnamefont {S.}~\bibnamefont {Lin}},\ and\
  \bibinfo {author} {\bibfnamefont {J.}~\bibnamefont {Mei}},\ }\bibfield
  {title} {\bibinfo {title} {Thermal diffusion and quantum chaos in neutral
  magnetized plasma},\ }\href {https://doi.org/10.1103/PhysRevD.100.046012}
  {\bibfield  {journal} {\bibinfo  {journal} {Physical Review D}\ }\textbf
  {\bibinfo {volume} {100}},\ \bibinfo {pages} {046012} (\bibinfo {year}
  {2019})}\BibitemShut {NoStop}%
\bibitem [{\citenamefont {Jeong}\ \emph {et~al.}(2018)\citenamefont {Jeong},
  \citenamefont {Ahn}, \citenamefont {Ahn}, \citenamefont {Niu}, \citenamefont
  {Li},\ and\ \citenamefont {Kim}}]{jeong_thermal_2018}%
  \BibitemOpen
  \bibfield  {author} {\bibinfo {author} {\bibfnamefont {H.-S.}\ \bibnamefont
  {Jeong}}, \bibinfo {author} {\bibfnamefont {Y.}~\bibnamefont {Ahn}}, \bibinfo
  {author} {\bibfnamefont {D.}~\bibnamefont {Ahn}}, \bibinfo {author}
  {\bibfnamefont {C.}~\bibnamefont {Niu}}, \bibinfo {author} {\bibfnamefont
  {W.-J.}\ \bibnamefont {Li}},\ and\ \bibinfo {author} {\bibfnamefont {K.-Y.}\
  \bibnamefont {Kim}},\ }\bibfield  {title} {\bibinfo {title} {Thermal
  diffusivity and butterfly velocity in anisotropic q-lattice models},\ }\href
  {https://doi.org/10.1007/JHEP01(2018)140} {\bibfield  {journal} {\bibinfo
  {journal} {Journal of High Energy Physics}\ }\textbf {\bibinfo {volume}
  {2018}},\ \bibinfo {pages} {140} (\bibinfo {year} {2018})}\BibitemShut
  {NoStop}%
\bibitem [{\citenamefont {Zhang}\ \emph {et~al.}(2017)\citenamefont {Zhang},
  \citenamefont {Levenson-Falk}, \citenamefont {Ramshaw}, \citenamefont {Bonn},
  \citenamefont {Liang}, \citenamefont {Hardy}, \citenamefont {Hartnoll},\ and\
  \citenamefont {Kapitulnik}}]{zhang_anomalous_2017}%
  \BibitemOpen
  \bibfield  {author} {\bibinfo {author} {\bibfnamefont {J.}~\bibnamefont
  {Zhang}}, \bibinfo {author} {\bibfnamefont {E.~M.}\ \bibnamefont
  {Levenson-Falk}}, \bibinfo {author} {\bibfnamefont {B.~J.}\ \bibnamefont
  {Ramshaw}}, \bibinfo {author} {\bibfnamefont {D.~A.}\ \bibnamefont {Bonn}},
  \bibinfo {author} {\bibfnamefont {R.}~\bibnamefont {Liang}}, \bibinfo
  {author} {\bibfnamefont {W.~N.}\ \bibnamefont {Hardy}}, \bibinfo {author}
  {\bibfnamefont {S.~A.}\ \bibnamefont {Hartnoll}},\ and\ \bibinfo {author}
  {\bibfnamefont {A.}~\bibnamefont {Kapitulnik}},\ }\bibfield  {title}
  {{\selectlanguage {english}\bibinfo {title} {Anomalous thermal diffusivity in
  underdoped {YBa2Cu3O6}+x}},\ }\href {https://doi.org/10.1073/pnas.1703416114}
  {\bibfield  {journal} {\bibinfo  {journal} {Proceedings of the National
  Academy of Sciences}\ }\textbf {\bibinfo {volume} {114}},\ \bibinfo {pages}
  {5378} (\bibinfo {year} {2017})}\BibitemShut {NoStop}%
\bibitem [{\citenamefont {Martelli}\ \emph {et~al.}(2018)\citenamefont
  {Martelli}, \citenamefont {Jiménez}, \citenamefont {Continentino},
  \citenamefont {Baggio-Saitovitch},\ and\ \citenamefont
  {Behnia}}]{martelli_thermal_2018}%
  \BibitemOpen
  \bibfield  {author} {\bibinfo {author} {\bibfnamefont {V.}~\bibnamefont
  {Martelli}}, \bibinfo {author} {\bibfnamefont {J.~L.}\ \bibnamefont
  {Jiménez}}, \bibinfo {author} {\bibfnamefont {M.}~\bibnamefont
  {Continentino}}, \bibinfo {author} {\bibfnamefont {E.}~\bibnamefont
  {Baggio-Saitovitch}},\ and\ \bibinfo {author} {\bibfnamefont
  {K.}~\bibnamefont {Behnia}},\ }\bibfield  {title} {\bibinfo {title} {Thermal
  transport and phonon hydrodynamics in strontium titanate},\ }\href
  {https://doi.org/10.1103/PhysRevLett.120.125901} {\bibfield  {journal}
  {\bibinfo  {journal} {Physical Review Letters}\ }\textbf {\bibinfo {volume}
  {120}},\ \bibinfo {pages} {125901} (\bibinfo {year} {2018})},\ \bibinfo
  {note} {arXiv: 1802.05868}\BibitemShut {NoStop}%
\bibitem [{\citenamefont {Behnia}\ and\ \citenamefont
  {Kapitulnik}()}]{behnia_lower_2019}%
  \BibitemOpen
  \bibfield  {author} {\bibinfo {author} {\bibfnamefont {K.}~\bibnamefont
  {Behnia}}\ and\ \bibinfo {author} {\bibfnamefont {A.}~\bibnamefont
  {Kapitulnik}},\ }\bibfield  {title} {\bibinfo {title} {A lower bound to the
  thermal diffusivity of insulators},\ }\href
  {https://doi.org/10.1088/1361-648X/ab2db6} {\bibfield  {journal} {\bibinfo
  {journal} {Journal of Physics: Condensed Matter}\ }\textbf {\bibinfo {volume}
  {31}},\ \bibinfo {pages} {405702}}\BibitemShut {NoStop}%
\bibitem [{\citenamefont {Zhang}\ \emph {et~al.}(2019)\citenamefont {Zhang},
  \citenamefont {Kountz}, \citenamefont {Behnia},\ and\ \citenamefont
  {Kapitulnik}}]{zhang_thermalization_2019}%
  \BibitemOpen
  \bibfield  {author} {\bibinfo {author} {\bibfnamefont {J.}~\bibnamefont
  {Zhang}}, \bibinfo {author} {\bibfnamefont {E.~D.}\ \bibnamefont {Kountz}},
  \bibinfo {author} {\bibfnamefont {K.}~\bibnamefont {Behnia}},\ and\ \bibinfo
  {author} {\bibfnamefont {A.}~\bibnamefont {Kapitulnik}},\ }\bibfield  {title}
  {{\selectlanguage {english}\bibinfo {title} {Thermalization and possible
  signatures of quantum chaos in complex crystalline materials}},\ }\href
  {https://doi.org/10.1073/pnas.1910131116} {\bibfield  {journal} {\bibinfo
  {journal} {Proceedings of the National Academy of Sciences}\ }\textbf
  {\bibinfo {volume} {116}},\ \bibinfo {pages} {19869} (\bibinfo {year}
  {2019})}\BibitemShut {NoStop}%
\bibitem [{\citenamefont {Martelli}\ \emph {et~al.}()\citenamefont {Martelli},
  \citenamefont {Abud}, \citenamefont {Jiménez}, \citenamefont
  {Baggio-Saitovich}, \citenamefont {Zhao},\ and\ \citenamefont
  {Behnia}}]{martelli_thermal_2021}%
  \BibitemOpen
  \bibfield  {author} {\bibinfo {author} {\bibfnamefont {V.}~\bibnamefont
  {Martelli}}, \bibinfo {author} {\bibfnamefont {F.}~\bibnamefont {Abud}},
  \bibinfo {author} {\bibfnamefont {J.~L.}\ \bibnamefont {Jiménez}}, \bibinfo
  {author} {\bibfnamefont {E.}~\bibnamefont {Baggio-Saitovich}}, \bibinfo
  {author} {\bibfnamefont {L.-D.}\ \bibnamefont {Zhao}},\ and\ \bibinfo
  {author} {\bibfnamefont {K.}~\bibnamefont {Behnia}},\ }\bibfield  {title}
  {\bibinfo {title} {Thermal diffusivity and its lower bound in orthorhombic
  {SnSe}},\ }\href {https://doi.org/10.1103/PhysRevB.104.035208} {\bibfield
  {journal} {\bibinfo  {journal} {Physical Review B}\ }\textbf {\bibinfo
  {volume} {104}},\ \bibinfo {pages} {035208}}\BibitemShut {NoStop}%
\bibitem [{\citenamefont {Tulipman}\ and\ \citenamefont
  {Berg}(2020)}]{tulipman_strongly_2020}%
  \BibitemOpen
  \bibfield  {author} {\bibinfo {author} {\bibfnamefont {E.}~\bibnamefont
  {Tulipman}}\ and\ \bibinfo {author} {\bibfnamefont {E.}~\bibnamefont
  {Berg}},\ }\bibfield  {title} {\bibinfo {title} {Strongly coupled quantum
  phonon fluid in a solvable model},\ }\href
  {https://doi.org/10.1103/PhysRevResearch.2.033431} {\bibfield  {journal}
  {\bibinfo  {journal} {Physical Review Research}\ }\textbf {\bibinfo {volume}
  {2}},\ \bibinfo {pages} {033431} (\bibinfo {year} {2020})}\BibitemShut
  {NoStop}%
\bibitem [{\citenamefont {Sachdev}\ and\ \citenamefont
  {Ye}(1993)}]{sachdev_gapless_1993}%
  \BibitemOpen
  \bibfield  {author} {\bibinfo {author} {\bibfnamefont {S.}~\bibnamefont
  {Sachdev}}\ and\ \bibinfo {author} {\bibfnamefont {J.}~\bibnamefont {Ye}},\
  }\bibfield  {title} {\bibinfo {title} {Gapless {Spin}-{Fluid} {Ground}
  {State} in a {Random} {Quantum} {Heisenberg} {Magnet}},\ }\href
  {https://doi.org/10.1103/PhysRevLett.70.3339} {\bibfield  {journal} {\bibinfo
   {journal} {Physical Review Letters}\ }\textbf {\bibinfo {volume} {70}},\
  \bibinfo {pages} {3339} (\bibinfo {year} {1993})},\ \bibinfo {note} {arXiv:
  cond-mat/9212030}\BibitemShut {NoStop}%
\bibitem [{\citenamefont {Maldacena}\ and\ \citenamefont
  {Stanford}(2016)}]{maldacena_remarks_2016}%
  \BibitemOpen
  \bibfield  {author} {\bibinfo {author} {\bibfnamefont {J.}~\bibnamefont
  {Maldacena}}\ and\ \bibinfo {author} {\bibfnamefont {D.}~\bibnamefont
  {Stanford}},\ }\bibfield  {title} {{\selectlanguage {english}\bibinfo {title}
  {Remarks on the {Sachdev}-{Ye}-{Kitaev} model}},\ }\bibfield  {journal}
  {\bibinfo  {journal} {Physical Review D}\ }\textbf {\bibinfo {volume} {94}},\
  \href {https://doi.org/10.1103/PhysRevD.94.106002}
  {10.1103/PhysRevD.94.106002} (\bibinfo {year} {2016})\BibitemShut {NoStop}%
\bibitem [{\citenamefont {Cugliandolo}\ \emph {et~al.}(2001)\citenamefont
  {Cugliandolo}, \citenamefont {Grempel},\ and\ \citenamefont
  {Santos}}]{cugliandolo_quantum_2001}%
  \BibitemOpen
  \bibfield  {author} {\bibinfo {author} {\bibfnamefont {L.~F.}\ \bibnamefont
  {Cugliandolo}}, \bibinfo {author} {\bibfnamefont {D.~R.}\ \bibnamefont
  {Grempel}},\ and\ \bibinfo {author} {\bibfnamefont {C.~A. d.~S.}\
  \bibnamefont {Santos}},\ }\bibfield  {title} {\bibinfo {title} {The {Quantum}
  {Spherical} p-{Spin}-{Glass} {Model}},\ }\href
  {https://doi.org/10.1103/PhysRevB.64.014403} {\bibfield  {journal} {\bibinfo
  {journal} {Physical Review B}\ }\textbf {\bibinfo {volume} {64}},\ \bibinfo
  {pages} {014403} (\bibinfo {year} {2001})}\BibitemShut {NoStop}%
\bibitem [{\citenamefont {Giombi}\ \emph {et~al.}()\citenamefont {Giombi},
  \citenamefont {Klebanov},\ and\ \citenamefont
  {Tarnopolsky}}]{giombi_bosonic_2017}%
  \BibitemOpen
  \bibfield  {author} {\bibinfo {author} {\bibfnamefont {S.}~\bibnamefont
  {Giombi}}, \bibinfo {author} {\bibfnamefont {I.~R.}\ \bibnamefont
  {Klebanov}},\ and\ \bibinfo {author} {\bibfnamefont {G.}~\bibnamefont
  {Tarnopolsky}},\ }\bibfield  {title} {\bibinfo {title} {{Bosonic tensor
  models at large $N$ and small $\varepsilon$}},\ }\href
  {https://doi.org/10.1103/PhysRevD.96.106014} {\bibfield  {journal} {\bibinfo
  {journal} {Physical Review D}\ }\textbf {\bibinfo {volume} {96}},\ \bibinfo
  {pages} {106014}}\BibitemShut {NoStop}%
\bibitem [{\citenamefont {Benedetti}\ and\ \citenamefont
  {Delporte}()}]{benedetti_remarks_2021}%
  \BibitemOpen
  \bibfield  {author} {\bibinfo {author} {\bibfnamefont {D.}~\bibnamefont
  {Benedetti}}\ and\ \bibinfo {author} {\bibfnamefont {N.}~\bibnamefont
  {Delporte}},\ }\bibfield  {title} {\bibinfo {title} {Remarks on a melonic
  field theory with cubic interaction},\ }\href
  {https://doi.org/10.1007/JHEP04(2021)197} {\bibfield  {journal} {\bibinfo
  {journal} {Journal of High Energy Physics}\ }\textbf {\bibinfo {volume}
  {2021}},\ \bibinfo {pages} {197}},\ \Eprint
  {https://arxiv.org/abs/2012.12238} {2012.12238} \BibitemShut {NoStop}%
\bibitem [{\citenamefont {Blake}\ \emph {et~al.}(2018)\citenamefont {Blake},
  \citenamefont {Lee},\ and\ \citenamefont {Liu}}]{blake_quantum_2018}%
  \BibitemOpen
  \bibfield  {author} {\bibinfo {author} {\bibfnamefont {M.}~\bibnamefont
  {Blake}}, \bibinfo {author} {\bibfnamefont {H.}~\bibnamefont {Lee}},\ and\
  \bibinfo {author} {\bibfnamefont {H.}~\bibnamefont {Liu}},\ }\bibfield
  {title} {\bibinfo {title} {A quantum hydrodynamical description for
  scrambling and many-body chaos},\ }\bibfield  {journal} {\bibinfo  {journal}
  {Journal of High Energy Physics}\ }\textbf {\bibinfo {volume} {2018}},\ \href
  {https://doi.org/10.1007/JHEP10(2018)127} {10.1007/JHEP10(2018)127} (\bibinfo
  {year} {2018})\BibitemShut {NoStop}%
\bibitem [{\citenamefont {Baggioli}\ and\ \citenamefont
  {Li}()}]{baggioli_universal_2020}%
  \BibitemOpen
  \bibfield  {author} {\bibinfo {author} {\bibfnamefont {M.}~\bibnamefont
  {Baggioli}}\ and\ \bibinfo {author} {\bibfnamefont {W.-J.}\ \bibnamefont
  {Li}},\ }\bibfield  {title} {\bibinfo {title} {Universal bounds on transport
  in holographic systems with broken translations},\ }\href
  {https://doi.org/10.21468/SciPostPhys.9.1.007} {\bibfield  {journal}
  {\bibinfo  {journal} {{SciPost} Physics}\ }\textbf {\bibinfo {volume} {9}},\
  \bibinfo {pages} {007}}\BibitemShut {NoStop}%
\bibitem [{\citenamefont {Wu}\ \emph {et~al.}()\citenamefont {Wu},
  \citenamefont {Baggioli},\ and\ \citenamefont {Li}}]{wu_universality_2021}%
  \BibitemOpen
  \bibfield  {author} {\bibinfo {author} {\bibfnamefont {N.}~\bibnamefont
  {Wu}}, \bibinfo {author} {\bibfnamefont {M.}~\bibnamefont {Baggioli}},\ and\
  \bibinfo {author} {\bibfnamefont {W.-J.}\ \bibnamefont {Li}},\ }\bibfield
  {title} {\bibinfo {title} {On the universality of {AdS}2 diffusion bounds and
  the breakdown of linearized hydrodynamics},\ }\href
  {https://doi.org/10.1007/JHEP05(2021)014} {\bibfield  {journal} {\bibinfo
  {journal} {Journal of High Energy Physics}\ }\textbf {\bibinfo {volume}
  {2021}},\ \bibinfo {pages} {14}}\BibitemShut {NoStop}%
\bibitem [{\citenamefont {Ziman}(1960)}]{Ziman_2001}%
  \BibitemOpen
  \bibfield  {author} {\bibinfo {author} {\bibfnamefont {J.~M.}\ \bibnamefont
  {Ziman}},\ }\href@noop {} {}\ (\bibinfo  {publisher} {Oxford university
  press},\ \bibinfo {year} {1960})\BibitemShut {NoStop}%
\bibitem [{\citenamefont {Prosen}\ and\ \citenamefont
  {Campbell}(2000)}]{prosen_momentum_2000}%
  \BibitemOpen
  \bibfield  {author} {\bibinfo {author} {\bibfnamefont {T.}~\bibnamefont
  {Prosen}}\ and\ \bibinfo {author} {\bibfnamefont {D.~K.}\ \bibnamefont
  {Campbell}},\ }\bibfield  {title} {\bibinfo {title} {Momentum conservation
  implies anomalous energy transport in 1d classical lattices},\ }\href
  {https://doi.org/10.1103/PhysRevLett.84.2857} {\bibfield  {journal} {\bibinfo
   {journal} {Physical Review Letters}\ }\textbf {\bibinfo {volume} {84}},\
  \bibinfo {pages} {2857} (\bibinfo {year} {2000})}\BibitemShut {NoStop}%
\bibitem [{\citenamefont {Gu}\ \emph {et~al.}(2018)\citenamefont {Gu},
  \citenamefont {Wei}, \citenamefont {Yin}, \citenamefont {Li},\ and\
  \citenamefont {Yang}}]{gu_colloquium_2018}%
  \BibitemOpen
  \bibfield  {author} {\bibinfo {author} {\bibfnamefont {X.}~\bibnamefont
  {Gu}}, \bibinfo {author} {\bibfnamefont {Y.}~\bibnamefont {Wei}}, \bibinfo
  {author} {\bibfnamefont {X.}~\bibnamefont {Yin}}, \bibinfo {author}
  {\bibfnamefont {B.}~\bibnamefont {Li}},\ and\ \bibinfo {author}
  {\bibfnamefont {R.}~\bibnamefont {Yang}},\ }\bibfield  {title} {\bibinfo
  {title} {Colloquium: Phononic thermal properties of two-dimensional
  materials},\ }\href {https://doi.org/10.1103/RevModPhys.90.041002} {\bibfield
   {journal} {\bibinfo  {journal} {Reviews of Modern Physics}\ }\textbf
  {\bibinfo {volume} {90}},\ \bibinfo {pages} {041002} (\bibinfo {year}
  {2018})}\BibitemShut {NoStop}%
\bibitem [{\citenamefont {Mezard}\ \emph {et~al.}(1986)\citenamefont {Mezard},
  \citenamefont {Parisi},\ and\ \citenamefont {Virasoro}}]{mezard_spin_1986}%
  \BibitemOpen
  \bibfield  {author} {\bibinfo {author} {\bibfnamefont {M.}~\bibnamefont
  {Mezard}}, \bibinfo {author} {\bibfnamefont {G.}~\bibnamefont {Parisi}},\
  and\ \bibinfo {author} {\bibfnamefont {M.}~\bibnamefont {Virasoro}},\ }\href
  {https://doi.org/10.1142/0271} {\emph {\bibinfo {title} {Spin {Glass}
  {Theory} and {Beyond}}}},\ \bibinfo {series} {World {Scientific} {Lecture}
  {Notes} in {Physics}}, Vol.\ \bibinfo {volume} {Volume 9}\ (\bibinfo
  {publisher} {World Scientific},\ \bibinfo {year} {1986})\BibitemShut
  {NoStop}%
\bibitem [{\citenamefont {Wu}\ and\ \citenamefont
  {Sau}(2021)}]{wu_classical_2021}%
  \BibitemOpen
  \bibfield  {author} {\bibinfo {author} {\bibfnamefont {H.-K.}\ \bibnamefont
  {Wu}}\ and\ \bibinfo {author} {\bibfnamefont {J.}~\bibnamefont {Sau}},\
  }\bibfield  {title} {\bibinfo {title} {A classical model for sub-planckian
  thermal diffusivity in complex crystals},\ }\href
  {https://doi.org/10.1103/PhysRevB.103.184305} {\bibfield  {journal} {\bibinfo
   {journal} {Physical Review B}\ }\textbf {\bibinfo {volume} {103}},\ \bibinfo
  {pages} {184305} (\bibinfo {year} {2021})},\ \Eprint
  {https://arxiv.org/abs/2101.05353} {2101.05353} \BibitemShut {NoStop}%
\bibitem [{\citenamefont {Liao}\ and\ \citenamefont
  {Galitski}(2018)}]{liao_nonlinear_2018}%
  \BibitemOpen
  \bibfield  {author} {\bibinfo {author} {\bibfnamefont {Y.}~\bibnamefont
  {Liao}}\ and\ \bibinfo {author} {\bibfnamefont {V.}~\bibnamefont
  {Galitski}},\ }\bibfield  {title} {\bibinfo {title} {Nonlinear sigma model
  approach to many-body quantum chaos: Regularized and unregularized
  out-of-time-ordered correlators},\ }\href
  {https://doi.org/10.1103/PhysRevB.98.205124} {\bibfield  {journal} {\bibinfo
  {journal} {Physical Review B}\ }\textbf {\bibinfo {volume} {98}},\ \bibinfo
  {pages} {205124} (\bibinfo {year} {2018})}\BibitemShut {NoStop}%
\bibitem [{\citenamefont {Grozdanov}\ \emph {et~al.}(2019)\citenamefont
  {Grozdanov}, \citenamefont {Schalm},\ and\ \citenamefont
  {Scopelliti}}]{grozdanov_kinetic_2019}%
  \BibitemOpen
  \bibfield  {author} {\bibinfo {author} {\bibfnamefont {S.}~\bibnamefont
  {Grozdanov}}, \bibinfo {author} {\bibfnamefont {K.}~\bibnamefont {Schalm}},\
  and\ \bibinfo {author} {\bibfnamefont {V.}~\bibnamefont {Scopelliti}},\
  }\bibfield  {title} {\bibinfo {title} {Kinetic theory for classical and
  quantum many-body chaos},\ }\href
  {https://doi.org/10.1103/PhysRevE.99.012206} {\bibfield  {journal} {\bibinfo
  {journal} {Physical Review E}\ }\textbf {\bibinfo {volume} {99}},\ \bibinfo
  {pages} {012206} (\bibinfo {year} {2019})}\BibitemShut {NoStop}%
\bibitem [{\citenamefont {Romero-Bermúdez}\ \emph {et~al.}(2019)\citenamefont
  {Romero-Bermúdez}, \citenamefont {Schalm},\ and\ \citenamefont
  {Scopelliti}}]{romero-bermudez_regularization_2019}%
  \BibitemOpen
  \bibfield  {author} {\bibinfo {author} {\bibfnamefont {A.}~\bibnamefont
  {Romero-Bermúdez}}, \bibinfo {author} {\bibfnamefont {K.}~\bibnamefont
  {Schalm}},\ and\ \bibinfo {author} {\bibfnamefont {V.}~\bibnamefont
  {Scopelliti}},\ }\bibfield  {title} {\bibinfo {title} {Regularization
  dependence of the {OTOC}. which lyapunov spectrum is the physical one?},\
  }\href {https://doi.org/10.1007/JHEP07(2019)107} {\bibfield  {journal}
  {\bibinfo  {journal} {Journal of High Energy Physics}\ }\textbf {\bibinfo
  {volume} {2019}},\ \bibinfo {pages} {107} (\bibinfo {year}
  {2019})}\BibitemShut {NoStop}%
\bibitem [{\citenamefont {Kobrin}\ \emph {et~al.}(2021)\citenamefont {Kobrin},
  \citenamefont {Yang}, \citenamefont {Kahanamoku-Meyer}, \citenamefont
  {Olund}, \citenamefont {Moore}, \citenamefont {Stanford},\ and\ \citenamefont
  {Yao}}]{kobrin_many-body_2021}%
  \BibitemOpen
  \bibfield  {author} {\bibinfo {author} {\bibfnamefont {B.}~\bibnamefont
  {Kobrin}}, \bibinfo {author} {\bibfnamefont {Z.}~\bibnamefont {Yang}},
  \bibinfo {author} {\bibfnamefont {G.~D.}\ \bibnamefont {Kahanamoku-Meyer}},
  \bibinfo {author} {\bibfnamefont {C.~T.}\ \bibnamefont {Olund}}, \bibinfo
  {author} {\bibfnamefont {J.~E.}\ \bibnamefont {Moore}}, \bibinfo {author}
  {\bibfnamefont {D.}~\bibnamefont {Stanford}},\ and\ \bibinfo {author}
  {\bibfnamefont {N.~Y.}\ \bibnamefont {Yao}},\ }\bibfield  {title} {\bibinfo
  {title} {{Many-Body Chaos in the Sachdev-Ye-Kitaev Model}},\ }\href
  {https://doi.org/10.1103/PhysRevLett.126.030602} {\bibfield  {journal}
  {\bibinfo  {journal} {Physical Review Letters}\ }\textbf {\bibinfo {volume}
  {126}},\ \bibinfo {pages} {030602} (\bibinfo {year} {2021})}\BibitemShut
  {NoStop}%
\bibitem [{\citenamefont {Chowdhury}\ and\ \citenamefont
  {Swingle}(2017)}]{chowdhury_onset_2017}%
  \BibitemOpen
  \bibfield  {author} {\bibinfo {author} {\bibfnamefont {D.}~\bibnamefont
  {Chowdhury}}\ and\ \bibinfo {author} {\bibfnamefont {B.}~\bibnamefont
  {Swingle}},\ }\bibfield  {title} {\bibinfo {title} {{Onset of many-body chaos
  in the $O(N)$ model}},\ }\href {https://doi.org/10.1103/PhysRevD.96.065005}
  {\bibfield  {journal} {\bibinfo  {journal} {Physical Review D}\ }\textbf
  {\bibinfo {volume} {96}},\ \bibinfo {pages} {065005} (\bibinfo {year}
  {2017})},\ \Eprint {https://arxiv.org/abs/1703.02545} {1703.02545}
  \BibitemShut {NoStop}%
\bibitem [{\citenamefont {Aleiner}\ \emph {et~al.}(2016)\citenamefont
  {Aleiner}, \citenamefont {Faoro},\ and\ \citenamefont
  {Ioffe}}]{aleiner_microscopic_2016}%
  \BibitemOpen
  \bibfield  {author} {\bibinfo {author} {\bibfnamefont {I.~L.}\ \bibnamefont
  {Aleiner}}, \bibinfo {author} {\bibfnamefont {L.}~\bibnamefont {Faoro}},\
  and\ \bibinfo {author} {\bibfnamefont {L.~B.}\ \bibnamefont {Ioffe}},\
  }\bibfield  {title} {\bibinfo {title} {Microscopic model of quantum butterfly
  effect: Out-of-time-order correlators and traveling combustion waves},\
  }\href {https://doi.org/10.1016/j.aop.2016.09.006} {\bibfield  {journal}
  {\bibinfo  {journal} {Annals of Physics}\ }\textbf {\bibinfo {volume}
  {375}},\ \bibinfo {pages} {378} (\bibinfo {year} {2016})}\BibitemShut
  {NoStop}%
\bibitem [{\citenamefont {Shenker}\ and\ \citenamefont
  {Stanford}(2015)}]{shenker_stringy_2015}%
  \BibitemOpen
  \bibfield  {author} {\bibinfo {author} {\bibfnamefont {S.~H.}\ \bibnamefont
  {Shenker}}\ and\ \bibinfo {author} {\bibfnamefont {D.}~\bibnamefont
  {Stanford}},\ }\bibfield  {title} {\bibinfo {title} {Stringy effects in
  scrambling},\ }\href {https://doi.org/10.1007/JHEP05(2015)132} {\bibfield
  {journal} {\bibinfo  {journal} {Journal of High Energy Physics}\ }\textbf
  {\bibinfo {volume} {2015}},\ \bibinfo {pages} {132} (\bibinfo {year}
  {2015})}\BibitemShut {NoStop}%
\bibitem [{\citenamefont {Banerjee}\ and\ \citenamefont
  {Altman}(2017)}]{banerjee_solvable_2017}%
  \BibitemOpen
  \bibfield  {author} {\bibinfo {author} {\bibfnamefont {S.}~\bibnamefont
  {Banerjee}}\ and\ \bibinfo {author} {\bibfnamefont {E.}~\bibnamefont
  {Altman}},\ }\bibfield  {title} {\bibinfo {title} {Solvable model for a
  dynamical quantum phase transition from fast to slow scrambling},\ }\href
  {https://doi.org/10.1103/PhysRevB.95.134302} {\bibfield  {journal} {\bibinfo
  {journal} {Physical Review B}\ }\textbf {\bibinfo {volume} {95}},\ \bibinfo
  {pages} {134302} (\bibinfo {year} {2017})}\BibitemShut {NoStop}%
\bibitem [{\citenamefont {Lunts}\ and\ \citenamefont
  {Patel}(2019)}]{lunts_many-body_2019}%
  \BibitemOpen
  \bibfield  {author} {\bibinfo {author} {\bibfnamefont {P.}~\bibnamefont
  {Lunts}}\ and\ \bibinfo {author} {\bibfnamefont {A.~A.}\ \bibnamefont
  {Patel}},\ }\bibfield  {title} {\bibinfo {title} {Many-body chaos in the
  antiferromagnetic quantum critical metal},\ }\href
  {https://doi.org/10.1103/PhysRevB.100.235104} {\bibfield  {journal} {\bibinfo
   {journal} {Physical Review B}\ }\textbf {\bibinfo {volume} {100}},\ \bibinfo
  {pages} {235104} (\bibinfo {year} {2019})}\BibitemShut {NoStop}%
\bibitem [{\citenamefont {Bera}\ \emph {et~al.}(2021)\citenamefont {Bera},
  \citenamefont {Lokesh},\ and\ \citenamefont {Banerjee}}]{bera_quantum_2021}%
  \BibitemOpen
  \bibfield  {author} {\bibinfo {author} {\bibfnamefont {S.}~\bibnamefont
  {Bera}}, \bibinfo {author} {\bibfnamefont {K.~Y.~V.}\ \bibnamefont
  {Lokesh}},\ and\ \bibinfo {author} {\bibfnamefont {S.}~\bibnamefont
  {Banerjee}},\ }\bibfield  {title} {\bibinfo {title} {Quantum to classical
  crossover in many-body chaos in a glass},\ }\href
  {http://arxiv.org/abs/2105.13376} {\bibfield  {journal} {\bibinfo  {journal}
  {{arXiv}:2105.13376 [cond-mat, physics:hep-th]}\ } (\bibinfo {year}
  {2021})},\ \Eprint {https://arxiv.org/abs/2105.13376} {2105.13376}
  \BibitemShut {NoStop}%
\bibitem [{\citenamefont {Anous}\ and\ \citenamefont
  {Haehl}(2021)}]{anous_quantum_2021}%
  \BibitemOpen
  \bibfield  {author} {\bibinfo {author} {\bibfnamefont {T.}~\bibnamefont
  {Anous}}\ and\ \bibinfo {author} {\bibfnamefont {F.~M.}\ \bibnamefont
  {Haehl}},\ }\bibfield  {title} {\bibinfo {title} {{The quantum $p$-spin glass
  model: A user manual for holographers}},\ }\href
  {http://arxiv.org/abs/2106.03838} {\bibfield  {journal} {\bibinfo  {journal}
  {{arXiv}:2106.03838 [cond-mat, physics:hep-th]}\ } (\bibinfo {year}
  {2021})},\ \Eprint {https://arxiv.org/abs/2106.03838} {2106.03838}
  \BibitemShut {NoStop}%
\bibitem [{\citenamefont {Feynman}\ and\ \citenamefont
  {Hibbs}(1965)}]{feynman_quantum_1965}%
  \BibitemOpen
  \bibfield  {author} {\bibinfo {author} {\bibfnamefont {R.~P.}\ \bibnamefont
  {Feynman}}\ and\ \bibinfo {author} {\bibfnamefont {A.~R.}\ \bibnamefont
  {Hibbs}},\ }\href@noop {} {{\selectlanguage {english}\emph {\bibinfo {title}
  {Quantum mechanics and path integrals}}}}\ (\bibinfo  {publisher}
  {McGraw-Hill},\ \bibinfo {year} {1965})\BibitemShut {NoStop}%
\bibitem [{\citenamefont {Mahan}(2000)}]{mahan_many-particle_2000}%
  \BibitemOpen
  \bibfield  {author} {\bibinfo {author} {\bibfnamefont {G.~D.}\ \bibnamefont
  {Mahan}},\ }\href {https://doi.org/10.1007/978-1-4757-5714-9} {\emph
  {\bibinfo {title} {Many-Particle Physics}}},\ \bibinfo {edition} {3rd}\ ed.,\
  Physics of Solids and Liquids\ (\bibinfo  {publisher} {Springer {US}},\
  \bibinfo {year} {2000})\BibitemShut {NoStop}%
\bibitem [{\citenamefont {Werman}\ \emph {et~al.}(2018)\citenamefont {Werman},
  \citenamefont {Chatterjee}, \citenamefont {Morampudi},\ and\ \citenamefont
  {Berg}}]{werman_signatures_2018}%
  \BibitemOpen
  \bibfield  {author} {\bibinfo {author} {\bibfnamefont {Y.}~\bibnamefont
  {Werman}}, \bibinfo {author} {\bibfnamefont {S.}~\bibnamefont {Chatterjee}},
  \bibinfo {author} {\bibfnamefont {S.~C.}\ \bibnamefont {Morampudi}},\ and\
  \bibinfo {author} {\bibfnamefont {E.}~\bibnamefont {Berg}},\ }\bibfield
  {title} {\bibinfo {title} {Signatures of fractionalization in spin liquids
  from interlayer thermal transport},\ }\href
  {https://doi.org/10.1103/PhysRevX.8.031064} {\bibfield  {journal} {\bibinfo
  {journal} {Physical Review X}\ }\textbf {\bibinfo {volume} {8}},\ \bibinfo
  {pages} {031064} (\bibinfo {year} {2018})}\BibitemShut {NoStop}%
\bibitem [{\citenamefont {Kamenev}(2011)}]{kamenev_field_2011}%
  \BibitemOpen
  \bibfield  {author} {\bibinfo {author} {\bibfnamefont {A.}~\bibnamefont
  {Kamenev}},\ }\href {https://doi.org/10.1017/CBO9781139003667} {\emph
  {\bibinfo {title} {Field Theory of Non-Equilibrium Systems}}}\ (\bibinfo
  {publisher} {Cambridge University Press},\ \bibinfo {year}
  {2011})\BibitemShut {NoStop}%
\end{thebibliography}%

\onecolumngrid

\appendix

\section{Imaginary time}
\label{Rep_app}
Here, we generalize the replica analysis of \cite{tulipman_strongly_2020} to the lattice model. Following our steps in \cite{tulipman_strongly_2020}, we obtain the effective action and corresponding SPEs of the two phases of the model. Let us consider $d=1$ for simplicity. We comment on higher dimensions later on. We integrate over the disorder and introduce the  $G,\Pi$ fields, such that the disorder-averaged replicated partition function can be expressed as a functional integral given by  
\begin{eqnarray}
\overline{Z^{n}}=\mathcal{D}\boldsymbol{G}\mathcal{D}\boldsymbol{\Pi}\exp\left(-\frac{nN}{2}S_{\text{eff}}\left(\boldsymbol{G},\boldsymbol{\Pi}\right)\right)
\end{eqnarray}

where $S_{\text{eff}}=S_{0}+S_{v}+S_{u}$ with 
\begin{eqnarray}
S_{0}	&=&\sum_{b}n_{b}\int_{k,k'}\ln\det\left(\delta_{\alpha\beta}\delta_{k+k'}\delta\left(\tau-\tau'\right)\left(-\partial_{\tau}^{2}+\Omega_{b}^{2}+4\Omega_{\text{d}b}^{2}\sin^{2}\left(\frac{k}{2}\right)\right)-\Pi_{b}^{\alpha\beta}\left(-k,\tau;-k',\tau'\right)\right); \nonumber \\
S_{v}	&=&\sum_{\alpha\beta}\sum_{r,r'}\int d\tau d\tau'\left(-\frac{v^{2}}{3}\widetilde{\mathcal{G}}^{\alpha\beta}\left(r,\tau;r',\tau'\right)^{3}+\sum_{b}n_{b}\Pi_{b}^{\alpha\beta}\left(r,\tau;r',\tau'\right)\widetilde{G}_{b}^{\alpha\beta}\left(r,\tau;r',\tau'\right)\right) ; \nonumber \\
S_{u}	&=&\frac{u}{2}\sum_{\alpha\beta}\sum_{r,r'}\int d\tau d\tau'\delta\left(\tau-\tau'\right)\delta_{r,r'}\delta_{\alpha\beta}\widetilde{\mathcal{G}}^{\alpha\beta}\left(r,\tau;r',\tau'\right)^{2}.
\label{eq:replicaS}
\end{eqnarray}
Here, we consider the most general case of $N_B$ branches, including acoustic phonons, such that $\widetilde{\mathcal{G}}\equiv \sum_b n_b \widetilde{G}_b$ where $\widetilde{G}_b$ are defined with respect to the generalized fields in (\ref{eq:generalized_phi}). $\alpha$ and $\beta$ are replica indices.

\subsection{Replica-diagonal saddle point and specific heat}

The replica-diagonal saddle point is defined as ${G}_{b}^{\alpha\beta}\left(\tau,r;\tau',r'\right)={G}_{b}\left(\tau-\tau',r-r'\right)\delta_{\alpha\beta}$. Substituting the solution in (\ref{eq:replicaS}), we obtain the SPEs that govern the thermodynamics of the disordered phase. The optical branches satisfy (\ref{eq:SPE_s_SOBM_imag1}),(\ref{eq:SPE_s_SOBM_imag2}) while acoustic branches satisfy 
\begin{eqnarray}
G_{\text{a}}\left(i\omega_{n},k\right)&=&\frac{1}{\omega_{n}^{2}+4\Omega_{\text{a}}^{2}\sin^{2}\left(\frac{k}{2}\right)-\Pi_{\text{a}}\left(i\omega_{n},k\right)}, \nonumber \\
\Pi_{\text{a}}\left(i\omega_{n},k\right)&=&4\sin^{2}\left(\frac{k}{2}\right)\Pi_{\text{o}}\left(i\omega_{n},k\right). 
\end{eqnarray}
We solve the SPEs to linear order in $n_{\rm a}$ by an iterative procedure similarly to \cite{tulipman_strongly_2020}. In particular, the small $n_{\rm a}$ and the weakly dispersive limit renders the self-energy of the optical phonons to be momentum-independent to leading order. This is a major simplification for the numerical solution of the SPEs. 

To compute the specific heat we obtain the internal energy $U$ by repeating the steps in \cite{tulipman_strongly_2020} where the only modification is the added summation over space/momentum. $U$ is given by 
\begin{eqnarray}
U&=\sum_{b}n_{b}\int_{k}\left(\Omega_{b}^{2}+4\Omega_{\text{d}b}^{2}\sin^{2}\left(\frac{k}{2}\right)\right)\widetilde{G}_{b}\left(\tau=0,k\right)-\sum_{r}\int d\tau\frac{v^{2}}{3}\widetilde{\mathcal{G}}\left(\tau,r\right)^{3}+\frac{3u}{4}\widetilde{\mathcal{G}}\left(\tau=0,r=0\right)^{2}.
\end{eqnarray}

From $U$, the specific heat $c=\partial_T U$ can be computed either numerically or analytically in the low- and high-$T$ limits similarly to \cite{tulipman_strongly_2020}.

\subsection{One-step replica symmetry breaking saddle point of optical modes}

Apart from the replica-diagonal saddle point, the one-step replica-symmetry breaking saddle point is the only other stable solution in replica space \cite{cugliandolo_quantum_2001,tulipman_strongly_2020}. We extend our previous analysis to the lattice model. This is a two step process. As a first step, we show that the SPEs are given as a sum over the SPEs of the different branches. This allows us to explicitly solve the SPEs in the case of multiple optical modes in the weakly dispersive limit, where we approximate the Green's functions to be $k$-independent, i.e. letting $\Omega_{\rm d} \to 0$. The reason for that is to avoid unnecessary complication related to $k$-labels that have essentially no effect on the phase diagram. As a second step, we introduce acoustic phonons and show that their contribution is continuous as a function of $n_{\rm a}$. Relying on this fact, in the $n_{\rm a} \to 0$ limit, staying sufficiently far away from the glass phase, we may consider systems with small $n_{\rm a} > 0$ without any risk of inducing a phase transition.  

We define the \textit{local} 1SRSB solution of the $b$th branch by
\begin{eqnarray}
G_{b}^{\alpha\beta}\left(i\omega_{n},k\right)\equiv\left(g_{b}^{d}\left(i\omega_{n},k\right)-g_{EA}\right)\delta_{\alpha\beta}+\epsilon_{\alpha\beta}g_{EA}
\label{1srsb_ansatz}
\end{eqnarray}
where $g_{EA}$ is the Edwards-Anderson order parameter, and $\epsilon_{\alpha\beta}=1$ if $\alpha,\beta$ are on a diagonal block of size $m$ and zero otherwise. As stated above, we consider the zeroth order in $\Omega_{\rm d}$, where the solution is $k$-independent: 
\begin{eqnarray}
G_{b}^{\alpha\beta}\left(i\omega_{n}\right)\equiv\left(g_{b}^{d}\left(i\omega_{n}\right)-g_{EA}\right)\delta_{\alpha\beta}+\epsilon_{\alpha\beta}g_{EA}.
\end{eqnarray}
To proceed, we follow our steps in \cite{tulipman_strongly_2020}, with the small modification of considering branch-dependent quantities. We arrive at the following SPEs,
\begin{eqnarray}
0	&=&\sum_{b}n_{b}\left\{ \frac{1}{g_{b}^{d}\left(i\omega_{n}\right)}-\left(\omega_{n}^{2}+\Omega_{b}^{2}-\Pi_{b}\left(i\omega_{n}\right)\right)\right\} \label{eq:SPE_a} \\
0	&=&\sum_{b}n_{b}\left\{ \frac{g_{b}^{d}\left(0\right)+\left(m-2\right)\beta g_{EA}}{g_{b}^{d}\left(0\right)^{2}+\left(m-2\right)\beta g_{EA}g_{b}^{d}\left(0\right)-\left(m-1\right)\beta^{2}g_{EA}^{2}}-\left(\Omega_{b}^{2}-\Pi_{b}\left(i\omega_{n}=0\right)\right)\right\} \label{eq:SPE_b} \\
0	&=&\sum_{b}n_{b}\left\{ \frac{1}{\left(g_{b}^{d}\left(0\right)-\beta g_{EA}\right)\left(g_{b}^{d}\left(0\right)+\left(m-1\right)\beta g_{EA}\right)}-g_{EA}\right\} \label{eq:SPE_c} \\
0	&=&\sum_{b}n_{b}\left\{ \frac{m\beta g_{EA}}{g_{b}^{d}\left(0\right)+\left(m-1\right)\beta g_{EA}}+\ln\left(\frac{g_{b}^{d}\left(0\right)-\beta g_{EA}}{g_{b}^{d}\left(0\right)+\left(m-1\right)\beta g_{EA}}\right)+\frac{1}{3}m^{2}\beta^{2}g_{EA}^{3}\right\} ,\label{eq:SPE_d}
\end{eqnarray}
corresponding to $\delta S_{\rm eff}/\delta \Theta = 0$ with $\Theta = \Pi_{b}^{\alpha=\beta}\left(i\omega_{n}\ne0\right),\Pi_{b}^{\alpha\ne\beta}\left(i\omega_{n}=0\right),G_{b}^{\alpha\ne\beta}\left(i\omega_{n}=0\right)$ and $m$, respectively. Fortunately, these equations are amenable to a similar treatment as the single branch model, for any number of branches. Let us consider the case of two branches for simplicity. Generalizing to any $N_B>2$ is straightforward. 

We define
\begin{eqnarray}
y_b \equiv \frac{\beta g_{EA}}{g_b^d(0)},\quad x_b \equiv \frac{my_b}{1-y_b}
\label{1srsb_new_vars}
\end{eqnarray}
for branches $b=1,2$. Substituting the above in (\ref{eq:SPE_c}) reads
\begin{eqnarray}
m^{2}\beta^{2}g_{EA}^{3}=n_{1}\frac{x_{1}^{2}}{1+x_{1}}+n_{2}\frac{x_{2}^{2}}{1+x_{2}},
\label{eq:SPE_final_a}
\end{eqnarray}
such that (\ref{eq:SPE_d}) can be recasted into 
\begin{eqnarray}
0&=n_{1}\left(\frac{x_{1}}{1+x_{1}}+\ln\left(\frac{1}{1+x_{1}}\right)+\frac{1}{3}\frac{x_{1}^{2}}{1+x_{1}}\right)+n_{2}\left(\frac{x_{2}}{1+x_{2}}+\ln\left(\frac{1}{1+x_{2}}\right)+\frac{1}{3}\frac{x_{2}^{2}}{1+x_{2}}\right).
\label{eq:SPE_final_b}
\end{eqnarray}
From here, it is useful to define regularized functions $G^r,\Pi^r$ according to 
\begin{eqnarray}
g_{b}^{d}\left(i\omega_{n}\right)=\beta g_{EA}\delta_{n,0}+G_{b}^{r}\left(i\omega_{n}\right),\quad \Pi\left(i\omega_{n}\right)=\beta g_{EA}^{2}\delta_{n,0}+\Pi_{}^{r}\left(i\omega_{n}\right).
\label{eq:reg_RSB_G_F}
\end{eqnarray}
Here we have already used the fact that the self-energy $\Pi$ is branch-independent. Note that consistency demands that $G_{b}^{r}\left(0\right) = \frac{m}{x_{b}}\beta g_{EA}$. Inserting (\ref{eq:reg_RSB_G_F}) into (\ref{eq:SPE_a}) and (\ref{eq:SPE_b}), we arrive that 
\begin{eqnarray}
0&=n_{1}\left\{ \frac{1}{G_{1}^{r}\left(i\omega_{n}\right)}-\left(\omega_{n}^{2}+\Omega_{1}^{2}-\Pi^{r}\left(i\omega_{n}\right)\right)\right\} +n_{2}\left\{ \frac{1}{G_{2}^{r}\left(i\omega_{n}\right)}-\left(\omega_{n}^{2}+\Omega_{2}^{2}-\Pi^{r}\left(i\omega_{n}\right)\right)\right\} 
\end{eqnarray}
where $\Pi^r\left( \tau \right) = \mathcal{G}\left( \tau \right)^2 + 2g_{EA}\mathcal{G}\left( \tau \right) - u\left(\mathcal{G}\left( \tau \right) + g_{EA} \right)\delta(\tau)$ with $\mathcal{G} = n_1 G_1 + n_2 G_2$. 

To solve the SPEs, we start by fixing $m$ and $x_2$. This determines $x_1$ implicitly according to (\ref{eq:SPE_final_b}). Given $x_1,x_2$ and $m$, we may use (\ref{eq:SPE_final_a}) to determine $g_{EA}$. Given $m$ and $g_{EA}$, the constraint $G_{b}^{r}\left(0\right) = \frac{m}{x_{b}}\beta g_{EA}$ is explicitly enforced by setting $\delta \Pi_b = \frac{x_b}{m\beta g_{EA}}$ where $\delta \Pi _b = \Omega^2_b - \Pi^r(0)$. This, in turn, determines the phonon frequency implicitly: $ \Omega^2_b  \equiv \delta \Pi _b + \Pi^r(0)$. 

Having solved for all components of the SPEs, we may evaluate the free-energy density:
\begin{eqnarray}
  2\beta \overline{f}	&=&\sum_b n_b \left[ -\left(\frac{m-1}{m}\right)\ln\left(\frac{1-y_b}{1-\left(1-m\right)y_b}\right)-\ln\left(1+\left(m-1\right)y_b\right)-\sum_{n}\ln\left(\left(\omega_{n}^{2}+\Omega_{b}^{2}\right){g}_{b}^{b}\left(i\omega_{n}\right)\right)
\right]	\nonumber\\
	&+&\sum_{n,b}n_b\left(\left(\omega_{n}^{2}+\Omega_{b}^{2}\right){g}_{b}^{d}\left(i\omega_{n}\right)-1\right)-\frac{v^2}{3}\left(\beta\int_{0}^{\beta}d\tau \left(\sum_b n_b g_{b}^{d}\left(\tau\right)\right)^{3}+\left(m-1\right)\beta^{2}g_{EA}^{3}\right) \nonumber\\
	&+&\frac{u}{2}\beta \left(\sum_b n_b g_{b}^d\left(\tau=0\right)\right)^{2}+C.	
	\label{eq:free_energy_glass}
\end{eqnarray}
Here, $C=\sum_{n,b} n_b \ln\left(\beta^2 \left(\omega_n^2 + \Omega_b^2\right)\right)=\sum_b n_b 2\ln\left(2\sinh\left(\frac{\beta\Omega_b}{2}\right)\right)$  \cite{feynman_quantum_1965,cugliandolo_quantum_2001}. For $N_B>2$, we simply replace the summation accordingly, while in the solution we start by fixing $m$ and $x_2,...,x_B$ and then extract $x_1$. 

One may extended this procedure to the case of $\Omega_{\rm d} > 0$. This will add $k$ as a label to the variables we defined above. In addition, a summation over $k$ will be added to the summation over the branches. The limit of weak dispersion implies that the corrections we neglected are small, such that our phase diagram, and in particular the $T\to 0$ glass boundary are accurate up to corrections of order $\Omega_{\rm d}/{\rm min} \left[\overline{\Omega} \right] \ll 1$.

\subsection{One-step replica symmetry breaking saddle point with acoustic modes}

Before proceeding to consider a system with acoustic phonons, it is useful to note that it is sufficient to compute the free-energy difference between the two phase in order to map out the phase diagram. By simply observing (\ref{eq:replicaS}), we notice that only the $k=0$ acoustic mode at $\omega_n = 0$ might introduce a divergence. Importantly, this mode is decoupled from the rest of the system in both phases, such that it drops out when one considers the difference in the free-energy of the different phases. We will now explicitly show this decoupling is consistent with the solution of the 1SRSB SPEs in the presence of acoustic phonons.  

Let us now describe the solution for a system with a single acoustic and a single optical branch. We obtain the SPEs as above:
\begin{eqnarray}
0&=&\sum_{b}n_{b}\int_{k}\left\{ \frac{1}{g_{b}^{d}\left(i\omega_{n},k\right)}-\left(\omega_{n}^{2}+\epsilon_{b}\left(k\right)-\Pi_{b}\left(i\omega_{n},k\right)\right)\right\}  \\0&=&\sum_{b}n_{b}\int_{k}\left\{ \frac{g_{b}^{d}\left(0,k\right)+\left(m-2\right)\beta g_{EA}}{g_{b}^{d}\left(0,k\right)^{2}+\left(m-2\right)\beta g_{EA}g_{b}^{d}\left(0,k\right)-\left(m-1\right)\beta^{2}g_{EA}^{2}}-\left(\epsilon_{b}\left(k\right)-\Pi_{b}\left(0,k\right)\right)\right\}  \\0&=&\sum_{b}n_{b}\int_{k}\left\{ \frac{1}{\left(g_{b}^{d}\left(0,k\right)-\beta g_{EA}\right)\left(g_{b}^{d}\left(0,k\right)+\left(m-1\right)\beta g_{EA}\right)}-b\beta^{2}v^{2}g_{EA}\right\} \\0&=&\sum_{b}n_{b}\int_{k}\left\{ \frac{m\beta g_{EA}}{g_{b}^{d}\left(0,k\right)+\left(m-1\right)\beta g_{EA}}+\ln\left(\frac{g_{b}^{d}\left(0,k\right)-\beta g_{EA}}{g_{b}^{d}\left(0,k\right)+\left(m-1\right)\beta g_{EA}}\right)+\frac{v^{2}}{3}bm^{2}\beta^{2}g_{EA}^{3}\right\} 
\end{eqnarray}
where we introduced the shorthand notation $\epsilon_{b}\left(k\right)\equiv\Omega_{b}^{2}\left(1+\delta_{b\text{a}}4\sin^{2}\left(\frac{k}{2}\right)\right)+4\Omega_{\text{d}b}\sin^{2}\left(\frac{k}{2}\right)$ and $b \equiv \int_k \left[q*q*q\right]\left(k\right)$ with $q(k) = n_{\rm o}+4n_{\rm a}\sin^{2}\frac{k}{2}$ ($*$ denotes convolution). The factor of $b$ is the main difference between these SPEs and the $n_{\rm a}=0$ SPEs. This factor will force us to use slightly different definitions for the regularized function $G^r,\Pi^r$. Similraly to the previous case, we define 
\begin{eqnarray}
g_{b}^{d}\left(i\omega_{n}\right)=\beta g_{EA}\delta_{n,0}+G_{b}^{r}\left(i\omega_{n}\right),\quad \Pi_b\left(i\omega_{n}\right)=p_b(k) \delta_{n,0}+\Pi_{b}^{r}\left(i\omega_{n}\right),
\end{eqnarray}
with $p_{\rm o} = \frac{b\beta g_{\text{EA}}^{2}}{\int_{k}\left(4n_{\text{a}}\sin^{2}\left(\frac{k}{2}\right)+n_{\text{o}}\right)}$ and $p_{\rm a}(k) = 4\sin^{2}\left(\frac{k}{2}\right) p_{\rm o}$, together with a momentum-dependent change of variables:
\begin{eqnarray}
y_{b}\left(k\right)=\frac{\beta g_{EA}}{g_{b}^{d}\left(0,k\right)},\quad x_{b}\left(k\right)=\frac{my_{b}\left(k\right)}{1-y_{b}\left(k\right)}.
\end{eqnarray}
With these definitions, the SPEs can be recasted into 
\begin{eqnarray}
0&=&\int_{k}\left[n_{\text{a}}\frac{x_{\text{a}}^{2}\left(k\right)}{1+x_{\text{a}}\left(k\right)}+n_{\text{o}}\frac{x_{\text{o}}^{2}\left(k\right)}{1+x_{\text{o}}\left(k\right)}-bm^{2}\beta^{2}g_{EA}^{3}\right] \label{glass_acoustic1} \\
0&=&\int_{k}\bigg[n_{\text{a}}\left(\frac{x_{\text{a}}\left(k\right)}{1+x_{\text{a}}\left(k\right)}+\ln\left(\frac{1}{1+x_{\text{a}}\left(k\right)}\right)+\frac{1}{3}\frac{x_{\text{a}}^{2}\left(k\right)}{1+x_{\text{a}}\left(k\right)}\right)\nonumber +n_{\text{o}}\left(\frac{x_{\text{o}}\left(k\right)}{1+x_{\text{o}}\left(k\right)}+\ln\left(\frac{1}{1+x_{\text{o}}\left(k\right)}\right)+\frac{1}{3}\frac{x_{\text{o}}^{2}\left(k\right)}{1+x_{\text{o}}\left(k\right)}\right) \bigg] \nonumber \\
\label{glass_acoustic2}\\
0&=&n_{\text{a}}\left\{ \frac{1}{G_{\text{a}}^{r}\left(i\omega_{n},k\right)}-\left(\omega_{n}^{2}+\varepsilon_{\text{a}}\left(k\right)-\Pi_{\text{a}}^{r}\left(i\omega_{n},k\right)\right)\right\} +n_{\text{o}}\left\{ \frac{1}{G_{\text{o}}^{r}\left(i\omega_{n},k\right)}-\left(\omega_{n}^{2}+\varepsilon_{\text{o}}\left(k\right)-\Pi_{\text{o}}^{r}\left(i\omega_{n},k\right)\right)\right\} 
\end{eqnarray}
where 
\begin{eqnarray}
\Pi_{\text{o}}^{r}\left(i\omega_{n},k\right)&=\delta_{n,0}\left(\left[q*q\right]\left(k\right)-p_{\text{o}}\right)\beta g_{\text{EA}}^{2}+2\left[G_{EA}*\widetilde{\mathcal{G}}\right]\left(i\omega_{n},k\right)+\widetilde{\mathcal{G}}*\widetilde{\mathcal{G}}\left(i\omega_{n},k\right),
\end{eqnarray} 
$\Pi_{\text{a}}^{r}\left(i\omega_{n},k\right)=4\sin^{2}\left(\frac{k}{2}\right)\Pi_{\text{o}}^{r}\left(i\omega_{n},k\right)$ and $G_{EA}\left(i\omega_{n},k\right)=q\left(k\right)\beta g_{EA}\delta_{n,0}$. 

Our main point in writing the above equations is that we can now explicitly observe that acoustic phonons can be treated in the same manner as optical modes, and in particular, small values of $n_{\rm a}$ correspond to a small smooth deformation of the phase boundary. Indeed, the above equations are solved component-wise and branch-wise. Then, for any $k\ne 0$ the solution is similar to the case of optical phonons. For $k=0$, we have that $G_{\rm a}^r\left(0,0\right) = \infty$. This implies that $y_{\rm a}\left(0\right) = 0$ and accordingly $x_{\rm a}\left(k\right) = 0$, which is a valid solution for the acoustic part of equations (\ref{glass_acoustic1}) and (\ref{glass_acoustic2}).

\section{Real time}
\label{SK_app}

We study the real-time dynamics, transport, and chaos in the disordered phase of the model using the Keldysh formalism. At the $N\to \infty$ limit, within the disordered, self-averaging phase, we obtain the SPEs by considering the disorder-averaged partition function, following, step by step, the procedure and definitions we presented in \cite{tulipman_strongly_2020}. This yields the SPEs as given in (\ref{eq:Keldysh_sc_eqs_AO_model}). 

Before discussing transport and chaos, let us quickly recall why $\tau_{\rm a}(k) \sim 1/k^2$. Consider a retarded Green's function of the form 
\begin{eqnarray}
-G_R\left(\omega,k\right)^{-1} = \omega^2 - a(k)^2 + ib\left(\omega,k\right) \label{approx_G_R}
\end{eqnarray}
and assume that $b(\omega=a(k),k) \ll a$. To extract the lifetime we note that the pole of $G_R$ (with positive frequency) are approximately given by 
\begin{eqnarray}
\omega = \sqrt{a(k)^2 - ib(a,k)} \approx a -i b(a,k)/2a.
\label{a_lifetime}
\end{eqnarray}
For acoustic phonons, at sufficiently small $k$, $a_{\rm a}(k) = \overline{\Omega}_{\rm a}k$ and $b_{\rm a}(\omega,k) = 2\gamma\omega k^2$. Hence, we can readily identify that $\tau_{\rm a}^{-1}(k) = b_{\rm a}(a,k)/2a_{\rm a}(k) = \gamma k^2$.

\subsection{Thermal transport}

Following the prescription in (\ref{kappa_formula}), we need to obtain the imaginary part of the retarded current-current correlation function. To do that, we must first derive the current operator. We begin by deriving the thermal current of the optical phonons in $d=1$ following \cite{mahan_many-particle_2000,werman_signatures_2018}. Consider the continuity equation for the energy density $E(r)$,
\begin{eqnarray}
\partial_t E(r) + \partial_r J_{\rm th} (r) = 0.
\end{eqnarray}
We multiply the above by $r$ and integrate over $r$. Integrating the current term by parts read
\begin{eqnarray}
\int_r \left(\partial_r  J_{\rm th}(r) r \right) &=& -\int_r  J_{\rm th}(r) +  J_{\rm th}(r) r|_{r=-\infty}^{r=\infty} \nonumber \\
&=& - J_{\rm th}
\end{eqnarray}
where we consider an isolated system with open boundary conditions, and hence the boundary term in the right-hand-side vanishes. We have defined $ J_{\rm th} \equiv \int_r  J_{\rm th}(r) $. Hence 
\begin{eqnarray}
 J_{\rm th} = \int_r r\partial_t E(r).
\end{eqnarray}
In discrete notation:  $ J_{\rm th} = \sum_r r\partial_t E(r)$. The energy density $E(r)$ is defined by a symmetrized version of $H_{r,0}+H_{r,{\rm int}}$ in (\ref{H_LM}). Since the interaction term is local, we simply rewrite
\begin{eqnarray}
H_{{r},0}  &=& \sum_{i=1}^{N}  \frac{\pi_{i,{r}}^{2}}{2}+\frac{\Omega_{i}^{2}}{2}\phi_{i,{r}}^{2}+\Omega_{\text{d},i}^{2} \left(\phi_{i,{r}}^2-\frac{1}{2} \left(\phi_{i,{r}+1}\phi_{i,{r}}+\phi_{i,{r}-1}\phi_{i,{r}}\right)\right).
\end{eqnarray}
such that $E(r) \equiv H_{r,0}+H_{r,{\rm int}}$ with the above form. The local energy density evolves in time according to the Heisenberg equation, $\partial_t E(r) = i \left[H, H_{r,0}+H_{r,{\rm int}}\right]$. One then has to evaluate the commutator and preform the summation. Eventually, we obtain that
\begin{eqnarray}
J_{\rm th}(t) = -i\sum_i \frac{\Omega_{{\rm d}i}^2}{2} \int_k  \left(\sin k \partial_t \phi_{i,{-k}}\phi_{i,{k}}- \sin k\phi_{i,{-k}}\partial_t\phi_{i,{k}}\right),
\label{eq:thermal_current_optical_beforeSK}
\end{eqnarray} 
where $\int_k = \int_{-\pi} ^{\pi} \frac{dk}{2\pi}$ and we are using the rescaled fields. 

Let us also describe a useful shortcut to obtain $J_{\rm th}$. Let us imagine that we take the continuum limit and consider the Energy-Momentum (EM) tensor:
\begin{eqnarray}
T_{\mu \nu} = \frac{1}{2}\left(\frac{\partial\mathcal{L}}{\partial\left(\partial_{\nu}\phi\right)}\partial_{\mu}\phi+\partial_{\mu}\phi\frac{\partial\mathcal{L}}{\partial\left(\partial_{\nu}\phi\right)}\right)-\delta_{\mu\nu}\mathcal{L}
\end{eqnarray}
with $\mu,\nu=t,r$ and $\mathcal{L}$ is the Lagrangian. Note that $H=\int_r T_{00}$ and $ J_{\rm th} = \int_r T_{01}$. 

In the ``continuum limit'', we may replace the discrete derivatives in (\ref{H_LM}) by derivatives: $\phi_{r+1} - \phi_r \to \partial_r \phi$. Then, for instance, 
\begin{eqnarray}
H_{{r},0}  &=& \sum_{i=1}^{N}  \frac{\pi_{i,{r}}^{2}}{2}+\frac{1}{2}\Omega_{i}^{2}\phi_{i,{r}}^{2}+\frac{1}{2}\Omega_{\text{d},i}^{2} \left(\partial_r\phi_{i,{r}}\right)^2,\nonumber \\
\mathcal{L}_{r,0} &=& \sum_{i=1}^{N}  \frac{\left(\partial_t\phi_{i,{r}}\right)^2}{2}-\frac{1}{2}\Omega_{i}^{2}\phi_{i,{r}}^{2}-\frac{1}{2}\Omega_{\text{d},i}^{2} \left(\partial_r\phi_{i,{r}}\right)^2.
\end{eqnarray}
We may use the EM tensor to derive the continuum $ J_{\rm th}$. To return to the discrete notation, we must replace $\partial_r\phi_r$ with its \textit{symmetrized} lattice derivative: $\partial_r\phi_r \mapsto \frac{1}{2}\left( \phi_{r+1}-\phi_{r-1} \right)$. This leads, again, to (\ref{eq:thermal_current_optical_beforeSK}).

\begin{figure}[t]
\centering

\includegraphics[width=\columnwidth]{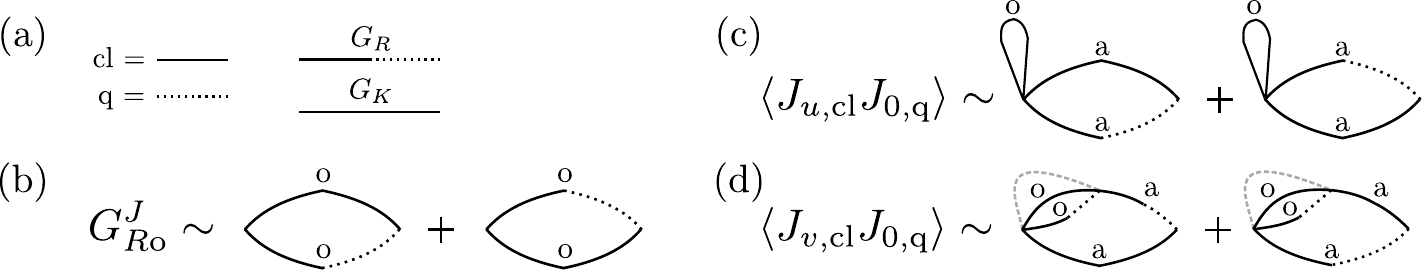}
\caption{{Thermal current in Keldysh space.} (a) Straight (dashed) lines correspond to cl(q) fields, giving the retarded/Keldysh components upon contractions, accordingly (see, e.g., \cite{kamenev_field_2011}). (b) Bare current correlator of optical phonons, corresponding to (\ref{G_RoJ}). (c) and (d) present examples of diagrams that contribute to the current correlator of acoustic phonons in Keldysh space. Dashed grey lines in (d) denote averaging over realizations of $v_{ijk}$.} 
\label{fig:KELDYSHdiagramsforkappa}
\end{figure}

Let us consider a single optical branch henceforth. To proceed we recall the Keldysh-rotated fields $\phi_{\rm cl/q} = \frac{1}{\sqrt{2}}\left(\phi_+ \pm \phi_-\right)$, where $\pm$ denotes the forward/backward contours in real-time, and cl/q are the classical/quantum components \cite{kamenev_field_2011} corresponding to straight and dashed lines in Fig.~\hyperref[fig:KELDYSHdiagramsforkappa]{\ref{fig:KELDYSHdiagramsforkappa}a}. These fields define the familiar Keldysh functions: $G_R(\omega) = i\langle \phi_{\rm cl} (\omega) \phi_{\rm q} (-\omega) \rangle $ and $G_K(\omega) = i\langle \phi_{\rm cl} (\omega) \phi_{\rm cl} (-\omega) \rangle $.  To obtain $\kappa$, we need to evaluate $G^J_R(\omega) = i\langle J_{\rm th,cl} (\omega) J_{\rm th,q} (-\omega) \rangle$. Using $\sqrt{2} J_{\rm cl/q} = J_+ \pm J_-$ (defined identically to the fields $\phi$) where $J_+ \sim \phi_+ \phi_+$ and similarly for $J_-$, we may insert the definitions above to obtain the bare current-current correlation function, given diagrammatically in Fig.~\hyperref[fig:KELDYSHdiagramsforkappa]{\ref{fig:KELDYSHdiagramsforkappa}b},
\begin{eqnarray}
-iG_{R}^{J}\left(\omega\right)=\left(\frac{\Omega_{\text{d}}^{2}}{4}\right)^{2}\int_{k}\int_{\nu}2\sin^{2}k\left(2\nu-\omega\right)^{2}\left(G_{K}\left(\nu,k\right)G_{R}\left(\omega-\nu,k\right)+G_{R}\left(\nu,k\right)G_{K}\left(\omega-\nu,k\right)\right).
\label{G_RoJ}
\end{eqnarray}
From here, we neglect the term $\propto \omega^2$, use the parity of the real and imaginary parts of $G_R(\omega)$ to null the term $\propto \nu \omega$ and expand the remaining term to leading order in $\omega$ to obtain (\ref{eq:kappa_o}). Since the thermal current is written as a sum $J_{\rm th} \sim \sum_i \phi_i \phi_i$, the generalization to multiple optical branches is straight forward, given below (\ref{eq:kappa_o}).

In the weakly dispersive limit, the corrections to the bare current correlator are suppressed by higher powers of $\Omega_{\rm d}/\overline{\Omega}_{\rm o} \ll 1$. To see this, we may write the Green's functions as $G\left(\omega,k\right) = G\left(\omega\right) + \delta G \left(\omega,k\right)$ where by construction $\delta G \left(\omega,k\right) \propto \Omega_{\rm d}^2/\overline{\Omega}^2_{\rm o}$ since we are expanding in powers of $\Omega_{\rm d}^2$. Vertex corrections will vanish at order $\mathcal{O}\left( \delta G^0 \right)$ as the $k$-integration now splits to two independent integrals: $\int_k \sin^2k \mapsto \int_{k,k'} \sin k \sin k' =0$. And indeed, non trivial orders in $\delta G$ are suppressed since $\delta G \left(\omega,k\right) \propto \Omega_{\rm d}^2/\overline{\Omega}^2_{\rm o} \ll 1$.

Consider the low-$T$ semiclassical regime, defined by $T\ll \Omega_v$. In this limit, the specific heat of a mode with energy $\nu$ is well-approximated by its harmonic expression: $c\left(\nu,\beta\right) \approx \frac{\left(\nu\beta\right)^2}{4}{\rm csch}^2\left(\frac{\beta\nu}{2}\right)$, which comes from the free-energy $\beta f = \ln \left( 2 \sinh \left( \frac{\beta \nu}{2}\right) \right)$. Then, we may write 
\begin{eqnarray} \label{schematic_deriv_of_kappa_o}
\kappa_{\rm o} &=& 2{\Omega_{\rm d}^{4}}\int_{\nu,k}\sin^{2}k\mathcal{A_{\rm o}}\left(\nu,k\right)^{2}c\left(\nu,\beta\right).
\end{eqnarray}
Using the weakly dispersive limit, we approximate $\mathcal{A}_{\rm o}\left(\nu,k\right) \approx \mathcal{A_{\rm o}}\left(\nu\right)$, such that
\begin{eqnarray}
\kappa_{\rm o} &=& {\Omega_{\rm d}^{4}}\int_{\nu}\mathcal{A_{\rm o}}\left(\nu\right)^{2}c\left(\nu,\beta\right).
\label{kappa_o_general}
\end{eqnarray}
Now, since $\tau_{\rm ph}$ is exponentially long at the semiclassical limit, the integrand above is concentrated around $\nu = \pm \overline{\Omega}_{\rm o}$, with width $\sim \tau_{\rm ph}^{-1}$ due to $\mathcal{A_{\rm o}}\left(\nu\right)^{2}$. Hence, we may approximate $\mathcal{A}_{\text{o}}\left(\nu\right)^{2}\sim\left(\delta\left(\nu-\overline{\Omega}_{\text{o}}\right)+\delta\left(\nu+\overline{\Omega}_{\text{o}}\right)\right)\tau_{\text{ph}}^{-1}\times\frac{1}{\overline{\Omega}_{\text{o}}^{2}\tau_{\text{ph}}^{-2}}$, in the spirit of (\ref{approx_G_R}), such that
\begin{eqnarray}
\kappa_{\rm o} &\approx & \overline{v}_{\rm o}^2 \tau_{\rm ph} c\left(\overline{\Omega}_{\rm o},\beta\right).
\end{eqnarray}

We may also consider the high-$T$ limit, defined by $T/\Omega_v \gg 1$. In (\ref{kappa_o_general}), we have that $c\sim 1$. Then, for example, by approximating $b\left(\omega,k\right) \approx 2 \omega/\tau_{\rm ph}$ and $a\left(\omega\right) \approx \overline{\Omega}_{\rm o}$ in (\ref{approx_G_R}) (as we did in \cite{tulipman_strongly_2020}), we may carry out the integration in (\ref{kappa_o_general}) to obtain $\kappa_{\rm o}\sim  \overline{v}_{\rm o}^2 \tau_{\rm ph}$, as before. Note that the approximated Green's function captures the scaling of $\kappa$ up to $\mathcal{O}(1)$ constants, whereas, in the actual computation of $\kappa$, there is an excellent agreement to the exact Boltzmann expression. 

%Consider (\ref{eq:kappa_o}) in the high-$T$ limit, where the support of $\mathcal{A}_{\rm o}$ is such that $\left(\beta\nu\right)^{2}\text{csch}^{2}\left(\frac{\nu\beta}{2}\right) \approx 4$ under the integral. We may use (\ref{spectral_function_approx_for_kappa}) such that  to approximate
%\begin{eqnarray} \label{schematic_deriv_of_kappa_o}
%\kappa &=& \frac{\Omega_{\rm d}^{4}}{2}\int_{\nu,k}\sin^{2}k\mathcal{A_{\rm o}}\left(\nu,k\right)^{2}\left(\beta\nu\right)^{2}\text{csch}^{2}\left(\frac{\nu\beta}{2}\right) \nonumber \\
%&\approx& \Omega_{\rm d}^{4}\int_{\nu}\mathcal{A_{\rm o}}\left( \nu \right)^{2} \nonumber \\
%&=& \frac{1}{4} \frac{\Omega_{\rm d}^4}{ a^2 b },
%\end{eqnarray}
%where, for optical phonons, $a\approx\overline{\Omega}_{\rm o}$, such that $\frac{\Omega_{\rm d}^{4}}{a^2}\approx\overline{v}_{\rm o}^2$. In addition, within the approximated Green's function, $b \sim \tau_{\rm ph}^{-1}$. Hence, $\kappa \approx \overline{v}_{\rm o}^2\tau_{\rm ph}$ for $T/\Omega_v \gg 1$.  

\subsubsection{Thermal current of acoustic phonons}

In the presence of acoustic phonons, the thermal current is modified by interactions. These extra terms capture the renormalized velocity of the acoustic modes. We will compute the thermal current related to the acoustic modes to linear order in $n_{\rm a}$ and $\Omega_{\rm d}^2$, as usual. Schematically, the current can be written as $J_{\rm th,a} = J_0 + J_v + J_u$ where $J_0,J_v$ and $J_u$ are related to the harmonic, cubic and quartic parts of the Hamiltonian, respectively. The current correlator is then given, schematically, by 
\begin{eqnarray}
\langle J_{\rm th,a}J_{\rm th,a} \rangle &=& \langle J_{0} J_{0} \rangle+\langle J_{u} J_{v} \rangle+\langle J_{v} J_{u} \rangle \nonumber \\
&+& \langle J_{v} J_{0} \rangle+\langle J_{0} J_{v} \rangle+\langle J_{v} J_{v} \rangle \nonumber \\
&+&\langle J_{u} J_{0} \rangle +\langle J_{0} J_{u} \rangle +\langle J_{u} J_{u} \rangle . 
\end{eqnarray}
The terms above correspond to the diagrams presented in Fig.~\ref{fig:thcurrent}. In all of the above we compute the disorder averaged correlators. Only terms with $J_v$ are contracted with interaction vertices. Insertions of interaction vertices in other terms are higher order in $n_{\rm a}$ and $\Omega_{\rm d}/\overline{\Omega}_{\rm o}$. Note also that in order to obtain $\kappa$ beyond the weakly dispersive limit, one must solve a Bethe-Salpeter-type equation for the current vertex. This is beyond the scope of this work. 

It is convenient to derive the thermal current operator using the shortcut we have previously introduced. Consider current associated with the quartic term, $J_{u}$. Before the Keldysh rotation, we have that 
\begin{eqnarray}
J_{u} = -\underbrace{\frac{u}{2}\sum_r \left( \frac{1}{N} \sum_{i}\tilde{\phi}_{i,r}^{2}\right)}_{A} \underbrace{ \sum_j \left( \frac{\phi_{j,r+1}-\phi_{j,r-1}}{2}\partial_{t}\phi_{j,r}1_{j\in I_{\rm a}} + \partial_{t}\phi_{j,r} \frac{\phi_{j,r+1}-\phi_{j,r-1}}{2}1_{j\in I_{\rm a}} \right)}_{B}.
\label{J_u}
\end{eqnarray}
Here the indicator function $1_{j\in I_{\rm a}} = 1$ if $j\in I_{\rm a}$ and zero otherwise. Namely, only acoustic phonons contribute to this term. In addition, we see that upon contracting $J_{u}$ with itself or with $J_0$, we will get a factor of $n_{\rm a}$ from the summation over $j$. Notice that $B$ in (\ref{J_u}) has the structure of the quadratic current operator, while $A$ is essentially the quartic renormalization to the bare acoustic frequency. Indeed, for example, the $\langle J_u J_0 \rangle$ term, presented in Fig.~\hyperref[fig:KELDYSHdiagramsforkappa]{\ref{fig:KELDYSHdiagramsforkappa}c}, is given by 
\begin{eqnarray}
\frac{-iu\Omega_{\text{d}}^{2}G_{K{\rm o}}(t=0,r=0)/2}{4^2}\int_{k}\int_{\nu}2\sin^{2}k\left(2\nu-\omega\right)^{2}\left(G_{K{\rm a}}\left(\nu,k\right)G_{R{\rm a}}\left(\omega-\nu,k\right)+G_{R{\rm a}}\left(\nu,k\right)G_{K{\rm a}}\left(\omega-\nu,k\right)\right).
\end{eqnarray}

Similarly, the current associated with the cubic term, before Keldysh rotation, is given by ($d=1$ for simplicity)
\begin{eqnarray}
J_{v}&=&\frac{1}{2N}\sum_{ijk,r}v_{ijk}\left(\left(\partial_{t}\phi_{i,r}1_{i\in I_{\text{a}}}\right)\tilde{\phi}_{j,r}\tilde{\phi}_{k,r}+\tilde{\phi}_{j,r}\tilde{\phi}_{k,r}\left(\partial_{t}\phi_{i,r}1_{i\in I_{\text{a}}}\right)\right)+\left(i\leftrightarrow j\right)+\left(i\leftrightarrow k\right).
\end{eqnarray}
Notice that $J_v$ contains no spatial derivatives in the leading order in $n_{\rm a}$. The contributions from this part of the current operator require contractions with interaction vertices. For example, the contribution of $\langle J_v J_0 \rangle$ to $\kappa$, corresponding to the Keldysh-space diagrams in Fig.~\hyperref[fig:KELDYSHdiagramsforkappa]{\ref{fig:KELDYSHdiagramsforkappa}d}, is given by
\begin{eqnarray}
\int_{\boldsymbol{k},\nu}\frac{{\Omega}_{\text{a}}^{2}{\rm Re}\Pi_{R{\rm o}}\left(\nu\right)}{2}\epsilon\left(\boldsymbol{k}\right)\mathcal{A}_{\rm a}^{2}\left(\nu,\boldsymbol{k}\right)\left(\nu\beta\right)^{2}\text{csch}^{2}\left(\frac{\nu\beta}{2}\right),
\end{eqnarray}
where $\Pi_{R{\rm o},v}(\nu) \equiv i v^2 n_{\rm o}^2 \int_{\nu'} G_{R{\rm o}}(\nu-\nu')G_{K{\rm o}}(\nu')$ is the self-energy of the optical phonons due to the cubic term. The remaining diagrams in (\ref{fig:thcurrent}) are similar to the examples above, giving in total (\ref{eq:kappa_a}). 

%given by 
%\begin{eqnarray}
%\frac{\Omega_{\text{d}}^{2}}{4^2}\int_{k}\int_{\nu}\Pi_{R{\rm o},v}(\nu)2\sin^{2}k\left(2\nu-\omega\right)^{2}\left(G_{K{\rm a}}\left(\nu,k\right)G_{R{\rm a}}\left(\omega-\nu,k\right)+G_{R{\rm a}}\left(\nu,k\right)G_{K{\rm a}}\left(\omega-\nu,k\right)\right), 
%\end{eqnarray}
%where $\Pi_{R{\rm o},v}(\nu) \equiv i v^2 n_{\rm o}^2 \int_{\nu'} G_{R{\rm o}}(\nu-\nu')G_{K{\rm o}}(\nu')$ is the self-energy of the optical phonons due to the cubic term. Note that only the real-part of $\Pi_{R{\rm o},v}(\nu)$ enters $\kappa$ since the term proportional to ${\rm Im}\Pi_{R{\rm o},v}(\nu)$ is subleading in $\omega$. Other diagrams in (\ref{fig:thcurrent}) are similar to the examples above, giving in total (\ref{eq:kappa_a}). 

Consider the contribution of the long-wavelength modes to the thermal conductivity. Similarly to (\ref{schematic_deriv_of_kappa_o}), we may approximate the contribution of these modes as
\begin{eqnarray}
\kappa_{\text{a},\text{long-wavelength}}&=&\int_{\left|\boldsymbol{k}\right|<k_{*},\nu}\frac{\overline{\Omega}_{\text{a}}^{4}\left(\nu\right)}{2}\epsilon\left(\boldsymbol{k}\right)\mathcal{A}_{\rm a}^{2}\left(\nu,\boldsymbol{k}\right)\left(\nu\beta\right)^{2}\text{csch}^{2}\left(\frac{\nu\beta}{2}\right) \nonumber\\
&\approx&2\overline{v}_{s}^{4}\int_{\left|\boldsymbol{k}\right|<k_{*}}k^{2}\int_{\nu}\mathcal{A}_{\rm a}^{2}\left(\nu,\boldsymbol{k}\right)\nonumber\\
&\sim&\overline{v}_{s}^{4}\int_{\left|\boldsymbol{k}\right|<k_{*}}\text{d}^{d}kk^{2}\frac{1}{b_{\text{a}}a_{\text{a}}^{2}}\\
&\sim&\overline{v}_{s}^{4}\int_{\left|\boldsymbol{k}\right|<k_{*}}k^{d+1}\frac{1}{\gamma\overline{v}_{s}^{2}k^{4}}\nonumber\\
&\sim&\overline{v}_{s}^{2}\gamma^{-1}\int_{k<k_{*}}k^{d-3}.
\end{eqnarray}
Here $k_*$ is some upper cutoff and we used the same approximations as below (\ref{a_lifetime}). We again see that the contribution of these modes diverge with the system size for $d\leq2$.

%Let us proceed to $J_v$, the current associated with the cubic term. Before the Keldysh rotation, we have
%\begin{eqnarray}
%J_{{\rm th},v} = - 
%\end{eqnarray}

%To proceed we denote the current correlator related to the acoustic phonon by $G_{R{\rm a}}^J\left(\omega \right)$. We preform the Keldysh rotation using the definition above and  we use the definitions above to obtain that  
%\begin{eqnarray}
%-iG_{R{\rm a}}^J\left(\omega \right) &= & n_{\rm a} \left(\frac{\widetilde{\Omega}^2}{4}\right)^{2}\int_{k}\int_{\nu}2\sin^{2}k\left(2\nu-\omega\right)^{2}\left(G_{K{\rm a}}\left(\nu,k\right)G_{R{\rm a}}\left(\omega-\nu,k\right)+G_{R{\rm a}}\left(\nu,k\right)G_{K{\rm a}}\left(\omega-\nu,k\right)\right), \\
%\widetilde{\Omega}^2& \equiv &  \Omega_{\rm a}^2 -\frac{iu}{2}\widetilde{\mathcal{G}}_{K}\left(t=0,r=0\right) , 
%\end{eqnarray}
%where this 

\subsection{Chaos}

Here we supply some extra details on our computations in Sec.~\ref{sec:chaos}. In particular, we give expressions for the matrices $A$ and $B$ with which we computed the chaos diffusivity and butterfly velocity. The expressions are presented for the case of a single optical and a single acoustic branches. The derivation of the corrections for systems with multiple optical branches are given by the `${\rm oo}$' components of the expressions below, with the relevant optical branches. Note that the (non-symmetric) matrices $A$ and $B$ and the eigenvectors of the retarded kernel are real, ensuring that $D_L$ is real. 

Consider $A$, the matrix related to the correction in $\lambda_L$ due to $|k_+|>0$. Note that the only $\lambda_L$ dependence in $K$ is coming from the exponent in the functions $h_{\lambda_L}^{ab}$ [Eq. \eqref{eq:eigen_value_eqn_for_f_AO}]:
\begin{eqnarray}
\partial_{\lambda_L} h_{\lambda_L}^{ab} = \left( \frac{t}{2} - \bar{t} \right) h_{\lambda_L}^{ab} \equiv \tilde{h}_{\lambda_L}^{ab}.
\end{eqnarray}
$A$ is then given by $K$ with replacing $h_{\lambda_L}^{ab}$ by $\tilde{h}_{\lambda_L}^{ab}$. 

$B$ is the matrix related to the leading behavior of $K$ with respect to $k_+$. The components of $B$ are obtained expanding $h_{\lambda_L}^{ab}$ and $s_{ab}$ to second order in $k_+$. Note that linear terms in $k_+$ vanish due to the integration over $k_-$. We obtain that 
\begin{eqnarray}
B=2v^{2}\int_{k_{-}}\widetilde{\mathcal{G}}_{W}\left(t'\right)\begin{pmatrix}n_{\text{a}}b_{\text{aa}}\left(\bar{t}-t,k_{-}\right) & n_{\text{o}}b_{\text{ao}}\left(\bar{t}-t,k_{-}\right)\\
n_{\text{a}}b_{\text{oa}}\left(\bar{t}-t,k_{-}\right) & n_{\text{o}}b_{\text{oo}}\left(\bar{t}-t,k_{-}\right)
\end{pmatrix}
\end{eqnarray}
where the components of $B$ are given by 
\begin{eqnarray}
b_{\text{oo}}\left(t,k_{-}\right)&=&\int_{\bar{t}}\left(I_{\text{o,}2}\left(t,\bar{t},k_{-}\right)-I_{\text{o,}1}\left(t,\bar{t},k_{-}\right)\right)\\
b_{\text{oa}}\left(t,k_{-}\right)&=&2\pi b_{\text{oo}}\left(t,k_{-}\right)-5\pi h_{\lambda_{L}}^{\text{oo}}\left(t,k_{+}=0,k_{-}\right)\\
b_{\text{ao}}\left(t,k_{-}\right)&=& 4\int_{\bar{t}}\sin^{2}\left(\frac{k_{-}}{4}\right)\left(I_{\text{a},2}\left(t,\bar{t},k_{-}\right)-I_{\text{a},1}\left(t,\bar{t},k_{-}\right)\right)\nonumber\\
&&+h_{\lambda_{L}}^{\text{ao}}\left(t,k_{+}=0,k_{-}\right) \\
b_{\text{aa}}\left(t,k_{-}\right)&=& 8\pi\int_{\bar{t}}\sin^{2}\left(\frac{k_{-}}{4}\right)\left(I_{\text{a},2}\left(t,\bar{t},k_{-}\right)-I_{\text{a},1}\left(t,\bar{t},k_{-}\right)\right) \nonumber \\
&&+4\pi\left(\frac{1}{2}-5\sin^{2}\left(\frac{k_{-}}{4}\right)\right)h_{\lambda_{L}}^{\text{aa}}\left(t,k_{+}=0,k_{-}\right)
\end{eqnarray}
with 
\begin{eqnarray}
I_{\text{o,}1}\left(t,\bar{t},k_{-}\right)&=&g_{R\text{o},1}\left(\bar{t},\frac{k_{-}}{2}\right)g_{A\text{o},1}\left(t-\bar{t},\frac{k_{-}}{2}\right)\\
I_{\text{o,}2}\left(t,\bar{t},k_{-}\right)&=&g_{R\text{o},2}\left(\bar{t},\frac{k_{-}}{2}\right)g_{A\text{o}}\left(t-\bar{t},\frac{k_{-}}{2}\right)+g_{R\text{o}}\left(\bar{t},\frac{k_{-}}{2}\right)g_{A\text{o},2}\left(t-\bar{t},\frac{k_{-}}{2}\right)\\
I_{\text{a},1}\left(t,\bar{t},k_{-}\right)&=&g_{R\text{a},1}\left(\bar{t},\frac{k_{-}}{2}\right)g_{A\text{a},1}\left(t-\bar{t},\frac{k_{-}}{2}\right)\\
I_{\text{a},2}\left(t,\bar{t},k_{-}\right)&=&g_{R\text{a},2}\left(\bar{t},\frac{k_{-}}{2}\right)g_{A\text{a}}\left(t-\bar{t},\frac{k_{-}}{2}\right)+g_{R\text{a}}\left(\bar{t},\frac{k_{-}}{2}\right)g_{A\text{a},2}\left(t-\bar{t},\frac{k_{-}}{2}\right)
\end{eqnarray}
where we denoted
\begin{eqnarray}
g_{Rb}\left(t,\frac{k_{-}}{2}\right)&\equiv& e^{-\lambda_{L}\frac{t}{2}}G_{Rb}\left(t,\frac{k_{-}}{2}\right), \quad b={\rm a,o}\\
g_{R\text{o},1}\left(t,\frac{k_{-}}{2}\right)&\equiv&-2\Omega_{\text{d}}^{2}\sin\left(\frac{k_{-}}{2}\right)e^{-\lambda_{L}\frac{t}{2}}\int_{\omega}e^{i\omega t}G_{R{\rm o}}\left(\omega,\frac{k_{-}}{2}\right)^{2}\\
g_{R\text{o},2}\left(t,\frac{k_{-}}{2}\right)&\equiv&4\Omega_{\text{d}}^{4}\sin^{2}\left(\frac{k_{-}}{2}\right)e^{-\lambda_{L}\frac{t}{2}}\int_{\omega}e^{i\omega t}G_{R{\rm o}}\left(\omega,\frac{k_{-}}{2}\right)^{3}\\
g_{R\text{a},1}\left(t,\frac{k_{-}}{2}\right)&\equiv&-2\sin\left(\frac{k_{-}}{2}\right)e^{-\lambda_{L}\frac{t}{2}}\int_{\omega}e^{i\omega t}\left(\Omega_{\text{a}}^{2}+\Pi_{R\text{o}}\left(\omega\right)\right)G_{R\text{a}}\left(\omega,\frac{k_{-}}{2}\right)^{2}\\
g_{R\text{a},2}\left(t,\frac{k_{-}}{2}\right)&\equiv&4\sin^{2}\left(\frac{k_{-}}{2}\right)e^{-\lambda_{L}\frac{t}{2}}\int_{\omega}e^{i\omega t}\left(\Omega_{\text{a}}^{2}+\Pi_{R\text{o}}\left(\omega\right)\right)^{2}G_{R\text{a}}\left(\omega,\frac{k_{-}}{2}\right)^{3}.
\end{eqnarray}

Note that $v_B=\sqrt{D_L\lambda_L}$ is determined by the two velocity scales in $I_{\rm a,o}$. Terms associated with the optical modes are proportional to $n_{\rm o}\Omega_{\rm d}^4$, corresponding to $n_{\rm o}\overline{v}_{\rm o}^2$, while terms associated with the acoustic modes are proportional to $n_{\rm a}\left( \Omega_{\rm a}^2 + \Pi_{R{\rm o}}(\omega)\right)^2$, corresponding to $n_{\rm a}\overline{v}_s^2$ (similarly to the case of thermal transport).

\section{Generalization to higher dimensions}
\label{high_d}

For $n_{\rm a}=0$, the generalization for $d>1$ is straightforward: In the harmonic part, we sum over spatial directions: 
\begin{eqnarray}
\left(\phi_{\rm }(r+1) - \phi_{\rm}(r)\right)^2 \to \sum_{\boldsymbol{\delta}} \left(\phi_{\rm }(\boldsymbol{r}+\boldsymbol{\delta}) - \phi_{\rm }(\boldsymbol{r})\right)^2.
\end{eqnarray}
One can also consider anisotropic couplings. 

Upon introducing acoustic branches, we need to make the following modifications. The harmonic part is modified similarly to optical phonons. In addition, we set the couplings containing different (discrete) derivatives of the acoustic modes to be uncorrelated. The random cubic coupling is modified to be  $v_{i(l)j(m)k(n)}$ where $l,m,n=x,y,z$ denotes spatial directions. $v_{i(l)j(m)k(n)}$ satisfies 
\begin{eqnarray}
\overline{v_{i(l)j(m)k(n)}v_{i'(l')j'(m')k'(n')}}=2v^2 \delta_{ii'}(1_{i\in I_{\rm a}}\delta_{ll'}+ 1_{i\in I_{\rm o}})(j\leftrightarrow i,m\leftrightarrow l)(k\leftrightarrow i,n\leftrightarrow l). 
\end{eqnarray}
Here $1_{i\in I_{\rm a}}$ is the indicator function: $1_{i\in I_{\rm a}} = 1$ if ${i\in I_{\rm a}}$ and zero otherwise. Essentially, the disorder remains the same for optical phonons and makes sure that acoustic phonons, inserted as discrete derivative, are only correlated when acting in the same spatial direction. Then, a summation over spatial directions is added to the flavor summation for the cubic term: \begin{eqnarray}
\sum_{ijk} \to \sum_{i(l)j(m)k(n)} \equiv \sum_{ijk} \left[ \left(1_{i\in I_{\rm a}} \sum_l + 1_{i\in I_{\rm o}} \right) \left( j\leftrightarrow i,m\leftrightarrow l \right)\left(k\leftrightarrow i,n\leftrightarrow l\right) \right].
\end{eqnarray}
The quartic term is modified similarly:
\begin{eqnarray}
\sum_{ij} \to \sum_{i(l)j(m)},
\end{eqnarray}
where $\sum_{ij}$ refers to $\sum_{ij}\phi_{i,r}^2\phi_{j,r}^2$ in (\ref{H_LM}).
These definitions simply imply that terms that contained, for example, $\sin^2\left(\frac{k}{2}\right)$ in the $d=1$ case, will contain $\sin^2\left( \frac{k_x}{2}\right)+\sin^2\left( \frac{k_y}{2}\right)+\sin^2\left( \frac{k_z}{2}\right)$ in the $d=3$ case, and similarly for $d=2$. 

\section{Numerics}
\label{Num_app}

The numerical solution of the SPEs in is done similarly to \cite{tulipman_strongly_2020}. In real-time, we account for the weakly dispersive limit by computing the momentum-integrated Green's function $G(\omega) = \int_k G(\omega,k)$ with every updating step, where the acoustic modes are multiplied by $4\sin^2\left(\frac{k}{2}\right)$  according to (\ref{eq:Keldysh_sc_eqs_AO_model}). The grid in frequency space is typically taken to be of size $2^{16}$, while the $k$-grid is typically taken to be $\sim 40$. Finite size effects are found to be negligible at the weakly dispersive limit, giving an error of roughly 5\%. 

To compute the chaotic properties of the system, we diagonalize a coarse-grained version of the retarded kernel in real-time, where the time data points are uniformly sampled such that the typical dimension of the retarded kernel is $\sim 2^{10}$. This is done \textit{after} we construct the relevant functions ($G_W,g_R,g_A$) using a much finer grid (as above). The dependence on the maximal time cutoff and coarse-graining gives an error of roughly 5\%.

\end{document}